\documentclass[11pt]{article}
\pdfoutput=1
\usepackage[dvipsnames]{xcolor}
\usepackage{amsmath,amsfonts,amssymb,amsthm,amssymb,bbm,bm}
\usepackage{pdfsync}
\usepackage{graphicx,import}
\usepackage[latin1]{inputenc}
\usepackage{hyperref}
\usepackage{empheq}
\usepackage[all]{xy}
\usepackage{stmaryrd}
\usepackage{rotating}
\usepackage{color}  
\usepackage{slashed}
\usepackage{cancel}
\usepackage{array, makecell} %
\usepackage{comment}
\usepackage{tikz-cd}
\usepackage{hhline}
\usepackage{comment}
\usepackage{mathrsfs}
\usepackage{enumitem}
\usepackage{MnSymbol}
\usepackage{cite}
\usepackage{tikz}
\usetikzlibrary{arrows.meta,decorations.markings,decorations.pathmorphing,calc}
\pgfdeclarelayer{particlelayer}
\pgfsetlayers{particlelayer,main}
\usepackage{subcaption}

\newcommand{\p}{\partial}

\newcommand{\bit}{\begin{itemize}}
\newcommand{\eit}{\end{itemize}}
\newcommand{\bd}{\begin{description}}
\newcommand{\ed}{\end{description}}

\newcommand{\bc}{\begin{center}}
\newcommand{\ec}{\end{center}}

\newcommand{\N}{{\mathbb N}}

\newcommand{\be}{\begin{equation}}
\newcommand{\ee}{\end{equation}}
\newcommand{\bea}{\begin{eqnarray}}
\newcommand{\eea}{\end{eqnarray}}
\newcommand{\bs}{\begin{subequations}}
\newcommand{\es}{\end{subequations}}

\newcommand{\sgn}{\mathrm{sgn}}

\newcommand{\f}{\frac}

\def\p{\partial}

\def\b{\beta}

\def\k{\kappa}

\def\r{\rho}

\def\L{\Lambda}

\newcommand{\pin}{p_i^{\rm in}}
\newcommand{\pout}{p_i^{\rm out}}
\newcommand{\qin}{Q_i^{\rm in}}
\newcommand{\qout}{Q_i^{\rm out}}

\numberwithin{equation}{section}

\begin{document}

\begin{titlepage}

\unitlength = 1mm
\ \\
\vskip 2cm
\begin{center}

{\LARGE \textsc{On memory effects in odd spacetime dimensions}}
\date{}

\vspace{0.9cm}
Ana-Maria Raclariu and Binsong Xiong
\vspace{0.7cm}\\
\textit{Department of Mathematics, King's College London, Strand, London,\vspace{5pt}\\
WC2R 2LS, United Kingdom} 
\vspace{30pt}

\begin{abstract}
\vspace{15pt}

We establish the existence of a new electromagnetic memory effect in all odd spacetime dimensions. This effect is a universal property of electromagnetic fields sourced by charged particles of constant momenta, appears at the Coulombic order and is directly related to the Weinberg soft theorem. We provide a unified derivation of this effect in all higher dimensions by evaluating the Maxwell constraint on a null plane at large, finite distance. We observe that in odd spacetime dimensions, the lightray integral computing the memory effect does not commute with the large-$r$ expansion.  Our results necessarily assume that the lightray integral is taken prior to the $r \rightarrow \infty$ limit. The effect vanishes in the opposite order of limits, in agreement with previous results. We compute the Coulombic order electromagnetic memory effect for both massless and massive sources in all higher dimensions, explain its relation to the Fourier transform of the leading soft photon theorem and present a protocol for its measurement in a finite time parametrically larger than the distance of a detector from the sources. We conclude with comments on the generalization to gravity.

\end{abstract}
\vspace{0.5cm}
\end{center}
\vspace{0.8cm}

\end{titlepage}
\tableofcontents

\newpage

\section{Introduction}

Accelerated charged particles emit electromagnetic radiation. In $3+1$-dimensional (4d)  Minkowski spacetime, the relation between radiation and sources is dictated by the Maxwell constraint 
\begin{equation}
\label{eq:constraint}
   n^{\nu} \p^{\mu} F_{\mu\nu} = n^{\nu} J_{\nu},
\end{equation}
where $n$ is the normal to a codimension-1 surface $\Sigma$ in spacetime.
Taking $\Sigma$ to a lightsheet near future time-like infinity $\mathcal{I}^+$, \eqref{eq:constraint} becomes\footnote{In this paper, we will work in lightcone coordinates in which the metric takes the form $ds^2 = -du dr + r^2 dy_{\perp}^2.$}
\begin{equation}
\label{eq:maxwell-constr-intro}
   2 \p_u  E_r^{(2)} - D\cdot E_{\perp}^{(0)} = -J_u^{(2)},
\end{equation}
where $u$ and $r$ are lightcone coordinates, $D$ is the covariant derivative on the unit $2$-sphere and $E_r^{(2)}, E_{\perp}^{(0)}$ are the leading components of the Coulomb and radiative fields in a large-$r$ expansion. Integrating \eqref{eq:maxwell-constr-intro} over the lightsheet or, more generally, \eqref{eq:constraint} over a timelike tube results in a flux-balance law relating the difference in the charge on two cuts of tube to the flux in between.  On the other hand, integrating \eqref{eq:constraint} over a spacelike surface $\Sigma$ results in the well-known Gauss law. In general, Maxwell's equations and the associated constraints carry strictly more information. 

For instance, a notable  consequence of \eqref{eq:maxwell-constr-intro} is that generic radiative fields carry electromagnetic memory \cite{Bieri:2013hqa},
\begin{equation}
\label{eq:em-mem}
    \mathcal{M}_A(x_{\perp}) \equiv \int_{-\infty}^{\infty} du E^{(0)}_{\perp, A}(u, x_{\perp}).
\end{equation}
The Maxwell constraint then implies that the memory effect is sourced by a non-vanishing change in the Coulomb field and/or the flux, namely 
\begin{equation}
    D^A \mathcal{M}_A = 2\Delta E_r^{(2)} + \int_{-\infty}^{\infty} du J_u^{(2)}.
\end{equation}
This effect is gauge invariant, and therefore observable. Indeed, the Lorentz force law implies that a probe particle of charge $q$ and mass $m$ will experience a velocity kick when exposed to radiation
\begin{equation}
\label{eq:vel-kick}
    \Delta v_{\perp}^A = \frac{q}{m} \mathcal{M}^A. 
\end{equation}

The electromagnetic memory effect admits a gravitational analog discovered much earlier \cite{Zeldovich:1974gvh,Christodoulou,Thorne}, whose origin is the gravity analog of \eqref{eq:maxwell-constr-intro}, namely the Hamiltonian constraint. Its definition is analogous to \eqref{eq:em-mem}, where $E_{\perp}$ is replaced by the Bondi news tensor (or the radiative Weyl tensor component $\Psi_4$), while the geodesic deviation equation implies that this imprints a net relative displacement of probe particles, instead of a velocity kick. 

More recently, these memory effects were shown to be the spacetime counterparts of the leading Weinberg soft theorem \cite{Weinberg:1965nx} which governs the low-energy limit of a scattering amplitude involving at least a massless particle, such as a photon or graviton \cite{Strominger:2014pwa,Pasterski:2015zua,Pasterski:2015tva}. In QED, the stationary phase approximation relates the momentum space zero mode of the electromagnetic field to a soft photon insertion, 
\begin{equation}
\label{eq:field-phot}
  \int du  D \cdot E_{\perp}^{(0)}(u, x_{\perp}) \propto \lim_{\omega \rightarrow 0}\omega D^A\left( a_{A}(\omega, x_{\perp}) + a^{\dagger}_{A}(\omega, x_{\perp}) \right) .
\end{equation}
 Equivalently, the Fourier transform of the Weinberg soft pole implies the radiative gauge field is proportional to a step function in retarded time, and therefore carries memory. On the other hand, the electromagnetic constraint \eqref{eq:maxwell-constr-intro} identifies the memory operator \eqref{eq:em-mem} with the soft component of the leading large-gauge charge, while the conservation of this large-gauge charge can be shown to imply the soft theorem \cite{He:2014cra}. The equivalence between memory effects, soft theorems and asymptotic symmetries has been widely established to be a universal property of theories with massless particles \cite{He:2014laa,Strominger:2014pwa,He:2015zea,Pasterski:2015tva,Mao:2017wvx,Campiglia:2017dpg, Mao:2017axa, Pate:2017fgt,Pate:2017vwa,Hamada:2017atr} (for a review, see \cite{Strominger:2017zoo}), and has to some extent been shown to generalize to subleading orders in the soft expansion \cite{Kapec:2014opa,Lysov:2014csa,Campiglia:2014yka,Campiglia:2016hvg,Campiglia:2016jdj,Laddha:2017vfh, Hirai:2018ijc,Campiglia:2018dyi,Hamada:2018vrw,Freidel:2021dfs} including logarithmic and loop corrections \cite{Laddha:2018vbn,Sahoo:2018lxl,Campiglia:2019wxe,AtulBhatkar:2019vcb,Sen:2024qzb,Choi:2024ygx,Choi:2024mac,Karan:2026jzs}.

In this paper we will be interested in the generalization of these ideas to higher-dimensional spacetimes. The same story was shown to carry through, subject to appropriate modifications that we will review shortly, for all even-dimensional spacetimes in both electromagnetism \cite{Mao:2017axa, He:2019jjk} and gravity \cite{Pate:2017fgt}. However, the generalization to odd-dimensional spacetimes  remained only partially understood. Although a relation between soft theorems and asymptotic symmetries was established in QED with massless particles in \cite{He:2019jjk}, the role of memory effects remained elusive. At the core of this puzzle lies the absence of the standard, velocity kick electromagnetic memory for all odd $d \geq 5$, including at the Coulombic order\footnote{In $d = 3$ there is an infinite momentum memory effect \cite{Satishchandran:2017pek}.} \cite{Satishchandran:2017pek, Satishchandran:2019pyc, He:2019jjk}. In this paper, we will show that, a Coulombic order memory effect in fact exists across all odd spacetime dimensions and bears a universal relation to soft theorems and the asymptotic constraints, just as in even-dimensional spacetimes and in particular $d = 4$. 

As we will explain, the key to establishing this result is the  observation that, in general, the asymptotic expansions employed in  previous related works fail to commute with the integral transform \eqref{eq:em-mem} that computes the memory. 
A non-commutativity of similar operations has been encountered in a wide range of other physical applications, notably in studies of lightray and detector operators in quantum field theory \cite{Kologlu:2019mfz,Caron-Huot:2022eqs,Jackson:2026eml}. These issues have the same mathematical origin since \eqref{eq:em-mem} are a special class of detector operators, see also \cite{Hu:2022txx,Hu:2023geb,Gonzalez:2025ene,Moult:2025njc,Oertel:2026wsm}.  To the extent of our knowledge, the relevance of this observation for electromagnetic and gravitational memory in odd spacetime dimensions has not been discussed elsewhere. The main result that we will establish in this work is that the Coulombic memory effect is present in all higher dimensions, for both massless and massive sources moving along straight trajectories, as long as the time integral of the Maxwell constraint on a codimension-1 surface at large, finite distance $r$ is evaluated \textit{before} taking the limit $r \rightarrow \infty$. 

We next briefly review the argument for the vanishing of the memory effect in odd dimensions \cite{Satishchandran:2017pek, Campoleoni:2017qot,Satishchandran:2019pyc}. This is puzzling from the perspective of the Weinberg soft photon theorem, which universally exists in all higher dimensions. Drawing from the $d = 4$ infrared equivalences described before, it is natural to expect that a universal memory effect exists in all higher dimensions. In even-dimensional spacetimes, such an effect indeed appears at the Coulombic order, namely for $F_{Au} \sim \mathcal{O}(r^{4-d})$. One can see this by considering the large-$r$ expansion of the radiative field, 
\begin{equation}
    E_{\perp, A}^{}(x) \propto r \int d^{d-2}\hat{q}\int d\omega \omega^{d - 2}\left(e^{i\omega \hat{q}\cdot x} a_A(\omega,\hat{q}) - e^{-i\omega \hat{q}\cdot x} a^{\dagger}_A(\omega,\hat{q})\right).
\end{equation}
The memory effect is present provided the powers of $\omega$ in the integrand, including those from the measure, cancel. At large $r$, the integral localizes around $\omega \sim 0$ where $a_{A}(\omega, \hat{q}) \sim \omega^{-1}$ is governed by the leading soft theorem, while the exponential supplies a power series in $(\omega r)^{-1}$ whose leading term is $\propto (\omega r)^{-\frac{d - 2}{2}}$. For even $d$,  we easily conclude that the memory effect appears in the Coulombic component of the radiative field $E_{\perp, A}^{(d - 4)}$. On the other hand, for odd $d$, the large-$r$ expansion of the integrand contains no $\mathcal{O}(\omega^0)$ terms and therefore, the memory effect vanishes order by order \cite{Mao:2017axa, Mao:2017wvx}. One arrives at a similar conclusion by considering the large-$r$ expansion of the Maxwell constraint \eqref{eq:maxwell-constr-intro} in higher dimensions \cite{Campoleoni:2017qot,Pate:2017fgt}, as reviewed in Section \ref{sec:no-memory}. Note that in these derivations, the large-$r$ expansion is performed before computing the zero-mode or the memory integral in \eqref{eq:em-mem}. 

Physically motivated by the fact that observables such as the memory effect are in practice measured at finite distance and in finite time, we consider \textit{exact} solutions to the Maxwell equations sourced by point particles moving at constant velocity and evaluate the associated Maxwell constraint \eqref{eq:maxwell-constr-intro} on a lightsheet at large, finite distance. Our main result is that in odd spacetime dimensions, the light-ray operator associated with the constraint evaluated on this surface
\begin{equation}
\label{eq:int-max}
    2\int du \p_u E_r - \int du \p_{\perp} \cdot E_{\perp} = -\int du J_u
\end{equation}
relates the memory operator defined by 
\begin{equation}
\label{eq:Coul-mem}
    \mathcal{M}(x_{\perp}) = \lim_{r \rightarrow \infty} r^{d - 4} \int du \p_{\perp} \cdot E_{\perp}(u,r, x_{\perp})
\end{equation}
to the particle flux through the lightsheet. Importantly, while $E_{\perp}$ contains no integer powers of $r^{-1}$, and in particular no Coulombic order term order by order in the large-$r$ expansion, its time integral does. We show in Section \ref{sec:odd-mem} that outgoing massless charged particles of constant momenta source a field schematically given by
\begin{equation}
 \lim_{r \rightarrow \infty} r^{d - 4} \int du \p_{\perp} \cdot E^i_{\perp}(u,r, x_{\perp}) \propto  \lim_{r\rightarrow \infty}
  r^{d - 4} \int_0^\infty du\, \p_{\perp}^A
   \frac{r^{3-\frac{d}{2}}u^{1-\frac d2}(x_{\perp}-x_{\perp}^i)_A}
   {u+r|x_{\perp}-x^i_{\perp}|^2} \propto \delta^{(d - 2)}(x_{\perp} - x_{\perp}^i).
\end{equation}
Furthermore, since the change in the radial electric field across the lightsheet vanishes, this effect is solely sourced by the non-vanishing flux through the lightsheet, in agreement with the Maxwell constraint \eqref{eq:int-max}. This result is in complete agreement with that obtained in even $d$. We emphasize however that for massless sources in even $d$, the lightray/memory integral commutes with the large-$r$ expansion, while in odd $d$ it does not. Physically, we expect the different orders of limits to give different results whenever the Maxwell constraint at $i^{\pm}$ is non-trivial. This is indeed the case in odd $d$, where even in the absence of massive particles, Huygens' principle is violated and electromagnetic waves disperse away from the lightcone \cite{Hadamard1923}. 
Evaluating the constraint at finite distance bypasses this problem, as it automatically captures the contribution from $i^+$, as illustrated in Figure \ref{fig:light-sheet-0}. 

All results can be generalized to massive sources which we establish in Section \ref{sec:massive}. Just as in the massless case, the memory integral and the large-$r$ expansions do not commute. For odd $d$, we show that the Coulombic memory \eqref{eq:Coul-mem} is again governed by the Green's function on $\mathbb{R}^{d-2}$, or equivalently the particle flux through the lightsheet, provided that the large-$r$ limit is taken last. The analysis in this case is technically more involved, since the field strength is expressed in terms of special functions, see \eqref{eq:frame-massive-odd} and \eqref{eq:phii}. Nevertheless, the memory integral at finite distance can be computed and, after regularization, the result takes the same form as for massless sources, see \eqref{eq:memory-massive-odd}. Along the way, we explain in detail the relation between Coulombic memory and the soft theorem, therefore establishing a universal equivalence between electromagnetic memory effects and soft theorems for all $d$. This discussion also allows us to generalize all results to theories of gravity in odd spacetime dimensions, which we outline in Section \ref{sec:disc}. 

In addition to filling a gap in our understanding of the IR sector of gauge theories and gravity in higher dimensions, we hope that our work will find applications to holographic dualities where the bulk theory is defined in odd $d$. One natural example is the KK reduction on $S^5$ of type IIB string theory on AdS$_5$ $\times S^5$ which contains gauge fields governed by the Maxwell action on AdS$_5$. Our results suggest the existence of both electromagnetic and gravitational memory effects in the limit where the AdS radius is large compared to the Planck scale. It would be particularly interesting to work these out and understand their holographic duals.

The paper is organized as follows. Section \ref{sec:review} reviews the status of the relation between asymptotic symmetries, soft theorems and memory effects in higher dimensions. In Section \ref{sec:as-st} we review the relation between higher-dimensional asymptotic symmetries and soft theorems established in \cite{He:2019jjk}. We show in Section \ref{sec:no-memory} that the solutions constructed therein support no memory  at both the radiative and the Coulombic orders, in agreement with prior results  \cite{Satishchandran:2017pek, Satishchandran:2019pyc}. In Section \ref{sec:em-gf} we review the derivation of the gauge fields sourced by massless particles of constant momenta in arbitrary dimensions and clarify the origin and regularization of divergences appearing in these solutions for odd $d$. We also discuss the higher-dimensional electromagnetic analogue of the Aichelburg-Sexl shockwaves \cite{Aichelburg1971} which will play an important role in our analysis of memory sourced by massless particles. In Section \ref{sec:Coulomb-mem} we evaluate the Maxwell constraint on a lightsheet at finite distance near the conformal boundary. For all $d \geq 5$, we show that the change in the radial Coulomb field vanishes and the constraint implies a universal relation between the Coulombic memory effect and the flux through the lightsheet. In Section \ref{eq:mc-st} we establish that the Ward identity associated with the Weinberg soft theorem is equivalent, to all orders in the large-$r$ expansion, with the Maxwell constraint. In Section \ref{sec:meas-mem} we show that the Coulombic memory effect in odd dimensions persists in practical, finite time measurements, provided the measurement time is parametrically larger than the distance from the sources. In Section \ref{sec:massive} we generalize our analysis to electromagnetic fields sourced by massive particles. In Section \ref{sec:disc} we comment on the relation of our results to higher-dimensional spin-1 conformal primary wavefunctions and explain how they generalize to linearized gravity. Several detailed derivations are relegated to Appendices \ref{app-large-r-exp}-\ref{app:massive-odd}.  

\section{The memory puzzle in odd dimensions}
\label{sec:review}

In this section we review the relation between asymptotic symmetries, soft theorems and memory effects in higher dimensions. The absence of a memory effect in odd spacetime dimensions \cite{Satishchandran:2017pek,Satishchandran:2019pyc} has so far precluded a complete understanding of this equivalence. A universal relation between soft theorems and asymptotic symmetries in electromagnetism in all higher dimensions was proposed in \cite{He:2019jjk}. After reviewing these results, we will show that the solutions constructed therein appear to carry memory only in even spacetime dimension, leaving the memory puzzle unresolved. Related constructions in even-dimensional Yang-Mills theory and gravity, including higher spins have been discussed in \cite{Pate:2017fgt,Campoleoni:2017qot,Laddha:2018rle,Aggarwal:2018ilg,Campoleoni:2019ptc,Freidel:2019ohg,Colferai:2020rte,Campoleoni:2020ejn,Fuentealba:2021yvo,He:2023bvv}.

Following \cite{He:2019jjk}, we let $d = 2 + 2m$ be the spacetime dimension and decompose the electromagnetic field into radiative ($F^R$) and Coulombic ($F^C$) components defined by
\begin{equation}
\label{eq:ME}
    \begin{split}
        \nabla^{\mu} F_{\mu\nu}^R &= 0, \\
        \nabla^{\mu} F_{\mu\nu}^C &= j_{\nu}.
    \end{split}
\end{equation}
We note that this split is not a priori (ie. in the absence of additional fall-off or boundary conditions) canonical, since $F^C$ is only defined up to solutions to the free Maxwell equations.
Working in light-cone coordinates in which the Minkowski metric takes the form
\begin{equation}
\label{eq:Mink0}
    ds^2 =  -du dr + r^2 dy_{\perp}^2,
\end{equation}
the large gauge charges are defined by 
\begin{equation}
\label{eq:lgc}
    Q^{\pm}_{\epsilon} \equiv \int_{\mathcal{I}^{\pm}_{\mp}} d^{2m} y_{\perp}~\epsilon (y_{\perp}) F_{ru}^{(2m)}(y_{\perp}).
\end{equation}
Here $F_{\mu\nu}^{(n)}$ are the $\mathcal{O}(r^{-n})$ components of the field strength in a large-$r$ expansion and $\epsilon(y)$ is an arbitrary function on $\mathbb{R}^{2m}$.
 The Maxwell equations reviewed in Appendix \ref{app-large-r-exp} then imply that 
\begin{equation}
\label{eq:rad-c}
    \begin{split}
        2\p_u F_{ru}^{(n + m + 1)} - D^A F_{Au}^{(m + n - 1)} &= 0, \quad  0 \leq n \leq m - 2, \\
        2\p_u F_{ru}^{(2m)} - D^A F_{Au}^{(2m - 2)} &= -j_u^{(2m)},
    \end{split}
\end{equation}
where we have set the charge unit $e = 1$.

Further assuming the following large-$r$ fall-offs for the radiative 
\begin{equation}
    F_{ru}^{R} \sim r^{-m - 1}, \quad F_{uA}^{R} \sim r^{1 - m},
\end{equation}
and Coulombic modes\footnote{We will see later on that the exact field strengths sourced by particles of constant momentum do not always obey these fall-offs.}
\begin{equation}
\label{eq:C-fall-off}
    F^{C}_{ru} \sim r^{-2m}, \quad F_{Au}^C \sim r^{1 - 2m},
\end{equation}
the second equation in \eqref{eq:rad-c} decomposes into 
\begin{equation}
\label{eq:constr}
\begin{split}2\p_u F_{ru}^{R(2m)} - D^A F_{Au}^{R(2m - 2)} &= 0,\\
    2 \p_u F_{ru}^{C(2m)} &= -j_u^{(2m)}.
    \end{split}
\end{equation}
As a result, the large gauge charges \eqref{eq:lgc} admit a split into radiative and Coulombic components
\begin{equation}
\label{eq:soft-hard}
    Q_{\epsilon}^{\pm} = \int_{\mathcal{I}^{\pm}_{\mp}} d^{2m} y_{\perp} ~\epsilon (y_{\perp}) F_{ru}^{R(2m)}(y_{\perp}) + \int_{\mathcal{I}^{\pm}_{\mp}} d^{2m} y_{\perp} ~\epsilon (y_{\perp}) F_{ru}^{C(2m)}(y_{\perp}) \equiv Q_{\epsilon}^{S\pm} + Q_{\epsilon}^{H\pm},
\end{equation}
or after applying the constraint \eqref{eq:constr}
\begin{equation}
\label{eq:hs-charges}
\begin{split}
    Q_{\epsilon}^{S+} &= \int_{\mathcal{I}^{+}_{-}} d^{2m} y_{\perp} ~\epsilon  F_{ru}^{R(2m)}(y_{\perp})= -\frac{1}{2} \int_{\mathcal{I}^+} du d^{2m}y_{\perp} ~\epsilon D^A F_{Au}^{R(2m - 2)} + \int_{\mathcal{I}^+_+} d^{2m}y_{\perp} ~\epsilon F_{ru}^{R(2m)} \\
    Q_{\epsilon}^{H+} &= \frac{1}{2}\int_{\mathcal{I}^+} du d^{2m}y_{\perp} ~\epsilon j_u^{(2m)} +\int_{\mathcal{I}^+_+} d^{2m}y_{\perp} ~\epsilon F_{ru}^{C(2m)}.
    \end{split}
\end{equation}
$Q_{\epsilon}^-$ obeys a similar decomposition. 

The relation between the soft theorems and large gauge charges is established by following the same reasoning as in $d = 4$ \cite{He:2014cra}. Inserting the antipodal matching condition 
\begin{equation}
\label{eq:matching}
    Q_{\epsilon}^+ = Q_{\epsilon}^-
\end{equation}
into an S-matrix element implies a relation between the soft and hard charges in \eqref{eq:soft-hard}. The Maxwell constraints relate the former to the soft radiative modes of the field strength and the latter to the sources, whose action on asymptotic states is encoded in the leading soft factor. This identification follows straightforwardly in  even dimensions, as a consequence of the asymptotic expansions \eqref{eq:rad-c} and the Bianchi identity. This approach fails in odd dimensions, but one can instead use the global properties of the solutions, in particular regularity of the field strength at the origin, to relate the Coulombic and radiative modes \cite{He:2019jjk}. As we review next, this establishes the bridge between asymptotic symmetries and soft theorems.

\subsection{From asymptotic charges to soft theorems}
\label{sec:as-st}

We first observe that the Bianchi identities allow us to trade the first equation in \eqref{eq:rad-c} for
\begin{equation}
\label{eq:rad-rec-rel}
    2 \p_u F_{ru}^{(n+m+1)} + \frac{1}{2 n} D^A \p_A F_{ru}^{(m+ n)} = 0, \quad 0 < n \leq m - 2.
\end{equation}
We review the proof of this fact in Appendix \ref{app-large-r-exp}. 
Therefore for $d \in 2\mathbf{Z}$ or $m \in \mathbf{Z}$, the subleading radiative modes are recursively related to the leading ones. This recursion relation allows for the Coulombic charge aspect at order $r^{-2m}$ to be related to the radiative data, and in turn, the soft photon operator upon canonical quantization. 
Importantly, this is not the case in odd spacetime dimensions, or for $m \in \frac{1}{2}\mathbf{Z} \setminus \mathbf{Z}$ since the half-integer asymptotic data does not mix, order by order in the large-$r$ expansion  with the integer data. It is easy to see that this conclusion remains unchanged upon allowing for both integer and half-integer powers or logarithmic terms in the large-$r$ expansion. It follows that unlike in even dimensions, in odd dimensions the recursion relations \eqref{eq:rad-rec-rel} do not relate the Coulombic charge aspect $F_{ru}^{(2m)}$ to the leading radiative mode $F_{ru}^{(m+1)}$.

This problem was bypassed in \cite{He:2019jjk} by considering the asymptotic expansions of solutions to the free Maxwell equations that are regular at the origin. In even dimensions, the  radiative solutions admit a large-$r$ expansion containing only integer powers (cf. eq. 2.25 in \cite{He:2019jjk}), while in odd dimensions, the expansion contains both an infinite series of half-integer powers and an infinite series of integer powers. The latter starts at Coulombic order (cf. eq. 2.23 in \cite{He:2019jjk}). In this case it not \eqref{eq:rad-rec-rel} but instead regularity at the origin of the exact plane-wave solutions that relates the two series. The component of the field strength relevant for the asymptotic charges is $F_{ru}^{(2m)}$ which was shown in \cite{He:2019jjk} to take the form
\begin{equation}
\label{eq:odd-fs}
    F_{ur}^{\pm R(2m)}(u, x_{\perp}) \propto \int d^{2m} y_{\perp} \frac{\p^A O_A^{\pm(0)}(y_{\perp}) + \p^A O_A^{\pm(0)\dagger}(y_{\perp})}{[(x_{\perp} - y_{\perp})^2]^{2m - 1}}
\end{equation}
in odd dimensions and 
\begin{equation}
\label{eq:even-field-str}
\begin{aligned}
F_{ur}^{\pm R(2m)}(u,x_{\perp})
={}& -\frac{1}{(4\pi)^{m+1}\Gamma\!\left(m\right)}
\sum_{n=1}^{\infty} 2^n \Gamma(n)
\left[
i(iu)^{-n}(-\partial^2)^{ m -1}
\partial^A O_A^{(\pm,n)}(x_{\perp})
+\mathrm{c.c.}
\right] \\
&+ \frac{i (-1)^{m}\Gamma(2m-1)}
{2^{4-2m}\pi}
\int d^d y_{\perp}\,
\frac{
\partial^A O_A^{(\pm,0)}(y_{\perp})
-\partial^A O_A^{(\pm,0)\dagger}(y_{\perp})
}{
\left[(x_{\perp}-y_{\perp})^2\right]^{2m-1}
} \\
&+ \frac{i \left[
\log\!\left(\frac{|u|e^{2\gamma_E}}{4}\right)
-H_{m-1}
\right]}
{(4\pi)^{m+1}\Gamma\!\left(m\right)}
(-\partial^2)^{m-1}
\left[
\partial^A O_A^{(\pm,0)}(x_{\perp})
-\partial^A O_A^{(\pm,0)\dagger}(x_{\perp})
\right] \\
&- \frac{\,[\Theta(u)-\Theta(r)]}
{8(4\pi)^{m}\Gamma\!\left(m \right)}
(-\partial^2)^{m-1}
\left[
\partial^A O_A^{(\pm,0)}(x_{\perp})
+\partial^A O_A^{(\pm,0)\dagger}(x_{\perp})
\right]
\end{aligned}
\end{equation}
in even dimensions, where $H_n$ is the $n$-th harmonic number. In the coordinates \eqref{eq:Mink0}, the expansions near $\mathcal{I}^{\pm}$ are obtained by taking $r \rightarrow \pm \infty$. 
Here $O_A^{(\pm, 0)}(y_{\perp})$ are the zero modes of the radiative field strength $F_{ru}^{R(m+1)} \propto \int du D^A F_{Au}^{(m-1)}$ in a plane wave basis. We note that one could have chosen to express the solutions to the free Maxwell equations in terms of a different set of wavefunctions instead of plane waves. One example is the higher-dimensional analog of the conformal primary basis constructed in \cite{Pasterski:2017kqt}. Mode by mode, the large-$r$ behavior of the different wavefunctions can differ, as we will discuss further in Section \ref{sec:rel-previous}. 

We observe that the even-dimensional solution \eqref{eq:even-field-str} has non-trivial dependence on the lightcone coordinate $u$, while the odd-dimensional one \eqref{eq:odd-fs} does not. This observation played no role in  \cite{He:2019jjk} who proceeded by eliminating the ${\rm log}(u)$ dependence\footnote{This would otherwise lead to divergences in the large gauge charges at $|u| \rightarrow \infty$} by imposing
\begin{equation}
\label{eq:match2}
    \p^A O_A^{(\pm, 0)}(x_{\perp}) =  \p^A O_A^{(\pm, 0) \dagger}(x_{\perp}). 
\end{equation}
As a result, only the zero mode in the last line of \eqref{eq:even-field-str} survives, and by taking $u \rightarrow -\infty, r \rightarrow \infty$, we see that up to normalization, the soft charges in odd and even dimensions can be put into the same form after an appropriate choice of $\epsilon$. The soft theorems then follow universally from the matching condition \eqref{eq:matching} \cite{He:2019jjk}.

\subsection{Absence memory in odd dimensions}
\label{sec:no-memory}

In order to establish a relation to the memory effect, one needs to further relate the (outgoing) soft insertion to the time integral  of an appropriate field strength component.
The different time-dependence of radiative solutions in even and odd dimensions leads to an obstruction in establishing a connection between soft theorems and memory effects in odd spacetime dimensions, as we discuss next.

In 4-dimensions, the memory effect is defined as the zero mode of the radiative electric field, namely
\begin{equation}
\label{eq:memory}
\mathcal{M}_2(y_{\perp}) \equiv \int du D^A F_{u A}^{(0)}(u, y_{\perp}).
\end{equation}
On the one hand, upon quantization, this mode can be directly related to the leading soft photon \cite{He:2014cra,Pasterski:2015zua}. On the other hand, the Maxwell constraint \eqref{eq:rad-c} relates this to the soft charge \eqref{eq:hs-charges}. 

The same argument carries through in higher even spacetime dimension, provided the memory operator is identified with the zero mode of the $\mathcal{O}(r^{-2m + 1})$ component of the radiative field
\begin{equation}
\label{eq:higher-even-memory}
\mathcal{M}_{2m}(y_{\perp}) \equiv \int du D^A F_{u A}^{(2m - 2)}(u, y_{\perp}).
\end{equation}
In this case, the free Maxwell constraint 
\begin{equation}
\label{eq:Maxwell-explicit}
2\p_u F_{ru}^{R(2m)} - D^A F_{Au}^{R(2m - 2)} = 0
\end{equation}
again relates \eqref{eq:higher-even-memory} to the soft charge, 
\begin{equation}
\label{eq:soft-expl-odd}
Q^{S+}_{\epsilon} = - \frac{1}{2} \int d^{2m} y_{\perp} \epsilon(y_{\perp}) \mathcal{M}_{2m}(y_{\perp}) + \int_{\mathcal{I}^+_+} d^{2m}y_{\perp} \epsilon(y_{\perp}) F_{ru}^{R (2m)}(y_{\perp}).
\end{equation}
From \eqref{eq:even-field-str}, we see that after imposing the condition \eqref{eq:match2}, the contribution from $\mathcal{I}^+_+$ vanishes and the soft charge is again controlled by the memory operator \eqref{eq:higher-even-memory}.

In odd spacetime dimensions however, the solution $F_{ru}^{R(2m)}$ constructed in \cite{He:2019jjk} is manifestly time-independent. As a result, \eqref{eq:Maxwell-explicit} together with the assumed fall-off conditions of the Coulombic $F_{Au}$ modes \eqref{eq:C-fall-off} imply the vanishing of the memory operator
\begin{equation}
D^A F_{Au}^{R(2m - 2)} = 0.
\end{equation}
The soft charge \eqref{eq:soft-expl-odd} is nevertheless non-vanishing due to the contribution from $F_{ru}^{(2m)}$ at $\mathcal{I}^+_+$ which equals that at $\mathcal{I}^+_-$. We emphasize that the memory is also absent at radiative order \cite{Satishchandran:2017pek,Satishchandran:2019pyc}. This can also be seen from the first line in eq. 2.23 of \cite{He:2019jjk} which shows that the time-dependence of $F_{ru}^{R(m+1)}$ is 
\begin{equation}
F_{ru}^{R(m+1)} \sim u^{-m + 1}  \rightarrow 0, \quad u \rightarrow \infty, \quad m > 1.
\end{equation}

The vanishing of the odd-dimensional analogue of \eqref{eq:higher-even-memory} is puzzling from the perspective of the Weinberg soft theorem. In even dimensions, the recursion relations \eqref{eq:rad-rec-rel}, together with the Maxwell constraint in \eqref{eq:rad-c}, imply the recursion relation
\begin{equation}
\label{eq:rec-rel}
\p_u D^A F_{A u}^{(m+k)}
 =
-\frac{1}{4(k+1)}\Box_{\perp} D^A F_{Au}^{(m+k-1)},
\qquad 0 \leq k\leq m-2,
\end{equation}
where $\Box_{\perp}$ is the transverse Laplacian.
Iterating \eqref{eq:rec-rel} gives
\begin{equation}
\label{eq:memory-rad-recursion}
\p_u^{m - 1} D^A F_{Au}^{(2m - 2)}
 =
\frac{(-1)^{m-1}}{4^{m-1}(m-1)!}\Box_{\perp}^{m - 1}D^A F_{Au}^{(m - 1)} .
\end{equation}
After Fourier transforming in $u$, we obtain
\begin{equation}
\label{eq:memory-rad-recursion-Fourier}
(i\omega)^{m - 1}D^A \widetilde{F}_{Au}^{(2m - 2)}
 =
\frac{(-1)^{m-1}}{4^{m-1}(m-1)!}\Box_{\perp}^{m - 1}D^A \widetilde{F}_{Au}^{(m - 1)},
\end{equation}
where $\widetilde{F}_{Au}^{(m-1)}$ is obtained from the leading term in a stationary-phase expansion in $d=2+2m$, namely
\begin{equation}
\label{eq:leading-radiative-fourier}
D^A\widetilde{F}_{Au}^{(m - 1)}(\omega,y_{\perp})
\propto
\omega^{m}\left[D^A a_A(\omega,y_{\perp})+ (-1)^{2m} D^A a_A^\dagger(\omega,y_{\perp})\right].
\end{equation}
It follows that  the memory mode defined by the zero mode of $D^AF_{Au}^{(2m-2)}$ is fixed by the Weinberg soft operator after applying $\Box_\perp^{m-1}$. In four dimensions this reduces to the familiar statement that the soft pole is the Fourier transform of a step function in retarded time. In higher even dimensions the same statement holds after the finite number of radial recursion steps displayed in \eqref{eq:memory-rad-recursion}.
As discussed before, in odd dimensions there is no finite recursion chain connecting the leading radiative half-integer series to $F_{Au}^{(2m-2)}$. 

The goal of this work is to identify the missing memory corner of the infrared triangle in higher odd dimensions and link it to asymptotic symmetries and soft theorems. The key observation which will allow us to achieve this is that in odd spacetime dimension the time integral appearing in the definition of memory \eqref{eq:higher-even-memory} does not commute with the large-$r$ expansion.  We will show that memory is in fact present if one evaluates it at large but finite radius $r$ before taking $r \rightarrow \infty$.

\section{Electromagnetic fields sourced by massless particles}
\label{sec:em-gf}

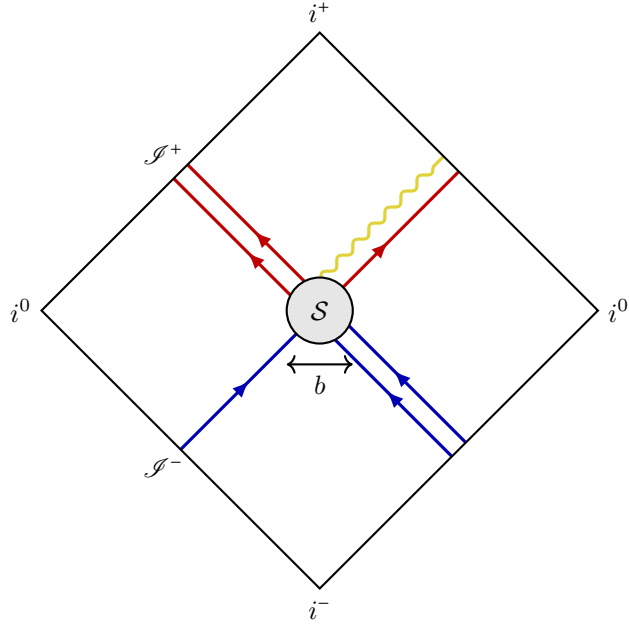
\begin{figure}[h!]
    \centering
\begin{tikzpicture}[
  scale=1.15,
  null/.style={thick, -{Latex[length=2.2mm]}},
  boundary/.style={thick},
 particle/.style={
  very thick,
  decoration={
    markings,
    mark=at position 0.5 with {\arrow{Latex[length=2.4mm]}}
  },
  postaction={decorate}
},radparticle/.style={
  very thick,
  color=yellow!85!black,
  decorate,
  decoration={
    snake,
    amplitude=0.5 mm,
    segment length=3 mm,
    post length=1 mm
  }},
  region/.style={draw, thick, fill=gray!20},
  label/.style={font=\small}
]

\def\M{3}
\def\N{3}

\coordinate (ip) at (0,3.2);
\coordinate (im) at (0,-3.2);
\coordinate (i0L) at (-3.2,0);
\coordinate (i0R) at (3.2,0);

\coordinate (C) at (0,0);
\def\r{0.38}

\draw[boundary] (im) -- (i0L) -- (ip) -- (i0R) -- cycle;

\draw[region] (C) circle (\r);
\node[label] at (0,0) {$\mathcal{S}$};

\draw[<->, thick] (-\r,-0.62) -- (\r,-0.62)
  node[midway, below, label] {$b$};

\def\epsIn{0.18}
\def\epsOut{0.18}

\def\Len{3.2}

\def\epsIn{0.16}
\def\epsOut{0.16}

\begin{pgfonlayer}{particlelayer}

\foreach \k in {1,...,\M} {
  \pgfmathsetmacro{\c}{-\epsIn + 2*\epsIn*(\k-1)/max(\M-1,1)}
  \pgfmathtruncatemacro{\side}{mod(\k,2)}

  \ifnum\side=0
    \coordinate (Sin) at ({(-\Len-\c)/2},{(-\Len+\c)/2});
    \coordinate (Tin) at (0,\c);
  \else
    \coordinate (Sin) at ({(\Len+\c)/2},{(\c-\Len)/2});
    \coordinate (Tin) at (0,\c);
  \fi

  \draw[particle, blue!70!black] (Sin) -- (Tin);
}

\foreach \k in {1,...,\N} {
  \pgfmathsetmacro{\c}{-\epsOut + 2*\epsOut*(\k-1)/max(\N-1,1)}
  \pgfmathtruncatemacro{\side}{mod(\k,2)}

  \ifnum\side=0
    \coordinate (Tout) at (0,\c);
    \coordinate (Sout) at ({(\Len-\c)/2},{(\Len+\c)/2});
  \else
    \coordinate (Tout) at (0,\c);
    \coordinate (Sout) at ({(\c-\Len)/2},{(\Len+\c)/2});
  \fi

  \draw[particle, red!75!black] (Tout) -- (Sout);
}

\pgfmathsetmacro{\crad}{0.35}
\coordinate (Trad) at (0,\crad);
\coordinate (Srad) at ({(\Len-\crad)/2},{(\Len+\crad)/2});

\draw[radparticle] (Trad) -- (Srad);
\end{pgfonlayer}

\node[label, above] at (ip) {$i^+$};
\node[label, below] at (im) {$i^-$};
\node[label, left] at (i0L) {$i^0$};
\node[label, right] at (i0R) {$i^0$};

\node[label, rotate=0] at (-1.8,1.8) {$\mathscr I^+$};
\node[label, rotate=0] at (-1.8,-1.8) {$\mathscr I^-$};


\end{tikzpicture}
\caption{$M = 3$ incoming particles (blue) interact with $N =3$ outgoing particles in a region of size $b$. Radiation (yellow) is emitted in the process.}
\label{fig:part-scatt}
\end{figure}

We consider the scattering of $M$ incoming and $N$ outgoing massless particles. The interactions are assumed to be restricted to a finite spacetime region with particles moving at constant velocity outside this region. The current density for this particle configuration is given by
\be\label{eq:current0}
J_\mu(x)=\sum_{i=1}^{M}\qin p^{{\rm in}}_{i\mu}\int_{-\infty}^0 d\tau \delta^{(d)}\left(x-b_i^{\rm in}- p_i^{\rm in} \tau \right)+\sum_{i=1}^{N}\qout p^{\rm out}_{i\mu} \int_0^\infty d\tau  \delta^{(d)}\left(x- b_i^{\rm out} -p_{i}^{\rm out}\tau\right).
\ee
Here $x$ is a point in $\mathbb{R}^{d-1,1},$ 
 $\pin$, $\pout$ and $\qin$, $\qout$ are the momenta and charges of the incoming and outgoing particles, while $b_i^{\rm in}, b_i^{\rm out}$ parameterize the starting points the straight line trajectories outside the interaction region. We illustrate our setup in the conformal compactification of Minkowski space in Figure \ref{fig:part-scatt}.

To avoid clutter, we will employ an all-outgoing convention where
    \begin{equation}
    \label{eq:all-outgoing}
    \begin{split}
        (p_i, Q_i) &= (-p_i^{\rm in}, - Q_i^{\rm in}), \quad 1 \leq i \leq M,\\
        (p_i, Q_i) &= (p_i^{\rm out}, ~Q_i^{\rm out}), \quad M+1 \leq i \leq M+ N.
        \end{split}
    \end{equation}
With this convention, the current \eqref{eq:current} takes the simple form
\be\label{eq:current}
J^\mu(x)=\sum_{i=1}^{M+N} Q_i p_{i}^{\mu}  \int_0^\infty d\tau  \delta^{(d)}\left(x- b_i -p_{i} \tau\right).
\ee
We work in the Lorenz gauge $\nabla_{\mu} A^{\mu} = 0$ in which case Maxwell's equations are given by\footnote{Note that the sign of $J$ in \eqref{eq:current}, \eqref{eq:Maxwell-intro} is fixed by the coupling of the gauge field to the source worldine and is such that the Lorentz force law takes the standard form $m p^{\mu} = Q F^{\mu\nu} p_{\nu}$ for particles of constant momentum. Therefore, $J_{\mu}$ here and $j_{\mu}$ used in Section \eqref{sec:review} differ by a sign which will be important when verifying the constraint later.}
\begin{equation}
\label{eq:Maxwell-intro}
    \Box A_{\mu}(x) = -J_{\mu}(x).
\end{equation}
The solutions to \eqref{eq:Maxwell-intro} are given by 
\bea
A^{\mu}(x)= -\int d^dy  
G(x-y) J^\mu(y)\, ,\label{eq:A_mu_gen}
\eea
where $G$ is a Green's function defined by the equation
\bea
\label{eq:box_G}
\Box G(x-y)=\delta^{(d)}(x-y).
\eea

We will be interested in gauge field configurations near future null infinity ($\mathcal{I}^+$) which are determined by the retarded Green's function obeying the boundary condition
\begin{equation}
 G_R(x-y)=0\quad \text{for}\quad x^0<y^0.
\end{equation}
Gauge field configurations near past null infinity ($\mathcal{I}^-$) are determined by advanced Green's functions that instead vanish for $x^0 > y^0$.
We review the construction of these Green's functions in Appendix \ref{app:Greensf} and simply quote the results for even and odd dimensions:
\bea
\label{eq:Gr}
G_R(x-y)
&\overset{}{=}&-\f{1}{2 \pi^\f{d-2}{2}}e^{-\epsilon(x^0-y^0)}\Theta(x^0-y^0) \delta^{\left(\f{d-4}{2}\right)}(-(x-y)^2), \quad d = 2k + 2, ~~ k \in \mathbb{N}
\eea
and
\begin{equation}
\label{eq:Godd}
\begin{split}
G_R(x-y)&\overset{}{=} (-1)^\f{d-1}{2}\f{1}{2 \pi^\f{d}{2}} \Gamma\left(\f{d-2}{2}\right)e^{-\epsilon(x^0-y^0)} \left(-(x-y)^2\right)^{-\f{d-2}{2}} \Theta(x^0-y^0-|\vec x-\vec y|), \quad d = 2k + 1, ~~ k \in \mathbb{N}.
\end{split}
\end{equation}
In \eqref{eq:Godd} we have discarded the divergent series localized at $(x - y)^2 = 0$ discussed in Appendix \ref{app:Greensf} which arises from operating with the differential operator in \eqref{eq:full-gf-odd} on the theta function (see also \cite{Satishchandran:2017pek}). 
This Green's function allows us to compute the gauge fields and associated field strengths near $\mathcal{I}^+$ which we do next. These solutions and their properties have been discussed in even dimensions in \cite{Mao:2017wvx,Garfinkle:2017fre} and in odd dimensions in \cite{Galtsov:2001iv, Hollands:2004ac,Satishchandran:2017pek}.

\subsection{Gauge field in even spacetime dimensions}
\label{sec:even-dim-gf}

Substituting \eqref{eq:current} and \eqref{eq:Gr} into \eqref{eq:A_mu_gen}, the  $y$ integral can be straightforwardly evaluated yielding
\begin{equation}
\label{eq:gauge-field-even}
\begin{split}
    A^{\mu}(x) &= \frac{1}{2\pi^{\frac{d - 2}{2}}} \sum_{i = 1}^{M+N}   Q_i p_i^{\mu} \int_0^{\infty} d\tau e^{-\epsilon(x^0 - p^0 \tau - b_i^0)} \Theta(x^0 - p_i^0 \tau - b_i^0) \delta^{(\frac{d - 4}{2})}\left(-(x - b_i - p_i \tau)^2 \right).
   \end{split}
\end{equation}
Since in even dimensions, the Green's function has support on the lightcone, the solution depends only on the root(s) of the equation
\begin{equation}
\label{eq:eq}
    (x - b_i - p_i \tau)^2 = 0.
\end{equation}
For massive sources $p_i^2 \neq 0$, these roots are distinct, while for massless sources $p_i^2 = 0$ they become degenerate. We will focus on the massless case and postpone the more involved discussion of massive particles to Section \ref{sec:massive}. 

For massless particles $(p^2 = 0)$ we have
\begin{equation}
    -(x - b_i - p_i\tau)^2 = 2 (\tau - \tau_i) (x - b_i)\cdot p_i ,
\end{equation}
where 
\begin{equation}
    \tau_i = \frac{(x-b_i)^2}{2 (x-b_i)\cdot p_i}.
\end{equation}
The integrals in \eqref{eq:gauge-field-even} can be evaluated 
by the change of variables
\begin{equation}
    \tau' = 2 (x - b_i) \cdot p_i (\tau - \tau_i)
\end{equation}
and integration by parts,
yielding \cite{Mao:2017wvx}, away from the locus $p\cdot x = 0$\footnote{These contributions will be computed in Section \ref{sec:shockwaves} and will play a crucial role in our discussion of memory.}
\begin{equation}
\label{eq:gf-massless-even}
\begin{split}
    A^{\mu}(x) = -\sum_{i = 1}^{M+N}\frac{1}{2 \pi^{\frac{d - 2}{2}}}  \frac{Q_i p_i^{\mu}}{2(x - b_i)\cdot p_i}  \delta^{(\frac{d - 6}{2})}(-(x - b_i)^2) .
    \end{split}
\end{equation}
We note that the theta function constraint is automatically obeyed for incoming particles, while for outgoing particles, on the support of $(x - b_i - p_i \tau)^2 = 0$, we also have $\tau \leq \frac{x^0 - b_i^0}{p_i^0}$.

\subsection{Gauge field in odd spacetime dimensions}
\label{sec:odd-dim-gf}

Substituting \eqref{eq:current} and \eqref{eq:Godd}  into \eqref{eq:A_mu_gen}, we obtain
\begin{equation}
\label{eq:odd-dim-int}
\begin{split}
    A^{\mu}(x) &= -\frac{(-1)^{\frac{d - 1}{2}}}{2 \pi^{\frac{d}{2}}} \Gamma\left(\frac{d - 2}{2} \right) \sum_{i = 1}^{M+N} Q_i p_i^{\mu} \int_{0}^{\infty} d\tau\left(-(x - b_i - p_i\tau)^2 \right)^{-\frac{d - 2}{2}}\Theta\left(x^0 - b_i^0 - p_i^0\tau \right)\Theta(-(x-p_i\tau)^2). 
    \end{split}
\end{equation}
It will be convenient to introduce Minkowski lightcone coordinates
\begin{equation}
\label{eq:coords}
    x = r q(y) + u n,  \quad q^2 = n^2 = 0, 
\end{equation}
where 
\begin{equation}
\label{eq:q}
    q(y) = \frac{1}{2}(1 + y^2, 2 y^A, 1 - y^2), \quad n = \frac{1}{2}(1, 0^A, -1)
\end{equation}
and $y^A$ are coordinates in $\mathbb{R}^{d - 2}$. 
In these coordinates, the Minkowski metric takes the form
\begin{equation}
\label{eq:Mink-lightcone}
    ds^2 =  -du dr + r^2 dy^A dy_A.
\end{equation}
We are interested in evaluating these integrals at large distance from the sources. We emphasize that a choice of coordinates is only introduced for convenience at this point to identify the correct integration limits imposed by the theta function in \eqref{eq:odd-dim-int}. The same coordinates will later be useful in comparing our results to the asymptotic expansions employed in the literature \cite{Satishchandran:2017pek, Satishchandran:2019pyc,He:2019jjk}. Nevertheless, we will see that none of our results depend on this choice.

The retarded support condition implies 
\begin{equation}
   x^0 - b_i^0 - p_i^0 \tau > |\vec{x} - \vec{b}_i - \vec{p}_i\tau|
    \quad\Longleftrightarrow\quad
    (x - b_i -p_i\tau)^2<0
    \quad\Longleftrightarrow\quad
   (x - b_i)^2 - 2(x - b_i)\cdot p_i \tau < 0.
\end{equation}
For a future-directed massless momentum (ie. an outgoing particle)  $p=\omega q(z)$  and 
\begin{equation}
    -2 x \cdot p = \omega\left(u+r|y-z|^2\right) > 0
\end{equation}
near future null infinity for fixed $u$ and sufficiently large $r$. For a past-directed massless momentum (ie. an incoming particle in our all-out convention) $p= -\omega q(z)$ and $-2 x\cdot p < 0.$ 
Therefore
\begin{equation}
\label{eq:tau}
\begin{split}
    \tau &> \tau_0^i \equiv \frac{(x - b_i)^2}{2(x - b_i)\cdot p_i}, \quad 1 \leq i \leq M, \\
  \tau &< \tau_0^i \equiv \frac{(x - b_i)^2}{2(x - b_i)\cdot p_i}, \quad M+1 \leq i \leq N+M.
    \end{split}
\end{equation}
Whether $\tau \in [0,\infty)$ or not then depends on the sign of $(x - b_i)^2$.

To evaluate the integral in \eqref{eq:odd-dim-int} for incoming momenta, the following formulas will be useful
\begin{equation}
    \begin{split}
       I_1 = \int_0^{\infty}d\tau ( - x^2 + 2 x\cdot p \tau)^{\frac{2 - d}{2}} &= \frac{(-x^2)^{2 - \frac{d}{2}}}{(d - 4) x \cdot p}, \quad \tau_0 < 0, \\
       I_2  =\int_{\tau_0 +  \epsilon}^{\infty} d\tau ( - x^2 + 2 x\cdot p \tau)^{\frac{2 - d}{2}} &= 
    \frac{(2x\cdot p \epsilon)^{2 - \frac{d}{2}}}{(d - 4)x\cdot p}, \quad \tau_0 \geq 0
    \end{split}
\end{equation}
while for outgoing momenta
\begin{equation}
    \int_0^{\tau_0 - \epsilon}d\tau ( - x^2 + 2 x\cdot p \tau)^{\frac{2 - d}{2}} = \frac{(-x^2)^{2 - \frac{d}{2}}}{(d - 4) x \cdot p} - \frac{(-2x\cdot p \epsilon)^{2 - \frac{d}{2}}}{(d - 4)x\cdot p} , \quad \tau_0 \geq 0.
\end{equation}
Using the expression for the Green's function in odd dimensions, we find away from the locus $p\cdot x = 0$
\begin{equation}
\label{eq:gf-odd}
\begin{split}
    N^{-1}A^{\mu}(x) &= -\sum_{i = 1}^M\left[\Theta(-(x-b_i)^2) \frac{Q_i p_i^{\mu}(-(x-b_i)^2)^{2 - \frac{d}{2}}}{(d - 4) (x-b_i) \cdot p_i}  + \Theta((x-b_i)^2) \frac{Q_i p_i^{\mu}(2 (x-b_i)\cdot p_i\epsilon_i)^{2 - \frac{d}{2}}}{(d - 4) (x-b_i)\cdot p_i} \right] \\
    &- \sum_{i = M+1}^{M+N}\Theta(-(x-b_i)^2)\left[ \frac{Q_i p_i^{\mu}(-(x-b_i)^2)^{2 - \frac{d}{2}}}{(d - 4) (x-b_i) \cdot p_i} - \frac{(-2(x-b_i)\cdot p_i \epsilon_i)^{2 - \frac{d}{2}}Q_i p_i^{\mu}}{(d - 4) (x-b_i) \cdot p_i} \right],
    \end{split}
\end{equation}
where we introduced the regulators $\epsilon_i$ in whose absence the integrals are divergent and the normalization
\begin{equation}
\label{eq:norm}
    N = \frac{(-1)^{\frac{d - 1}{2}}}{2\pi^{\frac{d}{2}}} \Gamma\left(\frac{d - 2}{2} \right).
\end{equation}
For simplicity, from now on we will further set $b_i = 0$. The dependence on $b_i$ in all formulas below can be restored by simply shifting $x \rightarrow x - b_i$ in each component of the gauge fields and field strengths sourced by the momentum $p_i.$ Unlike in even dimensions, the gauge field is non-vanishing inside the lightcone through the origin which is the well-known violation of the Huygens principle in odd $d$ \cite{Hadamard1923}.

The gauge field \eqref{eq:gf-odd} is divergent in the limit $\epsilon_i  \rightarrow 0$. One could proceed by employing a Hadamard regularization prescription \cite{Blanchet:2000nu} which amounts to discarding these terms. Alternatively, we can justify the regularization by observing that the divergences are a subtle consequence of taking the $\varepsilon \rightarrow 0$ limit in the definition of the Green's function before performing the integrals defining the gauge field. To see this, we instead revert to the expression for the odd dimensional Green's function given in \eqref{eq:finalG}, and substitute it into \eqref{eq:A_mu_gen}.  For outgoing particles, we find
\begin{equation}
\label{eq:finalG-main}
\begin{split}
A_{\mu}(x) &= \f{i}{4\pi^\f{d}{2}} \Gamma\left(\f{d-2}{2}\right)\sum_{i = M+1}^{M+N} Q_i p_{i\mu} \lim_{\varepsilon\rightarrow 0}\int_0^{\frac{x^0}{p_i^0}} d\tau \left[\f{1}{\left[-(x^0-p^0_i \tau-i\varepsilon)^2+|\vec x-\vec{p}_i\tau |^2\right]^\f{d-2}{2}}- h.c.\right]
\end{split}
\end{equation}
and observe that the integrand is a total derivative since
\begin{equation}
\label{eq:derivative-id}
    \f{1}{\left[-(x^0-p^0_i \tau-i\varepsilon)^2+|\vec x-\vec{p}_i\tau |^2\right]^\f{d-2}{2}} = \frac{1}{(d - 4)x\cdot p_i} \frac{d}{d\tau}\left[\f{1}{\left[-(x^0-p^0_i \tau-i\varepsilon)^2+|\vec x-\vec{p}_i\tau |^2\right]^\f{d-4}{2}} \right].
\end{equation}
Then
\begin{equation}
\label{eq:even-odd-bdry-gf}
  \begin{split}
      A_{\mu}(x) &= \f{i}{4\pi^\f{d}{2}} \Gamma\left(\f{d-2}{2}\right)\sum_{i = M+1}^{M+N}  \frac{Q_i p_{i\mu}}{(d - 4)x\cdot p_i} \lim_{\varepsilon\rightarrow 0}\left[ \f{1}{\left[-(x^0-p^0_i \tau-i\varepsilon)^2+|\vec x-\vec{p}_i\tau |^2\right]^\f{d-4}{2}} - h.c. \right] \Big|_0^{\frac{x^0}{p_i^0}}\\
      &= -\f{i}{4\pi^\f{d}{2}} \Gamma\left(\f{d-2}{2}\right)\sum_{i = M+1}^{M+N}  \frac{Q_i p_{i\mu}}{(d - 4)x\cdot p_i} \lim_{\varepsilon\rightarrow 0} \left[ \f{1}{\left[x^2 + i\varepsilon x^0 \right]^\f{d-4}{2}} - h.c.  \right],
  \end{split}  
\end{equation}
where the boundary term at $\tau = \frac{x^0}{p_i^0}$ vanishes\footnote{This statement holds away from $p\cdot x = 0$ and will be revisited in Section \ref{sec:shockwaves}.} because the denominator is positive in this case. Finally, using the identity \eqref{eq:id-ds} and after a similar computation for the incoming particles, we obtain the finite result 
\begin{equation}
\label{eq:reg-odd}
    \begin{split}
    N^{-1}A_{\rm fin.}^{\mu}(x) &= -\sum_{i = 1}^{M+N} \Theta(- x^2) \frac{Q_i p_i^{\mu}(-x^2)^{2 - \frac{d}{2}}}{(d - 4) x \cdot p_i} .
    \end{split}
\end{equation}
We will show in Section \ref{eq:mc-st} that \eqref{eq:reg-odd} coincides with the Fourier transform of the soft photon Ward identity.

\subsection{Shockwave contributions}
\label{sec:shockwaves}
In this section, we show that the gauge fields sourced by massless particles contain additional terms localized at $p \cdot x = 0$. These terms are the electromagnetic analogue of the Aichelburg-Sexl shockwaves in gravity \cite{Aichelburg1971}. Their relation to soft modes and memory in 4d Maxwell theory was discussed in \cite{Donnay:2018neh,Chen:2024kuq}.

These terms can be computed by carefully considering all boundary terms arising from the integrals of the form \eqref{eq:finalG-main} for both incoming and outgoing sources. As discussed in Section \ref{sec:odd-dim-gf}, the retarded support constraints allow for the net gauge field to be written as
\begin{equation}
\label{eq:in}
\begin{split}
    A_{\mu}(x) &= \frac{i}{4\pi^{\frac{d}{2}}} \Gamma\left(\frac{d - 2}{2} \right)\sum_{i = 1}^M Q_i p_{i,\mu} \left[\int_0^{\infty} d\tau e^{-\epsilon(x^0 - p^0_{i}\tau)} \Theta(-x^2) + \int_{\tau_0^i}^{\infty} d\tau e^{-\epsilon(x^0 - p^0_{i}\tau)} \Theta(x^2)\right] \Delta_d(x; p_i)\\
    &+  \frac{i}{4\pi^{\frac{d}{2}}} \Gamma\left(\frac{d - 2}{2} \right) \sum_{i = M+1}^{M+N} Q_i p_{i,\mu} \int_0^{\tau_0^i} d\tau e^{-\epsilon(x^0 - p^0_{i}\tau)} \Theta(-x^2) \Delta_d(x; p_i),\\
    \end{split}
\end{equation}
where $\tau_0^i$ was defined in \eqref{eq:tau} and
\begin{equation}
   \Delta_d(x; p_i) =  \lim_{\epsilon \rightarrow 0} \left[ \frac{1}{\left(-(x^0 - p_{i}^0\tau - i\epsilon)^2 + |\vec{x} - \vec{p}_{i}\tau|^2 \right)^{\frac{d - 2}{2}}} - \frac{1}{\left(-(x^0 - p_{i}^0\tau + i\epsilon)^2 + |\vec{x} - \vec{p}_{i}\tau|^2 \right)^{\frac{d - 2}{2}}} \right].
\end{equation}
For $x\cdot p \sim 0$, restoring the $i\epsilon$ dependence in \eqref{eq:derivative-id},
\begin{equation}
\begin{split}
  & \frac{1}{\left(-(x^0 - p_{i}^0\tau - i\epsilon)^2 + |\vec{x} - \vec{p}_{i}\tau|^2 \right)^{\frac{d - 2}{2}}} = \frac{1}{d - 4}\frac{1}{x\cdot p_i + i\epsilon p_i^0} \frac{d}{d\tau}\left(\frac{1}{((x-p_i\tau)^2 + 2 (x^0 - p_i^0 \tau)i\epsilon + \epsilon^2)^{\frac{d - 4}{2}}} \right)
   \end{split}
\end{equation}
and therefore the gauge field receives additional contributions localized at $x \cdot p = 0$ since
\begin{equation}
   \lim_{\epsilon \rightarrow 0}\left( \frac{1}{x\cdot p + i\epsilon p^0} - \frac{1}{x\cdot p - i\epsilon p^0} \right)= -2 i \pi \delta(x\cdot p).
\end{equation}
For $r > 0$, $\delta(x\cdot p) \Theta(-x^2) = 0$ and therefore only incoming particles can source such terms near $\mathcal{I}^+$. The boundary term from the endpoint of the particle trajectories at $\tau = \tau_0^i$ is divergent. This divergence is of the same nature as those discussed in Section \ref{sec:odd-dim-gf} and we will assume it can be discarded. We are left with the boundary term from $\tau \rightarrow \infty$ which contributes
\begin{equation}
\label{eq:shock}
    A_{\mu}(x) \supset \frac{1}{4\pi^{\frac{d}{2}-1}} \Gamma\left(\frac{d - 4}{2} \right)\sum_{i = 1}^M Q_i p_{i,\mu} \Theta(x^2) \delta(x\cdot p_i) (x^2)^{-\frac{d - 4}{2}}.
 \end{equation}
 These are precisely the higher dimensional counterparts of the shockwaves discussed in \cite{Chen:2024kuq}. We will see in the next section that they ensure that the electromagnetic constraint at $r = r_0 > 0$ is obeyed.

 Equation \eqref{eq:shock} then contributes the following term to the field strength at $r > 0$
\begin{equation}
\label{eq:fsshock}
\begin{split}
    F_{\mu\nu}(x)  &\supset -\frac{1}{2\pi^{\frac{d}{2}-1}} \Gamma\left(\frac{d - 2}{2} \right)\sum_{i = 1}^M Q_i (x_{\mu} p_{i,\nu} - x_{\nu} p_{i,\mu}) \Theta(x^2) \delta(x\cdot p_i) (x^2)^{-\frac{d - 2}{2}}\\
    &+\frac{1}{2\pi^{\frac{d}{2}-1}} \Gamma\left(\frac{d - 4}{2} \right)\sum_{i = 1}^M Q_i  (x_{\mu} p_{i,\nu} - x_{\nu} p_{i,\mu}) \delta(x^2) \delta(x\cdot p_i) (x^2)^{-\frac{d - 4}{2}}.
    \end{split}
\end{equation}
In the following, we will analyze the Maxwell constraint and show that it implies a memory effect at the Coulombic order in all spacetime dimensions. For simplicity, in the following we will restrict to $x \cdot p \neq 0$ and restore the contributions of \eqref{eq:fsshock} to the Coulombic order constraint at the end.

\section{Coulombic memory from Maxwell constraint}
\label{sec:Coulomb-mem}

By construction, the gauge fields computed in Sections \ref{sec:even-dim-gf}, \ref{sec:odd-dim-gf} ought to satisfy the Maxwell constraint. This is defined as the pull-back of (the Hodge dual of) Maxwell's equations
\begin{equation}
\label{eq:Maxwell-0}
 d \star F = \star j
\end{equation}
to any codimension-1 hypersurface in spacetime. We will be interested in the case where the codimension-1 surface is taken to be the lightsheet $\Sigma$ defined by constant $r = r_0$. The pull-back of \eqref{eq:Maxwell-0} to this surface is equivalent to the lower-$u$ component of Maxwell's equations derived in \eqref{eq:Maxwell}
\begin{equation}
\label{eq:Maxwell-constraint-fixed-r}
  2 r_0^{d - 2} \p_u F_{ru} - r_0^{d - 4} D^A F_{Au} = - r_0^{d - 2} j_u .
\end{equation}

In this section we will  study the interplay between evaluating the constraint on a family of bulk hypersurfaces labeled by $r = r_0$ and taking the limit $r_0 \rightarrow \infty$. We consider the even and odd dimensional cases separately. While the results in the former case are consistent with and essentially follow from previous studies \cite{Satishchandran:2017pek,Satishchandran:2019pyc}, the latter results present some novel features which are the main results of this work. In particular, we will uncover a universal memory effect at Coulombic order, in complete agreement with the expectation from the Weinberg soft theorem in odd dimensions. This effect appears if one computes the zero mode of the electromagnetic field at fixed $r = r_0$, \textit{before} taking $r_0 \rightarrow \infty$. In fact, this order of limits is important: we will see that in odd-dimensional spacetimes, the limit $r_0 \rightarrow \infty$ does \textit{not} commute with the pull-back of the Maxwell constraint to $\Sigma$! This observation seems to have been missed in prior works where the large-$r$ expansions are always evaluated before computing the memory effect. Our work therefore establishes the existence of a new observable in all higher odd spacetime dimensions, and resolves the puzzle outlined in Section \ref{sec:no-memory}.

\subsection{Even dimensions}

We start with the gauge field sourced by a single particle of momentum $p_i$
\begin{equation}
\label{eq:gf-massless-even}
\begin{split}
    A_i^{\mu}(x) &= -\frac{1}{2 \pi^{\frac{d - 2}{2}}}    \frac{Q_i p_i^{\mu}}{2x\cdot p_i}  \delta^{(\frac{d - 6}{2})}(-x^2) .
    \end{split}
\end{equation}
The associated field strength is given by
\begin{equation}
    F_i^{\mu\nu}(x) = -\frac{Q_i}{2\pi^{\frac{d - 2}{2}}} \frac{p_i^{\mu} x^{\nu} - p_i^{\nu} x^{\mu}}{x \cdot p_i} \delta^{(\frac{d - 4}{2})}(-x^2).
\end{equation}
It will be convenient to associate to the choice of coordinates \eqref{eq:coords} a null frame with components
\begin{equation}
\label{eq:null-frame}
    e_u^{\mu} = \p_u x^{\mu} = n^{\mu}, \quad e_r^{\mu} = q^{\mu}, \quad e_A^{\mu} = r \p_A q^{\mu}. 
\end{equation}
Taking the Minkowski metric $\eta_{\mu\nu} = {\rm diag}(-1,1, 1, \cdots)$, one has the following useful identities
\begin{equation}
\label{eq:null-frame-identities}
 q(y)\cdot n=-\frac{1}{2},\qquad q(y)\cdot q(z)=-\frac{1}{2}|y-z|^2,\qquad x^2=-ru 
\end{equation}
and
\begin{equation}
\label{eq:transverse-identities}
\p_A q\cdot \p_B q=\delta_{AB},\qquad
x\cdot \p_A q=0,\qquad
\p_A q_{\mu}\p^A q_{\nu}
=\eta_{\mu\nu}+2(n_{\mu}q_{\nu}+q_{\mu}n_{\nu}) .
\end{equation}
The projection of the field strength to the frame \eqref{eq:null-frame} is then 
\begin{equation}
\label{eq:fs-even}
    F^i_{ab}(x)= - \frac{Q_i}{2 \pi^{\frac{d - 2}{2}}} \frac{(e_a \cdot p_i) (e_b \cdot x) - (e_a \cdot x)(e_b \cdot p_i)}{x \cdot p_i} \delta^{(\frac{d - 4}{2})}(-x^2).
\end{equation}
Equivalently, $F_{ab}(x)$ are field strength components in the coordinates \eqref{eq:Mink-lightcone}. 

We would like to examine the Maxwell constraint \eqref{eq:Maxwell} on a surface of finite, large $r_0$. It is clear from \eqref{eq:fs-even} that the leading contribution to the electromagnetic field is $\mathcal{O}(r_0^{-\frac{d}{2} + 1})$, which agrees with the standard radiative fall-off in higher dimensions \cite{Sokolowski:1993,Ortaggio:2014ipa}. On the other hand,  $j_u \sim \mathcal{O}(r_0^{-d+2})$, which is easy to see by writing \eqref{eq:current} in the lightcone coordinates \eqref{eq:Mink-lightcone} and considering radial trajectories with $\tau \sim r$. As a result, we expect that \eqref{eq:fs-even} obey the free Maxwell constraint order by order in $r_0^{-1}$, up to the Coulombic order $\mathcal{O}(r_0^{-d+2})$ at which point the source kicks in. 

To confirm this, we introduce
\begin{equation}
  Y_i \equiv \frac{u\, n\cdot p_i}{r_0\,q\cdot p_i},\qquad s\equiv \frac{d-4}{2},
\end{equation}
in which case
\begin{equation}
\label{eq:massless-even-components}
\begin{split}
   F^i_{ru}(x)
    &= - \frac{Q_i}{2\pi^{\frac{d - 2}{2}}}(n\cdot q)
    \frac{1-Y_i}{1+Y_i}\,\delta^{(s)}(-x^2),\\
   F^i_{Au}(x)
    & = - \frac{Q_i}{2\pi^{\frac{d - 2}{2}}}
    \frac{r_0(\p_Aq\cdot p_i)(n\cdot q)}{q\cdot p_i}
    \frac{1}{1+Y_i}\,\delta^{(s)}(-x^2).
    \end{split}
\end{equation}
Furthermore,
\begin{equation}
\label{eq:delta-r-u}
   \delta^{(s)}(-x^2)=r_0^{-s-1}\delta^{(s)}(u).
\end{equation}
It is easy to see that the leading term in \eqref{eq:Maxwell-constraint-fixed-r} is proportional to $Q_i\delta^{(s+1)}(u)$ with a $p_i$-independent proportionality constant. As such, the net field strength from a collection of particles vanishes by charge conservation
\begin{equation}
    \sum_{i = 1}^{M+N} Q_i = 0.
\end{equation}

For the subleading orders, we will use the Taylor expansions
\begin{equation}
\frac{1-Y_i}{1+Y_i}=1+2\sum_{k=1}^{s+1}(-1)^kY_i^k+\mathcal{O}(Y_i^{s+2}),
\qquad
\frac{1}{1+Y_i}=\sum_{k=0}^{s}(-1)^kY_i^k+\mathcal{O}(Y_i^{s+1}),
\end{equation}
and the distributional identity
\begin{equation}
\label{eq:u-delta-identity}
   u^\alpha \delta^{(\beta)}(u)=(-1)^\alpha\frac{\beta!}{(\beta-\alpha)!}\delta^{(\beta-\alpha)}(u),\qquad 0\leq\alpha\leq\beta .
\end{equation}
For $1\leq k\leq s$, order by order in the large-$r$ expansion the Maxwell constraint is source-free and proportional to $r_0^{s+1-k}\delta^{(s+1-k)}(u)$. The coefficient is given by
\begin{equation}
\label{eq:even-order-cancel}
\frac{4}{(s-k)!}\left(\frac{n\cdot p_i}{q\cdot p_i}\right)^k
-
\frac{1}{(s-k+1)!}
D^A\left[
\frac{p_i\cdot \p_Aq}{q\cdot p_i}
\left(\frac{n\cdot p_i}{q\cdot p_i}\right)^{k-1}
\right]=0 .
\end{equation}
The vanishing of this coefficient follows from the identity
\begin{equation}
\label{eq:massless-divergence-identity}
D^A\left[
\frac{p_i\cdot \p_Aq}{q\cdot p_i}
\left(\frac{n\cdot p_i}{q\cdot p_i}\right)^{k-1}
\right]
=2(d-2-2k)\left(\frac{n\cdot p_i}{q\cdot p_i}\right)^k
=4(s+1-k)\left(\frac{n\cdot p_i}{q\cdot p_i}\right)^k .
\end{equation}
It will be useful to rewrite this identity in terms $p_i =\omega q(z_i)$, namely
\begin{equation}
D^A\left(\frac{2(y-z_i)_A}{|y-z_i|^{2k}}\right)
=2(d-2-2k)|y-z_i|^{-2k},
\qquad y\neq z_i .
\end{equation}
We conclude that the free Maxwell constraint is satisfied order by order in a large-$r$ expansion up to the Coulombic order $\mathcal{O}(r^{-d+2})$, as expected.

Furthermore, at Coulombic order, namely for $k=s+1=\frac{d-2}{2}$, the $\p_uF_{ru}$ term in \eqref{eq:Maxwell-constraint-fixed-r} vanishes distributionally since
\begin{equation}
(s+1)u^s\delta^{(s)}(u)+u^{s+1}\delta^{(s+1)}(u)=0 .
\end{equation}
On the other hand, the angular term is non-vanishing and has non-trivial support at $y=z$ 
\begin{equation}
\label{eq:massless-divergence-delta}
D^A\left[
\frac{p_i\cdot \p_Aq}{q\cdot p_i}
\left(\frac{n\cdot p_i}{q\cdot p_i}\right)^s
\right]
= 2\Omega_{d-3}\delta^{(d-2)}(y-z_i),\qquad
\Omega_{d-3}=\frac{2\pi^{\frac{d-2}{2}}}{\Gamma(\frac{d-2}{2})}.
\end{equation}
In lightcone coordinates, this identity follows upon observing that
\begin{equation}
 \frac{p_i\cdot \p_Aq}{q\cdot p_i}
\left(\frac{n\cdot p_i}{q\cdot p_i}\right)^s=  2 \frac{(y - z_i)_A}{|y - z_i|^{d - 2}} = -2 D_A\left[\frac{1}{(d - 4)} \frac{1}{|y - z_i|^{d - 4}}\right] = -2 \Omega_{d-3} D_A G_{d-2}(y;z_i),
\end{equation}
where $G_{d - 2}(y;z_i)$ is the Green's function for the Laplacian on $\mathbb{R}^{d-2}$,
\begin{equation}
\label{eq:Gfdm2}
    \Box_{\perp}G_{d-2}(y;z) = -\delta^{(d-2)}(y,z), \quad G_{d - 2}(y;z) = \frac{1}{(d - 4) \Omega_{d - 3}} \frac{1}{|z - y|^{d-4}}.
\end{equation}
 As a result, there is a non-trivial memory effect in even spacetime dimensions measured by the observable
\begin{equation}
\label{eq:even-memory-observable}
 D^A \mathcal{M}_A^{\rm even}(y)\equiv \int_{-\infty}^{\infty}du\,D^A F_{Au}^{(d-4)}(u,y) = \sum_{i} Q_i \delta^{(d - 2)}(y - z_i).
\end{equation}
The sum in \eqref{eq:even-memory-observable} runs over both incoming and outgoing particles. On the other hand, we expect from the Maxwell constraint at $r > 0$ that only outgoing particles source memory, since there is no incoming flux on this lightsheet. This apparent puzzle is resolved by accounting for the shockwave contributions discussed in Section \ref{sec:shockwaves}.

It is not hard to see that the projection of \eqref{eq:fsshock} onto the null frame \eqref{eq:null-frame} yields\footnote{Recall that $(y - z_i)_A \delta(|y - z_i|) = 0$.}
\begin{equation}
\label{eq:shock-memory}
    \begin{split}
        \Delta F_{ru}^{\rm shock} \Big|_{r = r_0} &\equiv F_{ru}(u = \infty) - F_{ru}(u = -\infty) = 0, \\
        F_{Au}^{\rm shock}\Big|_{r = r_0} &= -\frac{r_0^{4 - d}}{2\pi^{\frac{d}{2} - 1}}\Gamma\left( \frac{d- 2}{2}\right) \sum_{i = 1}^M Q_i  \frac{(y - z_i)_A}{|y - z_i|^{d - 2}} \Theta(-u) \delta(u + r_0 |y - z_i|^2).
    \end{split}
\end{equation}
Therefore 
\begin{equation}
\int_{-\infty}^{\infty} du D^A F_{Au}^{\rm shock} = -\sum_{i = 1}^M Q_i \delta^{(d - 2)}(y - z_i),
\end{equation}
which precisely cancels the memory contribution due to incoming particles in \eqref{eq:even-memory-observable}.

We therefore recover the Coulombic order component of the constraint \eqref{eq:Maxwell-constraint-fixed-r} 
\begin{equation}
\label{eq:Coulombfau-field}
  D^A F_{Au}^{(d - 4)} = j_u^{(d - 2)},
\end{equation}
where 
\begin{equation}
    j_u = -\lim_{r \rightarrow \infty} r^{d - 2} J_u
\end{equation}
and $J_u = n \cdot J$ is the outgoing particle current in \eqref{eq:current}.\footnote{At large-$r$, $\lim r^{d - 2} n\cdot J = -\frac{Q_i}{2} 2 \delta(u) \delta^{(d - 2)}(y - z_i)$ with $-\frac{1}{2}$ from $n \cdot p$ and $2$ from the Jacobian multiplying the delta function in our coordinates. }  
Equivalently, the Coulombic memory is controlled by the particle flux through the lightsheet defined by $r = r_0$, namely
\begin{equation}
\label{eq:even-memory-constraint}
  D^A\mathcal{M}_A^{\rm even}(y)=\int_{-\infty}^{\infty}du\,j_u^{(d-2)}(u,y).
\end{equation}

For more general field configurations, the memory can contain a further contribution from the change of $F_{ru}$, or equivalently the large-gauge charge aspects, across $u$. The latter vanishes for the simple particle scattering configurations considered here, but can be non-vanishing more generally. The local constraint \eqref{eq:even-memory-constraint} remains non-trivial and will be shown in Section \ref{eq:mc-st} to follow from the Weinberg soft theorem.

\subsection{Odd dimensions}
\label{sec:odd-mem}

In odd spacetime dimension the retarded Green's function \eqref{eq:Godd} has support inside the light cone. After subtracting the divergent terms in the gauge potential \eqref{eq:gf-odd}, the field strength of a single source of momentum $p_i$, is given by
\begin{equation}
\label{eq:odd-finite-fieldstrength}
(F_i^{\rm fin})^{\mu\nu}(x)
= -\,N Q_i\,
\Theta(-x^2)
\frac{x^{\mu}p_i^{\nu}-x^{\nu}p_i^{\mu}}{x\cdot p_i}
(-x^2)^{1-\frac d2},
\qquad
N =\frac{(-1)^{\frac{d-1}{2}}}{2\pi^{\frac d2}}\Gamma\!\left(\frac{d-2}{2}\right),
\end{equation}
where $F_i^{\rm fin}$ excludes the divergent contact terms at $x^2=0$. One can show that these terms will not contribute to the Coulombic order constraint. We present the derivation of this fact in Appendix \ref{sec:div-odd}. We will henceforth restrict our attention to the finite terms.
 
As before, the relevant field strength components are 
\begin{equation}
\label{eq:odd-components}
\begin{split}
        F_{ru}^{{\rm fin},i}(x)
        &=  -N Q_i\,\Theta(-x^2)(q\cdot n)
        \frac{u(n\cdot p_i)-r(q\cdot p_i)}{x\cdot p_i}
        (-x^2)^{1-\frac d2},\\
        F^{{\rm fin},i}_{Au}(x)
        &= -N Q_i \,\Theta(-x^2)
        \frac{r^2(n\cdot q)(\p_Aq\cdot p_i)}{x\cdot p_i}
        (-x^2)^{1-\frac d2}.
    \end{split}
\end{equation}
At fixed $r=r_0$, $\Theta(-x^2)=\Theta(u)$ and $F_{ru}\rightarrow0$ at both endpoints of the $u$ integral for all odd $d\geq 3$. We next show that integrating \eqref{eq:Maxwell-constraint-fixed-r} in retarded time again yields a local relation between the memory
\begin{equation}
\label{eq:memory-odd-d0}
  \mathcal{M}_A^{\rm odd}(y)\equiv \lim_{r_0 \rightarrow \infty} r_0^{d - 4} \int_{-\infty}^{\infty} du\, F_{Au}^{\rm fin}(u,r_0,y)
\end{equation}
and the flux through the fixed-$r_0$ hypersurface. In particular, this memory operator will be shown to be finite and take precisely the same form as in even dimensions, provided that the $u$ integral is evaluated prior to taking $r_0 \rightarrow \infty$.

For a massless momentum $p_i =\omega q(z_i)$, the scalar products in \eqref{eq:odd-components} evaluate to
\begin{equation}
x\cdot p_i=-\frac{\omega}{2}\left(u+r_0|y-z_i|^2\right),
\qquad
\p_Aq(y)\cdot p_i =-\omega(y-z_i)_A .
\end{equation}
In these coordinates, the integral that appears in \eqref{eq:memory-odd-d0} is
\begin{equation}
\label{eq:memory-obs-odd}
\begin{split}
 \mathcal{M}^{\rm odd}_A(y)
   &\equiv
 - N \sum_{i} Q_i \lim_{r_0 \rightarrow \infty} r_0^{d - 4} \int_0^\infty du\,
   \frac{r_0^{3-\frac d2}u^{1-\frac d2}(y-z_i)_A}
   {u+r_0|y-z_i|^2}\\
   &=
  - N \sum_{i} Q_i \pi(-1)^{\frac{d+1}{2}}
   \frac{(y-z_i)_A}{|y-z_i|^{d-2}}
   = -\sum_{i} Q_i \p_A G_{d-2}(y;z_i).
   \end{split}
\end{equation}
Here the integral is defined by analytic continuation from the retarded $i\epsilon$ prescription, and $G_{d-2}(y,z_i)$ is the Green's function in $\mathbb{R}^{d-2}$ defined in \eqref{eq:Gfdm2}. 
\begin{figure}
\centering
\begin{tikzpicture}[
  scale=1.15,
  boundary/.style={thick},
  lightsheet/.style={thick,color=green!65!black},
  null/.style={thick, dashed},
  region/.style={fill=gray!25, draw=black, thick},
  timelike/.style={fill=yellow!18, draw=none, fill opacity= 0.3},
  label/.style={font=\small},particle/.style={
  very thick,
  decoration={
    markings,
    mark=at position 0.5 with {\arrow{Latex[length=2.3mm]}}
  },
  postaction={decorate}
}
]

\def\L{3.4}
\def\b{0.42}

\coordinate (ip) at (0,\L);
\coordinate (im) at (0,-\L);
\coordinate (i0L) at (-\L,0);
\coordinate (i0R) at (\L,0);


\fill[timelike]
  (0,0)
  -- (1.65,1.65)
  -- (ip)
  -- (-1.65,1.65)
  -- cycle;

\draw[boundary] (im) -- (i0L) -- (ip) -- (i0R) -- cycle;

\draw[boundary] (-1.65,1.65) .. controls (-0.8,1.3) and (0.8,1.3)  .. (1.65,1.65);

\draw[null] (-1.65,1.65) .. controls (-0.5,1.9) and (0.5,1.9)  .. (1.65,1.65);

\draw[null] (-1.65,-1.65) .. controls (-0.8,-1.3) and (0.8,-1.3)  .. (1.65,-1.65);

\draw[boundary] (-1.65,-1.65) .. controls (-0.5,-1.9) and (0.5,-1.9)  .. (1.65,-1.65);


\draw[region]
  (0,\b) -- (\b,0) -- (0,-\b) -- (-\b,0) -- cycle;
\node[label] at (0,0) {$\mathcal S$};

\draw[lightsheet]
  (-\b,\L-\b) -- (\L-\b,-\b);

  \begin{pgfonlayer}{particlelayer}
  \draw[boundary] (-1.65,-1.65) -- (0,0);
\draw[particle, blue!70!black] (-1.45,-1.7) -- (0,0);
\draw[particle, blue!70!black, dashed]  (0.8,-1.45) -- (0,0);
\end{pgfonlayer}

\begin{pgfonlayer}{particlelayer}
\draw[particle, red!75!black] (0,0) -- (-0.8,1.45);
\draw[boundary] (-1.65,1.65) -- (0,0);
\draw[boundary] (1.65,-1.65) -- (0,0);
\draw[boundary] (0,0) -- (1.65,1.65);
\draw[particle, red!75!black, dashed] (0,0) -- (1.45,1.7);
\end{pgfonlayer}

\node[label, above] at (ip) {$i^+$};
\node[label, below] at (im) {$i^-$};
\node[label, left] at (i0L) {$i^0$};
\node[label, right] at (i0R) {$i^0$};

\node[label, rotate=0] at (2.5,1.8) {$\mathscr I^+$};
\node[label, rotate=0] at (2.5,-1.8) {$\mathscr I^-$};

\end{tikzpicture}
\caption{Outgoing massless particles crossing a null plane at finite distance (green) source an electromagnetic memory at the Coulombic order.  In odd $d$ the radiative field is non-vanishing inside the future (yellow) and past lightcones.}
\label{fig:light-sheet-0}
\end{figure}
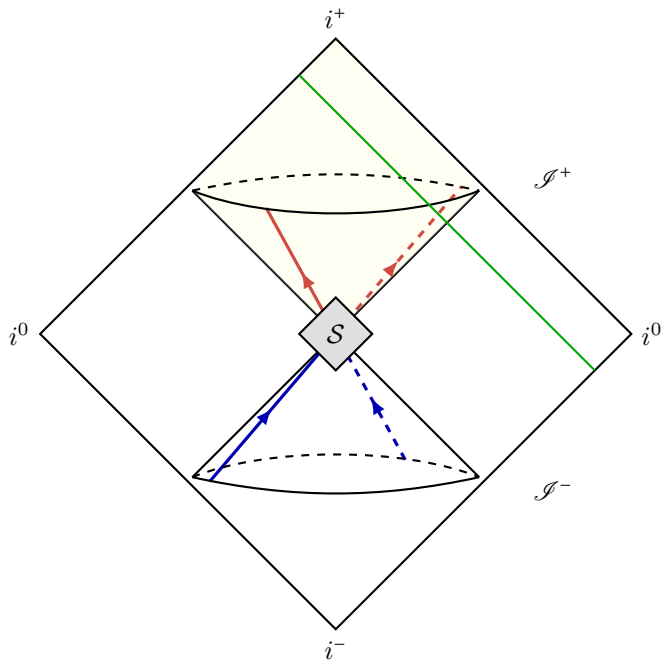
The shockwave contributions to the Coulombic memory take precisely the same form as in even $d$ (see \eqref{eq:shock-memory}). They again cancel the incoming terms in \eqref{eq:memory-obs-odd}, in agreement with the expectation from the Maxwell constraint. Consequently, summing over the electromagnetic fields sourced by all particles, the Coulombic order Maxwell constraint \eqref{eq:Maxwell-constraint-fixed-r} implies
\begin{equation}
\label{eq:odd-memory-constraint}
 \lim_{r_0 \rightarrow \infty} r_0^{d-4} \int_{-\infty}^{\infty}du\,D^AF_{Au}^{\rm fin}
  =
  \int_{-\infty}^{\infty}du\,j_u^{(d - 2)}
\end{equation}
at every point of the transverse plane. 
Thus the net large gauge charge again vanishes for the simple configuration of sources considered here, while the local Maxwell constraint produces the same non-zero memory profile governed by the flux through the lightsheet at $r = r_0$, just as in even spacetime dimension, see \eqref{eq:even-memory-constraint}. We note that only outgoing particles cross the light-sheet at $r = r_0$, as illustrated in Figure \ref{fig:light-sheet-0}. In spherical Bondi coordinates, the lightsheet would be replaced by a timelike cylinder at large radius which is clearly crossed by both the incoming and the outgoing massless particles. In this case, compatibility with the constraint implies that the memory effect receives contributions from both incoming and outgoing particles. We hence expect that the Coulombic shockwaves will not contribute to memory on this surface since it is outside the shockwave locus $x\cdot p = 0$.

We conclude this section by highlighting the key difference with respect to even dimensions. Expanding the integrand in \eqref{eq:memory-obs-odd} at large $r_0$ first, one obtains a half-integer power series in $r^{-1}$, namely
\begin{equation}
\label{eq:odd-noncommuting-expansion}
\frac{r_0^{3-\frac{d}{2}}u^{1-\frac{d}{2}}}{u+r_0|y-z_i|^2}
=
r_0^{2-\frac{d}{2}}
\sum_{\ell=0}^{\infty}
\frac{(-1)^\ell u^{\ell+1-\frac{d}{2}}}
{r_0^{\ell}|y-z_i|^{2\ell+2}}.
\end{equation}
 The memory integral \eqref{eq:memory-obs-odd} vanishes term by term up to the Coulombic order (which is absent in the series) and diverges for all subleading orders. The Coulombic contribution \eqref{eq:memory-obs-odd} is only recovered  \textit{after resumming the entire series} before performing the $u$ integral. In other words, the large-$r$ expansion and the memory integral fail to commute in odd spacetime dimensions. We will encounter the same phenomenon for massive sources in Section \ref{sec:massive}.

\section{Maxwell constraint from soft theorem}
\label{eq:mc-st}

In this section we show that the Maxwell constraint is equivalent to the soft theorem. Furthermore, we will identify the leading soft photon insertion with the memory operators \eqref{eq:even-memory-observable} and \eqref{eq:memory-obs-odd} in both even and odd spacetime dimension. As we will see, this relation is universally valid and can be established by computing the exact (i.e. to all orders in $r^{-1}$) Fourier transform of the Weinberg soft factor. This closes the gap regarding the relation between asymptotic symmetries, soft theorems and memory observables in electromagnetism in all spacetime dimensions.

 Since we are interested in establishing a relation to the Maxwell constraint \eqref{eq:Maxwell-constraint-fixed-r}, it is natural to consider the following S-matrix element
\begin{equation}
\label{eq:A-S-matr}
     \langle {\rm out}| A^{\rm out}_{\mu}(x) S -  S A^{\rm in}_{\mu}(x)|{\rm in}\rangle.
\end{equation}
Here $A^{\rm in/out}_{\mu}(x)$ are asymptotic (free) gauge fields which admit the mode expansion \cite{He:2014cra}
\begin{equation}
      A_{\mu}^{\rm in/out}(x) = \int \frac{d^{d-1}q}{2 q^0 (2\pi)^{d-1}} \left[e^{-i q\cdot x} a^{\dagger}_{\mu}(q)  + e^{i q \cdot x} a_{\mu}(q) \right]
\end{equation}
and 
\begin{equation}
    a_{\mu}(q) = \sum_{\alpha} \epsilon_{\mu}^{\alpha}(\hat{q}) a_{\alpha}(q), \quad q = \omega \hat{q},
\end{equation}
where $\alpha$ are the $(d-2)$ polarization indices. 
Substituting this expansion into \eqref{eq:A-S-matr}, we obtain
\begin{equation}
\label{eq:A-comm-0}
     \langle {\rm out}| A^{\rm out}_{\mu}(x) S -  S A^{\rm in}_{\mu}(x)|{\rm in}\rangle =  \int \frac{d^{d-1}q}{2 q^0 (2\pi)^{d-1}} e^{i q\cdot x}\langle {\rm out}| a_{\mu}(q) S |{\rm in}\rangle - \int \frac{d^{d-1}q}{2 q^0 (2\pi)^{d-1}} e^{-i q\cdot x}\langle {\rm out}| Sa^{\dagger}_{\mu}(q) |{\rm in}\rangle .
\end{equation}
The building blocks of \eqref{eq:A-S-matr} are therefore scattering amplitudes involving a photon emission. We will consider the rawest approximation to these amplitudes given by the  Weinberg soft theorem  \cite{Weinberg:1965nx}. This states that for low photon energies, the amplitude factorizes as
\begin{equation}
\label{eq:Weinberg}
     \langle {\rm out}| a_{\pm}(q) S|{\rm in}\rangle = \sum_{i = 1}^{n} Q_i\frac{p_i \cdot \epsilon^{\pm}(\hat{q})}{ \omega p_i \cdot \hat{q} - i\epsilon} \langle {\rm out}| S|{\rm in}\rangle + \mathcal{O}(\omega^0).
\end{equation}
Here the $i\epsilon$ prescription is inherited from the Feynman propagator. We continue working with the all outgoing convention \eqref{eq:all-outgoing}, from which we infer that 
\begin{equation}
     \langle {\rm out}| S a_{\pm}^{\dagger}(q) |{\rm in}\rangle = -\sum_{i = 1}^{n} Q_i\frac{p_i \cdot \epsilon^{\pm}(\hat{q})}{ \omega p_i \cdot \hat{q} + i\epsilon} \langle {\rm out}| S|{\rm in}\rangle + \mathcal{O}(\omega^0).
\end{equation}
Substituting these formulas into \eqref{eq:A-comm-0} and using the completeness relation for the polarization tensors, we obtain 
\begin{equation}
\label{eq:soft-fact0}
\begin{split}
     \langle {\rm out}| A^{\rm out}_{\mu}(x) S  - S A^{\rm in}_{\mu}(x)|{\rm in}\rangle &\simeq \sum_{i = 1}^n \int \frac{d^{d-1}q}{2 q^0 (2\pi)^{d-1}} \left(e^{iq\cdot x}\frac{ Q_i p_{i,\mu}}{p_i \cdot q - i\epsilon} + e^{-iq \cdot x}\frac{Q_i p_{i,\mu}}{p_i \cdot q + i\epsilon} \right)  \langle {\rm out}|S|{\rm in}\rangle.
     \end{split}
\end{equation}
The Fourier transforms of the soft factors can be evaluated using the identities
\begin{equation}
\label{eq:FT-soft-factor}
\begin{split}
   \int \frac{d^{d-1} q}{2 q^0} e^{i q \cdot x} \frac{p_i^{\mu}}{p_i \cdot q-i\epsilon}
    &=
    C_d\,\frac{(x^2 + i\epsilon\,x^0)^{2-\frac d2}}{x\cdot p_i},\\
    \int \frac{d^{d-1} q}{2q^0} e^{-iq \cdot x} \frac{p_i^{\mu}}{p_i \cdot q +i\epsilon}
    &=
    -C_d\,\frac{(x^2 - i\epsilon\,x^0)^{2-\frac d2}}{x\cdot p_i},\\
   C_d & \equiv -\frac{i\,2^{d-3}\pi^{\frac d2-1}\Gamma(\frac d2-1)}{d-4}.
    \end{split}
\end{equation}
We provide a derivation of these formulae in Appendix \ref{app:sf}. 

As a result, we obtain 
\begin{equation}
\label{eq:A-comm}
      \langle {\rm out}| A^{\rm out}_{\mu}(x) S  - S A^{\rm in}_{\mu}(x)|{\rm in}\rangle \simeq \frac{C_d}{(2\pi)^{d-1}}\sum_{i=1}^n \frac{Q_i p_{i,\mu}}{x \cdot p_i} \left((x^2 + i\epsilon x^0)^{2 - d/2} -  (x^2 - i\epsilon x^0)^{2 - d/2} \right)\langle {\rm out}|S|{\rm in}\rangle.
\end{equation}
Note that the results contain all order terms at large $r$, and not just the leading radiative $\mathcal{O}(r^{-\frac{d}{2} + 1})$ component. On the other hand, the application of the soft theorem picks out the component of the gauge field sourced by freely moving particles of constant momenta $p_i$.
Interactions will introduce acceleration, as reflected by the subleading corrections in $\omega$ to \eqref{eq:Weinberg}. As we are interested in establishing a relation to the Coulombic component of the Maxwell constraint and the leading electromagnetic memory effect, \eqref{eq:A-comm}  will suffice. 
Indeed, we will show next that the RHS of \eqref{eq:A-comm} agrees \textit{precisely} with the retarded electromagnetic potentials computed in sections \ref{sec:even-dim-gf} and \ref{sec:odd-dim-gf}. As before, the even and odd-dimensional cases have to be treated separately. 

For even $d = 2m + 2$, $m > 1$, the RHS of \eqref{eq:A-comm} vanishes up to contact terms. Using the identity 
\begin{equation}
\label{eq:even-discontinuity}
\begin{split}
   ( x^2+i\epsilon\,x^0)^{1-m}-(x^2-i\epsilon\,x^0)^{1-m}
    &= \frac{(-1)^{m-2}}{\Gamma(m-1)}
    \p_{x^2}^{m-2}
    \left[\frac{1}{x^2+i\epsilon\,x^0}-\frac{1}{x^2-i\epsilon\,x^0}\right]\\
    &= -\frac{2\pi i}{\Gamma(m-1)}\p_{x^2}^{m-2}\delta(-x^2), \quad x^0 > 0,
    \end{split}
\end{equation}
we obtain
\begin{equation}
\label{eq:soft-even-position}
    \frac{\langle {\rm out}| A^{\rm out}_{\mu}(x) \mathcal{S}  - \mathcal{S} A^{\rm in}_{\mu}(x)|{\rm in}\rangle}{\langle{\rm out}|\mathcal{S}|{\rm in}\rangle}
     \simeq
    - \frac{1}{2 \pi^{\frac d2-1}}\sum_{i=1}^n
     \frac{Q_i p_{i,\mu}}{2 x\cdot p_i}
     \delta^{(\frac d2-3)}(-x^2).
\end{equation}
This result agrees precisely with \eqref{eq:gf-massless-even}. 

For odd $d$, the identity
\begin{equation}
   (x^2+i\epsilon\,x^0)^{2-\frac d2}
    =
    \Theta(x^2)(x^2)^{2-\frac d2}
    +
    \Theta(-x^2)(-x^2)^{2-\frac d2}e^{-\frac{i\pi d}{2}\sgn(x^0)}
\end{equation}
implies that
\begin{equation}
\label{eq:theta-id}
(x^2+i\epsilon\,x^0)^{2-\frac d2}
-(x^2-i\epsilon\,x^0)^{2-\frac d2}
=-2i\Theta(-x^2)(-x^2)^{2-\frac d2} \sin \left(\frac{\pi d}{2}\right), \quad x^0 > 0.
\end{equation}
Therefore
\begin{equation}
\label{eq:soft-odd-position}
   \frac{\langle {\rm out}| A^{\rm out}_{\mu}(x) S  - S A^{\rm in}_{\mu}(x)|{\rm in}\rangle}{\langle{\rm out}|S|{\rm in}\rangle}
     \simeq
     -\frac{(-1)^{\frac{d-1}{2}}}{2 \pi^{d/2}}
     \frac{\Gamma(\frac d2-1)}{d-4}
     \sum_{i=1}^n \frac{Q_i p_{i,\mu}}{2x\cdot p_i}
     \Theta(-x^2)(-x^2)^{2-\frac d2}.
\end{equation}
This expression again agrees with the finite ($\epsilon_i$-independent) component of the odd-dimensional gauge field \eqref{eq:gf-odd}. More simply, we could have observed that \eqref{eq:A-comm} agrees precisely with the formula \eqref{eq:even-odd-bdry-gf} which holds for both even and odd $d.$ We note that the use of Feynman propagators in the scattering amplitudes computation should relate to the use of a time-ordered, as opposed to a retarded/advanced Green's function in the computation of the classical gauge fields in Section \ref{sec:em-gf}. For massless sources, this difference is inconsequential, in part due to having computed a matrix element of the commutator a photon with the S-matrix. We will see this difference manifest for massive sources in Section \ref{sec:soft-thm-massive}. 

As a result, we obtain the S-matrix identity
\begin{equation}
\label{eq:Ward-id}
   \Box \langle {\rm out}| A_{\mu}^{\rm out}(x) S - S A_{\mu}^{\rm in}(x)|{\rm in}\rangle = -\langle {\rm out}| J_{\mu}(x)S - S J_{\mu}(x)|{\rm in}\rangle
\end{equation}
where 
\begin{equation}
\label{eq:current-Sm}
\langle {\rm out}| J^{\rm out}_{\mu}(x)S - S J^{\rm in}_{\mu}(x)|{\rm in}\rangle \simeq \sum_{i = 1}^n Q_i p_i^{\mu} \int_0^{\infty}d \tau \delta^{(d)}(x - p_i\tau) \langle {\rm out}|S|{\rm in}\rangle.
\end{equation}
We confirm \eqref{eq:current-Sm} from an independent computation involving strictly the action of matter currents in the eikonal approximation
\begin{equation}
\label{eq:eik}
    J^{\rm in}_{\mu}(x)|p_i\rangle \simeq Q_i \int \frac{d^{d-1}k}{(2\pi)^{d-1}2k^0 } (k+p_i)_{\mu} e^{-i(k-p_i)\cdot x} |k\rangle \simeq \frac{p_{i,\mu}}{p_i^0} \delta^{(d-1)}\Big(\vec{x} - \frac{\vec{p}_i}{p_i^0}x^0\Big) |p_i\rangle.
\end{equation}
We derive \eqref{eq:eik} and therefore \eqref{eq:current-Sm} in Appendix \ref{app:current}. 

We have verified \eqref{eq:Ward-id} in the eikonal/soft limit, however it holds more generally and, in particular, for arbitrary currents. Indeed, \eqref{eq:Ward-id} is closely related to the photon Ward identity first derived by Yang and Feldman \cite{YangFeldman1950}. The soft/eikonal approximation first appeared in \cite{Bloch:1937pw}, and was built upon in \cite{Low:1958sn, Weinberg:1965nx,Kibble1968}. More recently, these ideas were shown to imply a relation between soft theorems and asymptotic symmetries, see \cite{Strominger:2017zoo} for a review.  In these works, the eikonal/soft approximation was shown to be implemented by a large $r$ expansion of the matching condition for asymptotic charges across spatial infinity. While not explicitly stated in these works, it is clear that the Ward identity \eqref{eq:Ward-id} underlies this more modern perspective on soft theorems.

\begin{figure}
\centering
\begin{tikzpicture}[
  scale=1.15,
  boundary/.style={thick},
  lightsheet/.style={thick,color=green!65!black},
  null/.style={thin, dashed},
  region/.style={fill=gray!25, draw=black, thick},
  spacelike/.style={fill=teal!18, draw=none},
  label/.style={font=\small},particle/.style={
  very thick,
  decoration={
    markings,
    mark=at position 0.5 with {\arrow{Latex[length=2.3mm]}}
  },
  postaction={decorate}
}
]

\def\L{3.4}
\def\b{0.42}

\coordinate (ip) at (0,\L);
\coordinate (im) at (0,-\L);
\coordinate (i0L) at (-\L,0);
\coordinate (i0R) at (\L,0);

\fill[spacelike]
  (\b,0)
  -- ({(\L+\b)/2},{(\L-\b)/2})
  -- (i0R)
  -- ({(\L+\b)/2},{-(\L-\b)/2})
  -- cycle;

\fill[spacelike]
  (-\b,0)
  -- ({-(\L+\b)/2},{(\L-\b)/2})
  -- (i0L)
  -- ({-(\L+\b)/2},{-(\L-\b)/2})
  -- cycle;

\draw[boundary] (im) -- (i0L) -- (ip) -- (i0R) -- cycle;

\draw[null] (-\b,0) -- ({-(\L+\b)/2},{(\L-\b)/2});
\draw[null] (\b,0)  -- ({(\L+\b)/2},{(\L-\b)/2});
\draw[null] (-\b,0) -- ({-(\L+\b)/2},{-(\L-\b)/2});
\draw[null] (\b,0)  -- ({(\L+\b)/2},{-(\L-\b)/2});

\draw[region]
  (0,\b) -- (\b,0) -- (0,-\b) -- (-\b,0) -- cycle;
\node[label] at (0,0) {$\mathcal S$};

\draw[lightsheet]
  (\L-\b,-\b) -- (-\b,\L-\b);

  \begin{pgfonlayer}{particlelayer}
\draw[particle, blue!70!black] (-1.65,-1.65) -- (0,0);
\draw[particle, blue!70!black] (1.65,-1.65) -- (0,0);

\draw[particle, red!75!black] (0,0) -- (-1.65,1.65);
\draw[particle, red!75!black] (0,0) -- (1.65,1.65);
\end{pgfonlayer}

\node[label, above] at (ip) {$i^+$};
\node[label, below] at (im) {$i^-$};
\node[label, left] at (i0L) {$i^0$};
\node[label, right] at (i0R) {$i^0$};

\node[label, rotate=0] at (2.5,1.8) {$\mathscr I^+$};
\node[label, rotate=0] at (2.5,-1.8) {$\mathscr I^-$};


\end{tikzpicture}
\caption{Massless particles interacting in region $\mathcal{S}$. The electric field in the regions (blue) spacelike separated from $\mathcal{S}$, and, in particular, in the intersection of the lightsheet at $r = r_0$  (green) with this region commutes with the S-matrix. Note that for generic massless source configurations, only outgoing particles cross the lightsheet.}
\label{fig:matching}
\end{figure}
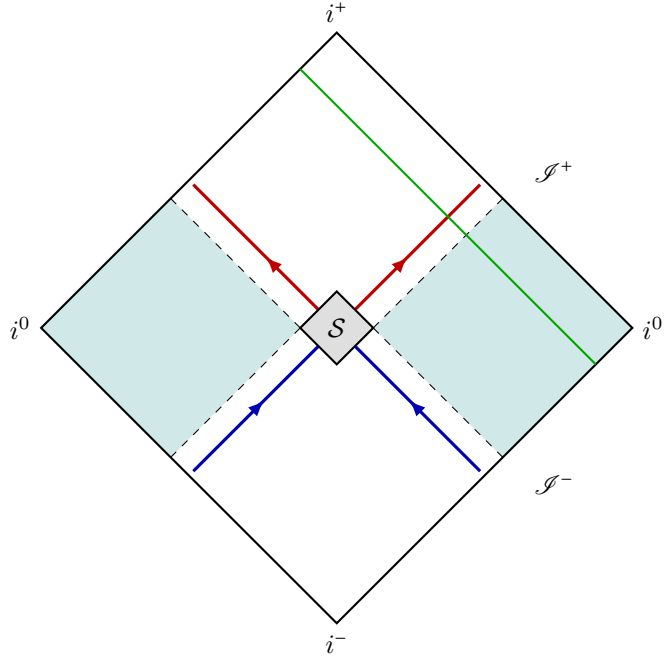

To see the relation explicitly, we start by assuming the conservation of large-gauge charge aspects in higher dimensions
\begin{equation}
\label{eq:Matching-large-r-0}
   \lim_{r \rightarrow r_0} r^{d - 2} \langle {\rm out}|  F_{ru}(x) S - S F_{ru}(x)|{\rm in}\rangle = 0
\end{equation}
which should hold at spacelike separation from the interaction region. Assuming that $F_{ru}(x)$ vanishes at early/late times, we can extend the local conservation law \eqref{eq:Matching-large-r-0} to 
\begin{equation}
\label{eq:Matching-large-r}
   \lim_{r \rightarrow r_0} r^{d - 2} \int^{\infty}_{-\infty} du \langle {\rm out}| \p_u F_{ru}(x) S - S \p_u F_{ru}(x)|{\rm in}\rangle = 0
\end{equation}
and simply apply the constraint \eqref{eq:Maxwell-constraint-fixed-r}. We illustrate the setup in Figure \ref{fig:matching}. Note that the antipodal matching condition is not necessary in this derivation since, unlike the Bondi coordinates, the lightcone coordinates \eqref{eq:coords} cover the whole of Minkowski space.

Equation \eqref{eq:Matching-large-r} then becomes
\begin{equation}
\label{eq:constr-m}
    \lim_{r \rightarrow r_0} r^{d - 4} \int_{-\infty}^{\infty} du \langle {\rm out}| D^A F_{Au}(x) S - S D^A F_{Au}(x)|{\rm in}\rangle =  \lim_{r \rightarrow r_0} r^{d - 2} \int_{-\infty}^{\infty} du \langle {\rm out}| j_{u}(x) S - S j_u(x)|{\rm in}\rangle.
\end{equation}
This equation is nothing but the gauge-invariant version of \eqref{eq:Ward-id} projected onto a large $r = r_0$ hypersurface and its zero-frequency component. Again, the subtle point is that this $u$ integral in general may not commute with either the mode expansion of $F_{uA}(x)$, or the large $r$ limit $r_0 \rightarrow \infty$. 

The natural order of limits, based on the all orders in $r^{-1}$ relation \eqref{eq:Ward-id} is to compute the $u$ integral of \eqref{eq:Ward-id} or equivalently \eqref{eq:constr} at finite distance $r_0$ and then take $r_0 \rightarrow \infty$. This is precisely the limit in which we showed in Section \ref{sec:Coulomb-mem} that a memory effect arises universally at Coulombic order in all spacetime dimensions. This  completes the identification of an infrared triangle in electromagnetism in all spacetime dimensions. The generalization to gravity is straightforward and will be outlined in Section \ref{sec:gravity}.

\subsection{Relation to previous works}
\label{sec:rel-previous}

In this section, we explain and resolve the apparent tension between our results and those of \cite{He:2019jjk}. Equation \eqref{eq:Coulombfau-field} implies that 
\begin{equation}
   \Delta F_{ur}^{C(2m)} = \int du \p_u F_{ur}^{C(2m)} = 0,
\end{equation}
and instead $F_{Au}^C \sim r^{-2m}$ in all higher spacetime dimensions. These results are naively inconsistent with those of \cite{He:2019jjk} who assume the Coulombic field strength fall-offs \eqref{eq:C-fall-off} at large $r$.

To reconcile these results,  
in this section, we show that \eqref{eq:odd-fs} relies on the assumption that the large-$r$ expansion commutes with the Fourier transform, which does not hold in odd spacetime dimensions. In particular, using the results of Sections \ref{sec:odd-mem} and \ref{eq:mc-st}, we will see that computing the Fourier transform prior to performing the large-$r$ expansions will result in a time-dependent Coulombic order electric field and a vanishing radiative field at the Coulombic order in odd spacetime dimensions.

We start with the definition eq. 2.15 in \cite{He:2019jjk}
\begin{equation}
\label{eq:Frus}
    F_{ru}(x) \propto \frac{1}{r} \int d^{2m} z \int_0^{\infty} d \omega \omega^{2m - 2} e^{-\frac{i\omega}{2}(u + r z^2)} \p^A O_A^{(0)}(y + z) + h.c. + \cdots,
\end{equation}
where $\cdots$ involve subleading terms in the small $\omega$ expansion of the $O_A$ modes. Evaluating the $\omega$ integrals, we obtain
\begin{equation}
\label{eq:Fru}
    F_{ru}(x) \propto \frac{1}{r} \int d^{2m} z \frac{1}{(u + r z^2 -i\epsilon)^{2m - 1}} \p^A O_A^{(0)}(y + z) + h.c. + \cdots .
\end{equation}
In \cite{He:2019jjk}, $O_A^{(0)}$ is then expressed in terms of its Fourier transform, then the $z$ integrals are evaluated and the result is expanded at large-$r$ to obtain \eqref{eq:odd-fs}. In even dimensions, it is easy to see that a radiative field component $F_{Au}^{(2m - 2)} \propto \delta(u)$ can indeed be ``sourced'' by the Coulomb field $F_{ru}^{(2m)} \propto \Theta(u)$. This generalizes the observation of \cite{Kapec:2017tkm} to higher even dimensions. In odd dimensions, the finite part of the Coulomb field is found to be time-independent and the free Maxwell constraint is trivially obeyed at the Coulombic order. 

To compare our results with those of \cite{He:2019jjk}, we first observe that \eqref{eq:Frus} is nothing but the Fourier transform of the soft factor computed in Section \ref{eq:mc-st}. To see this, we shift $z \rightarrow z - y$ and recall that
\begin{equation}
\label{eq:soft-fact}
    \p^A O_A^{(0)}(z) \propto Q_i \frac{\p^A \epsilon_A(z) \cdot p_i}{p_i \cdot q(z)}
\end{equation}
where we used the completeness of the polarization tensors, $\p^A \epsilon_A \propto n$ and $p_i^2 = 0$. Since $n$ is independent on $z$, we see that \eqref{eq:Frus} is nothing but Fourier transform of the soft factor, which was evaluated in \eqref{eq:FT-soft-factor}. In particular, we have seen that these expressions agree with the exact field strength \eqref{eq:odd-components}  sourced by point particles moving along straight line trajectories. 

We can now explicitly show that the large-$r$ expansion does not commute with the memory integral. Using \eqref{eq:odd-components}, we find
\begin{equation}
\label{eq:exp-fs-odd}
\begin{split}
    F_{ru}(x) &\propto \Theta(u) \frac{\frac{u}{r} (n \cdot p_i) - q\cdot p_i}{\frac{u}{r}+ |y - z_i|^2} (ur)^{1 - \frac{d}{2}} \propto \Theta(u) r^{2-d} \frac{\epsilon^{4 - d} (n\cdot p_i) - \epsilon^{2 - d} q\cdot p_i}{\epsilon^2+ |y - z_i|^2},\\
    F_{Au}(x) &\propto \Theta(u) r^{3 - d} \frac{\p_Aq \cdot p_i}{\frac{u}{r} + |y - z_i|^2}(ur)^{1 - \frac{d}{2}} \propto \Theta(u) r^{3 - d} \frac{\p_Aq \cdot p_i}{\epsilon^2 + |y - z_i|^2} \epsilon^{2 - d}
    \end{split}
\end{equation}
where we defined
\begin{equation}
    \epsilon^2 \equiv \frac{u}{r}.
\end{equation}
Then taking $\epsilon \rightarrow 0$, or equivalently expanding at large-$r$ first and using \cite{deGioia:2024yne}
\begin{equation}
\begin{split}
   \frac{\epsilon^{4 - d}}{\epsilon^2 + |y - z_i|^2} &= \pi^{\frac{d - 2}{2}}\Gamma\left(2 - \frac{d}{2} \right) \delta^{(d-2)}(y - z_i) + \cdots,\\
    \frac{\epsilon^{2 - d}}{\epsilon^2 + |y - z_i|^2} &= \pi^{\frac{d - 2}{2}}\frac{\Gamma\left(2 - \frac{d}{2} \right)}{\epsilon^2}\delta^{(d-2)}(y - z_i) + \cdots 
   \end{split}
\end{equation}
we see that $F_{ru}(x)$ contributes to the Coulombic order Maxwell constraint, while $r^{-2}F_{Au}(x)$ is one power of $r$ subleading. In this order of limits, the field-strength components obey the fall-offs \eqref{eq:C-fall-off}, the Coulombic order electric field $F_{ru}^{(d-2)}$ is sourced by the flux and the associated memory effect vanishes, in agreement with the results of \cite{He:2019jjk}. We emphasize that while \eqref{eq:Frus} is a linear superposition of plane waves, it only solves the source-free wave equation provided that the coefficients are analytic in $q$. Indeed, we have just seen that upon substituting \eqref{eq:soft-fact} into \eqref{eq:Frus}, one obtains the sourced solutions \eqref{eq:exp-fs-odd}.

As discussed before, these fall-offs are in sharp contrast to those we obtained in Section \ref{sec:odd-mem} in a different order of limits, where $r$ is kept fixed and the time integral involved in the definition of the memory effect is performed first. In that case, the flux through a surface at finite-distance instead sources a Coulombic memory effect and no change in the Coulomb field. That asymptotic expansions don't always commute with integrals is well known and has been observed in other physical applications, including \cite{Kologlu:2019mfz,Belin:2020lsr,Korchemsky:2021htm}. 

We next outline an experimental protocol that would allow for the leading memory effect to be measured in higher dimensions.

\section{Measuring the memory effect}
\label{sec:meas-mem}

In this section we identify the regime in which the odd-dimensional memory effect derived in Section \ref{sec:odd-mem} can be measured. In practice, all measurements are conducted at finite distance and in a finite amount of time. We first consider an idealized setup where a null detector/probe is placed at distance $r_0$ from the scattering center. We assume that the detector is switched on during the time-interval $u \in [-u_0, u_0]$ and measures the radiative flux through the light-sheet at $r = r_0$. This is obtained by simply adjusting the integration limits in \eqref{eq:memory-obs-odd}, namely
\begin{equation}
\label{eq:mem-meas}
  \widehat{\mathcal{M}}^i_{A} \equiv \int_{-u_0}^{u_0} du F_{Au}^{\rm fin, i} =  -N Q_i \int_0^{u_0} du \frac{r_0^{3-\frac d2}u^{1-\frac d2}(y-z_i)_A}
   {u+r_0|y-z_i|^2} = -NQ_i r_0^{3 - \frac{d}{2}} u_0^{1 - \frac{d}{2}}(y - z_i)_A I\left(\frac{u_0}{r_0 |y - z_i|^2}\right),
\end{equation}
where 
\begin{equation}
    I(x) = 2\frac{x}{4 - d} {}_2 F_1 \left[1, 2 - \frac{d}{2}; 3 - \frac{d}{2}; -x \right], \quad x \equiv \frac{u_0}{r_0|y - z_i|^2}.
\end{equation}
The small and large-$x$ expansions of the hypergeometric function are given by
\begin{equation}
\label{eq:hg-exp}
\begin{split}
    {}_2 F_1 \left[1, 2 - \frac{d}{2}; 3 - \frac{d}{2}; -x \right] & \sim \sum_{n = 0}^{\infty} (-1)^n \frac{4 - d}{4 - d + 2n} x^n, \quad x \sim 0, \\
     {}_2 F_1 \left[1, 2 - \frac{d}{2}; 3 - \frac{d}{2}; -x \right] & \sim \Gamma(3 - \frac{d}{2})\Gamma(\frac{d}{2} - 1) x^{\frac{d}{2} - 2} - (2 - \frac{d}{2}) \sum_{n = 0}^{\infty} \frac{(-1)^n}{n + \frac{d}{2} - 1}x^{-n-1}, \quad x \sim \infty.
    \end{split}
\end{equation}
Importantly, 
\begin{equation}
    \frac{1}{4 - d} \Gamma(3 - \frac{d}{2})\Gamma(\frac{d}{2} - 1) = -\frac{\pi}{2} \frac{1}{\sin \pi \frac{d}{2}} = (-1)^{\frac{d+1}{2}} \frac{\pi}{2}, \quad d ~ {\rm odd}
\end{equation}
so the leading term in the large-$x$ expansion contains a branch cut.

This allows us to identify the leading terms in the large/small-$x$ asymptotic expansions of the memory operator \eqref{eq:mem-meas}
\begin{equation}
\label{eq:memory-regimes}
    \begin{cases}
      \widehat{\mathcal{M}}^i_A \sim -NQ_i r_0^{4 - d} (x|y - z_i|^2)^{1 - \frac{d}{2}} (y - z_i)_A \dfrac{2x}{4 - d}, \quad  x \ll 1, \\
      \\
      \widehat{\mathcal{M}}^i_A \sim -(-1)^{\frac{d+1}{2}}  NQ_i \pi r_0^{4-d} ( |y - z_i|^2)^{1 - \frac{d}{2}}  (y - z_i)_A, \quad x \gg 1.
    \end{cases}
\end{equation}
At large $x$, $\mathcal{M}_A = r_0^{d - 4} \sum_i  \widehat{\mathcal{M}}_A^i$ is independent on $x$ and agrees with \eqref{eq:memory-obs-odd}. Consequently, both the odd-dimensional soft charge aspect and the soft charge
\begin{equation}
\label{eq:std-memory}
   Q_s(\epsilon) \equiv  \lim_{r_0 \rightarrow \infty}  \int d^{d-2}y \epsilon(y) D^A\mathcal{M}_A(y) = \sum_iQ_i \epsilon(z_i)
\end{equation}
are non-vanishing. We conclude that observers in \textit{any} odd dimension $d$ at fixed (large) radius $r_0$ will always observe the standard memory effect if they wait for a long enough time. For two charged probes at different transverse separation around $r = r_0$, the memory effect will induce a velocity kick in the Coulombic order trajectories, just as in even dimensional spacetimes \cite{Bieri:2013hqa,Mao:2017wvx,Campoleoni:2019ptc,Satishchandran:2019pyc}. While expected from the universal soft theorem, for odd dimensional spacetimes this effect is new and, in particular,  distinct from the tail memory discussed in \cite{Satishchandran:2017pek, Satishchandran:2019pyc, Sahoo:2026ulx}. 

To clarify why this effect was found to be absent in prior works \cite{Satishchandran:2017pek, Satishchandran:2019pyc}, consider the series expansion \eqref{eq:hg-exp} around $x \sim 0$. This amounts to performing the large-$r$ expansion prior to computing the memory integral. We observe once again that this series does not contain a term at the Coulombic order. As a result, the memory effect is given by the first case in \eqref{eq:memory-regimes}, 
\begin{equation}
\label{eq:naive}
    \widehat{\mathcal{M}}_A^i \sim -N Q_i \frac{2}{4 - d}\frac{u_0^{2 - \frac{d}{2}}}{r_0^{\frac{d}{2} - 3}} \frac{(y - z_i)_A}{|y - z_i|^2}.
\end{equation}
This is in agreement with the results of \cite{Satishchandran:2017pek} (cf. eq. 38 and the associated memory integral therein). Indeed, if one further takes $u_0 \rightarrow \infty$, \eqref{eq:naive} vanishes, up to interesting tail terms recently discussed in \cite{Sahoo:2026ulx}. It is nevertheless clear from \eqref{eq:memory-regimes} that this conclusion is only valid if one performs the large $r$ expansion prior to computing the memory integral, or implicitly assumes that the first equation in \eqref{eq:memory-regimes} continues to be valid much outside its regime of validity for $x \gg 1$. In other words, for long enough times, the infinite series of half-integer tails needs to be resumed and generates the standard memory observable in \eqref{eq:std-memory} in all higher odd dimensions. As discussed before, this conclusion remains valid if the boundary is taken to be a timelike cylinder at large distance, or in the context of this section, for timelike probes. We expect the effect will persist in a finite region of higher-dimensional Minkowski space such as a Rindler wedge, a causal diamond or in the vicinity of higher-dimensional black holes. Relatedly, it will revisit the results of \cite{Danielson:2022tdw,Danielson:2022sga,Danielson:2024yru,Biggs:2024dgp} in higher dimensions, and especially odd $d$ in the light of our results and their relation to soft theorems and asymptotic symmetries.

\section{Massive sources}
\label{sec:massive}

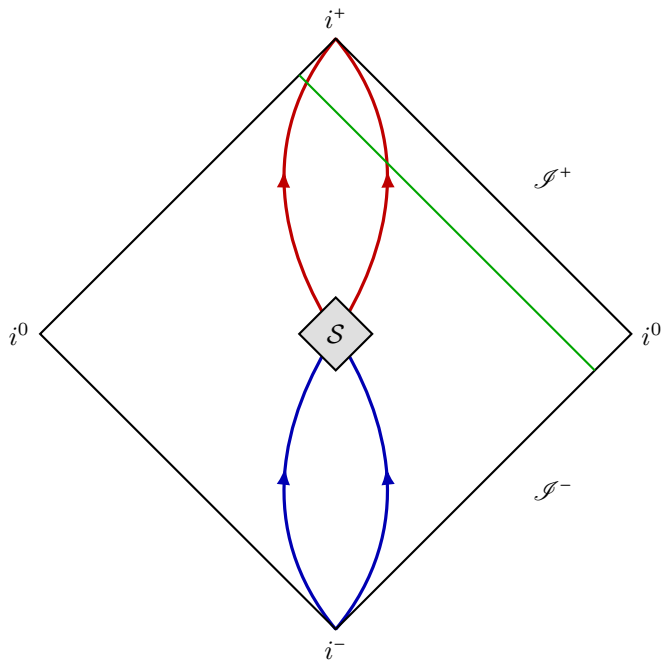
\begin{figure}
\centering
\begin{tikzpicture}[
  scale=1.15,
  boundary/.style={thick},
  lightsheet/.style={thick,color=green!65!black},
  massive/.style={
  very thick,
  decoration={
    markings,
    mark=at position 0.55 with {\arrow{Latex[length=2.4mm]}}
  },
  postaction={decorate}
},
  region/.style={fill=gray!25, draw=black, thick},
  spacelike/.style={fill=teal!18, draw=none},
  label/.style={font=\small},particle/.style={
  very thick,
  decoration={
    markings,
    mark=at position 0.5 with {\arrow{Latex[length=2.3mm]}}
  },
  postaction={decorate}
}
]

\def\L{3.4}
\def\b{0.42}

\coordinate (ip) at (0,\L);
\coordinate (im) at (0,-\L);
\coordinate (i0L) at (-\L,0);
\coordinate (i0R) at (\L,0);
\coordinate (S) at (0,0);

\draw[boundary] (im) -- (i0L) -- (ip) -- (i0R) -- cycle;

\draw[region]
  (0,\b) -- (\b,0) -- (0,-\b) -- (-\b,0) -- cycle;
\node[label] at (0,0) {$\mathcal S$};

\draw[lightsheet]
 (-\b,\L-\b) -- (\L-\b,-\b);

  \begin{pgfonlayer}{particlelayer}
\draw[massive, blue!70!black]
  (im) .. controls (-0.95,-2.25) and (-0.62,-0.95) .. (S);

\draw[massive, blue!70!black]
  (im) .. controls (0.95,-2.25) and (0.62,-0.95) .. (S);

\draw[massive, red!75!black]
  (S) .. controls (-0.62,0.95) and (-0.95,2.25) .. (ip);

\draw[massive, red!75!black]
  (S) .. controls (0.62,0.95) and (0.95,2.25) .. (ip);
\end{pgfonlayer}


\node[label, above] at (ip) {$i^+$};
\node[label, below] at (im) {$i^-$};
\node[label, left] at (i0L) {$i^0$};
\node[label, right] at (i0R) {$i^0$};

\node[label, rotate=0] at (2.5,1.8) {$\mathscr I^+$};
\node[label, rotate=0] at (2.5,-1.8) {$\mathscr I^-$};


\end{tikzpicture}
\caption{Massive particles do not reach $\mathcal{I}^{\pm}$ but outgoing massive particles cross the light-sheet at $r = r_0$. The Maxwell constraint on this hypersurface will therefore receive a non-vanishing contribution from the outgoing flux. This flux sources a universal memory effect at Coulombic order in both even and odd spacetime dimension.}
\label{fig:lightsheet-massive}
\end{figure}

In this section we generalize our analysis to massive sources. Massive worldlines do not reach $\mathcal{I}^{+}$ at a fixed retarded time \cite{Kapec:2015ena,Campiglia:2015kxa,Campiglia:2015qka, Strominger:2017zoo}. In even spacetime dimensions, we can evaluate the Maxwell constraint directly at $\mathcal{I}^{\pm}$, where the flux term is absent. The source data instead enters the constraint through the Coulombic boundary value of $F_{ru}$ at the future (past) endpoints of null infinity $\mathcal{I}^{+}_+$ ($\mathcal{I}^-_-$) which allows for a non-vanishing Coulombic memory effect. In odd spacetime dimensions, we will instead consider the Maxwell constraint evaluated on a lightsheet at finite, large distance $r = r_0$. It is not hard to see that outgoing massive sources do cross this lightsheet, as illustrated in Figure \ref{fig:lightsheet-massive}. Consequently, just as in the massless case, we will see that this flux sources a non-vanishing Coulombic memory effect, despite the fact that $\lim_{r_0 \rightarrow \infty} r_0^{d-2} \Delta F_{ru} = 0$ in this case. 
As before, we will consider pointlike particles moving along straight trajectories, so the same formulae for the current \eqref{eq:current0}, \eqref{eq:current} apply in this case. We will continue using the all outgoing conventions \eqref{eq:all-outgoing}. 

\subsection{Kinematics}

The formulas \eqref{eq:Gr}, \eqref{eq:Godd} continue to apply in this case, however the expressions for the gauge fields will change. For a source of momentum $p$, setting $b = 0$ as before, the lightcone condition
\begin{equation}
\label{eq:massive-roots-eq}
-(x-p\tau)^2=(\tau-\tau_+)(\tau-\tau_-) = 0,
\end{equation}
now has two distinct roots
\begin{equation}
\label{eq:tpm}
\tau_{\pm}=-x\cdot p \pm\sqrt{(x\cdot p)^2+x^2}.
\end{equation}
A useful parametrization of an outgoing momentum is \cite{Campiglia:2015kxa,Campiglia:2015qka}
\begin{equation}
\label{eq:massive-velocity-param}
p(\rho,z)=\rho^{-1}q(z)+\rho n,\qquad \rho>0,\qquad p^2=-1 .
\end{equation}
Here $\rho$ is a radial coordinate on the hyperbolic resolution of $i^+$.
Using \eqref{eq:null-frame-identities},
\begin{equation}
\label{eq:massive-dots}
q(y)\cdot p(\rho,z)=-\frac{\rho^2+|y-z|^2}{2\rho},\qquad
n\cdot p(\rho,z)=-\frac{1}{2\rho},\qquad
x\cdot p=-\frac{u+r(\rho^2+|y-z|^2)}{2\rho}.
\end{equation}
At large $r$ and fixed $u$,
\begin{equation}
\label{eq:massive-root-large-r}
\begin{split}
\tau_-&=\frac{\rho\,u}{\rho^2+|y-z|^2}+\mathcal{O}(r^{-1}),\qquad
\tau_+ =\frac{r(\rho^2+|y-z|^2)}{\rho}+ \frac{u}{\rho}\frac{|y-z|^2}{\rho^2+|y-z|^2} + \mathcal{O}(r^{-1}),\\
\tau_+-\tau_-&=\frac{r(\rho^2+|y-z|^2)}{\rho}+ \left( \frac{u}{\rho} - \f{2\rho u}{\rho^2 + |y-z|^2} \right) +\mathcal{O}(r^{-1}).
\end{split}
\end{equation}
The formulas for the incoming momenta are obtained by letting $p \rightarrow -p.$ Note that the large $r$ expansions become interchanged, namely $\tau_+ \leftrightarrow \tau_-$ for incoming particles.

\subsection{Even spacetime dimensions}
\label{sec:even-massive}

We start by evaluating the gauge field  sourced by a massive particle in even spacetime dimensions. The integral in \eqref{eq:gauge-field-even} is non-vanishing if and only if the roots $\tau_{\pm}$ obey the constraints imposed by the theta functions. These constraints are equivalent to demanding that the null vector
\begin{equation}
   N^{\mu} \equiv  x^{\mu} - p^{\mu} \tau_{\pm}, \quad N^2 = 0
\end{equation}
is future directed. Since this condition is preserved by orthochronous Lorentz transformations, we can verify it in the rest frame of the particle in which case
\begin{equation}
\label{eq:pm-rest-frame}
    \tau_{\pm} = x^0 {\rm sgn}(p^0) \pm |\vec{x}|.
\end{equation}
For outgoing particles, $p^0 > 0$ and demanding that $N^{\mu}$ is future-directed implies $\tau < x^0$.
Assuming $x^0 > 0,$ this is only satisfied by $\tau_-$.
 Furthermore, is it easy to see from \eqref{eq:pm-rest-frame} that
\begin{equation}
\label{eq:tau-ranges}
    \tau_- > 0 \iff x^2 < 0.
\end{equation}
For incoming particles, $p^0 < 0$ and demanding $N^{\mu}$ is future directed is equivalent to $ \tau > -x^0.$
Assuming $x^0> 0$, this condition can only be obeyed by $\tau_+$. Furthermore, 
\begin{equation}
    \tau_+ > 0 \iff x^2 > 0.
\end{equation}

To evaluate the integral \eqref{eq:gauge-field-even}, it will be convenient to define
\begin{equation}
    \label{eq:def-fd}
    \begin{split}
    f &\equiv (\tau - \tau_+) (\tau - \tau_-), \\
    \Delta &\equiv \tau_+ - \tau_-
    \end{split}
\end{equation}
and use the identity
\begin{equation}
\label{eq:df-id}
    \int_{\mathcal{I}} d\tau g(\tau) \delta^{(n)}(f) = \p_{\lambda}^n \sum_{\alpha = \pm, \tau_{\alpha}(\lambda) \in \mathcal{I}} \frac{g(\tau_{\alpha}(\lambda))}{\sqrt{\Delta^2 - 4 \lambda}} \left. \right|_{\lambda = 0},
\end{equation}
where $\mathcal{I}$ is the set of roots of the quadratic equation $f + \lambda = 0$, 
\begin{equation}
\label{eq:tau-lpm}
    \tau_{\pm}(\lambda) = \frac{\tau_+ + \tau_-}{2} \pm \frac{1}{2} \sqrt{\Delta^2 - 4 \lambda}.
\end{equation}
We prove this identity in Appendix \ref{app:delta-fid}.

Setting $n = \frac{d-4}{2}$, the gauge field \eqref{eq:gauge-field-even} in even dimensions becomes
\begin{equation}
\label{eq:gf}
    A^{\mu}(x) = \frac{1}{2\pi^{\frac{d-2}{2}}}\left[ \sum_{i = 1}^M Q_i p_i^{\mu}  \p_{\lambda}^n\left(\frac{1}{\sqrt{\Delta^2 - 4\lambda}} \Theta(\tau^i_+(\lambda))\right) + \sum_{i = M+1}^{M+N} Q_i p_i^{\mu} \p_{\lambda}^n\left(\frac{1}{\sqrt{\Delta^2 - 4\lambda}} \Theta(\tau^i_-(\lambda))\right)\right] \Big|_{\lambda = 0},
\end{equation}
where we recall that the first term corresponds to the contribution from incoming sources while the second term is due to the outgoing sources and the superscript on $\tau^i_{\pm}$ appears because $\tau_{\pm}$ depend on the source momenta. This formula can be further simplified using the identities
\begin{equation}
    \p_{\lambda}^n \frac{1}{\sqrt{\Delta^2 - 4\lambda}} \left. \right|_{\lambda = 0}= \frac{(2n)!}{n!(\tau_+ - \tau_-)^{2n + 1}}
\end{equation}
and hence
\begin{equation}
    -\frac{1}{2} \p_{\lambda}^n\sqrt{\Delta^2 - 4 \lambda} \left. \right|_{\lambda = 0} = \frac{(2n - 2)!}{(n - 1)!} (\tau_+ - \tau_-)^{1 - 2n}.
\end{equation}
Therefore
\begin{equation}
\label{eq:prod-id}
    \p_{\lambda}^n\left(\frac{1}{\sqrt{\Delta^2 - 4\lambda}} \Theta(\tau_{\pm}(\lambda))\right)\Big|_{\lambda = 0} = \sum_{p = 0}^n \left(\begin{matrix}
        n\\
        p
    \end{matrix} \right) \frac{(2p)!}{p!(\tau_+ - \tau_-)^{2p + 1}}\p_{\lambda}^{n - p}\Theta(\tau_\pm(\lambda)) \Big|_{\lambda = 0}.
\end{equation}
The $p < n$ terms in the series are contact terms  at $\tau_{\pm} = 0$. They are explicitly computed using the formula
\begin{equation}
\label{eq:theta-der-m}
    \p_{\lambda}^m \Theta(\tau_{\pm}(\lambda)) = \sum_{j = 0}^{m - 1} (\mp 1)^{j+1} \frac{(2m-j-2)!}{j! (m - j - 1)!} \frac{\delta^{(j)}(\tau_{\pm}(\lambda))}{(\sqrt{\Delta^2 - 4\lambda})^{2m - j - 1}}, \quad m \geq 1
\end{equation}
proven in Appendix \ref{sec:theta-der}. 
As a result,
\begin{equation}
\label{eq:prod-id}
\begin{split}
    \p_{\lambda}^n\left(\frac{1}{\sqrt{\Delta^2 - 4\lambda}} \Theta(\tau_{\pm}(\lambda))\right)\Big|_{\lambda = 0} &= \sum_{p = 0}^n \left(\begin{matrix}
        n\\
        p
    \end{matrix} \right) \frac{(2p)!}{p!(\tau_+ - \tau_-)^{2p + 1}}\p_{\lambda}^{n - p}\Theta(\tau_\pm(\lambda)) \Big|_{\lambda = 0}\\
    &=  \frac{(2n)!}{n!(\tau_+ - \tau_-)^{2n + 1}}\Theta(\tau_\pm)  \\
    &+ \sum_{p = 0}^{n-1} \sum_{j = 0}^{n - p - 1}\left(\begin{matrix}
        n\\
        p
    \end{matrix} \right) \frac{(\mp)^{j+1}(2p)!}{p!(\tau_+ - \tau_-)^{2n - j}}\frac{(2n-2p-j-2)!}{j! (n-p - j - 1)!} \delta^{(j)}(\tau_{\pm}).\\
    \end{split}
\end{equation}
Switching the order of the sums and performing the sum on $p$ using the identity
\begin{equation}
    \sum_{p = 0}^{n - 1 - j}\left(\begin{matrix}
        n\\
        p
    \end{matrix} \right) \frac{(2p)!}{p!}\frac{(2n-2p-j-2)!}{j! (n-p - j - 1)!} = \frac{(2n - j - 1)!}{(j+1)!(n - j - 1)!},
\end{equation}
we obtain the simple formula
\begin{equation}
\label{eq:useful-identity}
\begin{split}
      \p_{\lambda}^n\left(\frac{1}{\sqrt{\Delta^2 - 4\lambda}} \Theta(\tau_{\pm}(\lambda))\right)\Big|_{\lambda = 0} &= \frac{(2n)!}{n!(\tau_+ - \tau_-)^{2n + 1}}\Theta(\tau_\pm) + \sum_{j = 0}^{n - 1}\frac{(2n - j - 1)!}{(j+1)!(n - j - 1)!} \frac{(\mp 1)^{j+1}}{(\tau_+ - \tau_-)^{2n - j}} \delta^{(j)}(\tau_{\pm})\\
      &= \sum_{j = 0}^{n}\frac{(2n - j)!}{j!(n - j)!} \frac{(\mp 1)^{j}}{(\tau_+ - \tau_-)^{2n - j+1}} \delta^{(j-1)}(\tau_{\pm}),
      \end{split}
\end{equation}
where 
\begin{equation}
    \delta^{(-1)}( x) \equiv \Theta(x).
\end{equation}
The even-dimensional gauge field is then obtained by plugging \eqref{eq:useful-identity} into \eqref{eq:gf},
\begin{equation}
\label{eq:gf-even-massive}
A^\mu(x) = \frac{1}{2\pi^{\frac{d-2}{2}}} \sum_{j=0}^n \frac{(2n-j)!}{j!(n-j)!}\Bigg[\sum_{i=1}^M Q_ip_i^\mu \f{(-1)^j}{(\tau_+^i-\tau_-^i)^{2n-j+1}}\delta^{(j-1)}(\tau_+^i)+\sum_{i=M+1}^{M+N}Q_ip_i^\mu \f{1}{(\tau_+^i-\tau_-^i)^{2n-j+1}}\delta^{(j-1)}(\tau_-^i)\Bigg].\\
\end{equation}

\subsubsection{Field strength}

To compute the field strength, we use the identities
\begin{equation}
\begin{split}
   & p_{[\mu} \partial_{\nu]} \Delta = \frac{4  p_{[\mu} x_{\nu]}}{\Delta}, \\
    &p_{[\mu} \partial_{\nu]} (\tau_+ + \tau_-) = 0 .
  \end{split}  
\end{equation}
obtained using
    \begin{equation}
    p_{[\mu} \partial_{\nu]} \tau_{\pm} = \pm \frac{2 p_{[\mu} x_{\nu]}}{\Delta}.
\end{equation}
Here we defined 
\begin{equation}
    p_{[\mu} x_{\nu]} \equiv p_{\mu} x_{\nu} - x_{\mu} p_{\nu}.
\end{equation}
Using these formula together with \eqref{eq:gf-even-massive}, the field strength component sourced by momentum $p_i$ takes the form
\begin{equation}
\label{eq:fmunumassive}
\begin{split}
    F_{\mu\nu}(p_i) &= \frac{Q_ip^i_{[\mu} x_{\nu]}}{\pi^{\frac{d-2}{2}}}\left[2 \frac{(2n+1)!}{n!}\frac{\Theta(\tau_{\pm}^i)}{\Delta_i^{2n + 3}} + \sum_{j = 0}^{n-1} \frac{(\mp 1)^{j - 1}(2n - j +1)!}{(j+1)!(n - j)!} \frac{\delta^{(j)}(\tau_{\pm}^i)}{\Delta_i^{2n - j + 2}} + \frac{(\mp 1)^{n-1}\delta^{(n)}(\tau_{\pm}^i)}{\Delta_i^{n+2}}\right]\\
    &= \frac{Q_ip^i_{[\mu} x_{\nu]}}{\pi^{\frac{d-2}{2}}} \sum_{j = -1}^n a_j \frac{\delta^{(j)}(\tau_{\pm}^i)}{\Delta_i^{2n - j + 2}}, \quad a_j = (\mp 1)^{j-1}\frac{(2n-j+1)!}{(j+1)!(n-j)!},
    \end{split}
\end{equation}
where the upper branch $(a_j \sim (-1)^{j - 1}\cdots )$ applies to incoming while the lower one applies to outgoing momenta.
Working with the null frame introduced in \eqref{eq:null-frame}, one can verify explicitly that \eqref{eq:fmunumassive} obey the source-free\footnote{This is only because massive particles do not reach $\mathscr{I}$. The sources will reappear if $\mathscr{I}$ is replaced by a surface at finite distance. We will see this explicitly in the discussion of odd-dimensional spacetimes in Section \ref{sec:odd-massive}. } Maxwell constraint at $\mathcal{I}^+$. We do this in Appendix \ref{sec:Maxwell-sol-mass}.

\subsubsection{Maxwell constraint}

For the Coulombic memory effect, we are interested in the $\mathcal{O}(r^{2 - d})$ component of the Maxwell constraint \eqref{eq:Maxwell-constraint-fixed-r}. We show in Appendix \ref{sec:large-r} that the relevant order $\mathcal{O}(r^{2-d})$ components of \eqref{eq:fmunumassive} are
\begin{equation}
    \label{eq:rdm2}
    \begin{split}
    F_{ru}^{(d - 2)}(p_i) &=  \frac{Q_i}{2\pi^{\frac{d - 2}{2}}}\frac{(d - 3)!}{\left( \frac{d - 4}{2}\right)!} \frac{\Theta(u)}{(-2p_i \cdot q)^{d - 2}},\\
    F_{Au}^{(d - 4)}(p_i) &=   \frac{Q_i}{2 \pi^{\frac{d-2}{2}}} \frac{p_i \cdot \p_A q}{(-2p_i\cdot q)^{d - 3}} \mathcal{P}_{\frac{d - 4}{2}}\left(-(2p_i\cdot q) (p_i\cdot n)\right)\delta(u),
    \end{split}
\end{equation}
where 
\begin{equation}
    \mathcal{P}_n(z) = \sum_{k = 0}^n (-1)^{k+1} 2^k \frac{(2n - k)!}{(n - k)!} z^k.
\end{equation}
That the Maxwell constraint is obeyed, follows from the identity
\begin{equation}
\begin{split}
    D^A F_{Au}^{(d - 4)} &=  \frac{Q_i}{2\pi^{\frac{d - 2}{2}}}\frac{\delta(u)}{(-2p_i\cdot q)^{d - 2}} 2\left[z(1 + 2z) \mathcal{P}'_{n}(z) - (2n +2nz +1 )\mathcal{P}_n(z)\right],\\ n &= \frac{d - 4}{2}, \qquad
    z = -(2p_i\cdot q) (p_i\cdot n),
    \end{split}
\end{equation}
together with
\begin{equation}
 z(1 + 2z) \mathcal{P}'_{n}(z) - (2n +2nz +1 )\mathcal{P}_n(z) = \frac{(2n + 1)!}{n!}.
\end{equation}

The Coulombic order memory effect due to an outgoing massive particle is therefore non-vanishing and given by
\begin{equation}
    \mathcal{M}_A^{\rm massive}(q) \equiv \int_{-\infty}^{\infty} du F_{Au}^{(d - 4)}  = \frac{Q_i}{2\pi^{\frac{d - 2}{2}}}\frac{p_i\cdot \p_A q}{(-2p_i\cdot q)^{d - 3}} \mathcal{P}_{\frac{d - 4}{2}}\left(-(2p_i\cdot q) (p_i\cdot n)\right).
\end{equation}
While there is no flux through $\mathcal{I}^+$, the memory effect is controlled, via the Maxwell constraint, by the non-vanishing value of $F_{ru}^{(d - 2)}$ at $\mathcal{I}^+_+$. Indeed, it is easy to check, by linearity, that 
\begin{equation}
    2\partial_uF_{ru}^{(d-2)}-D^AF_{Au}^{(d-4)}=0 \iff 2\Delta F_{ru}^{(d-2)}-D^A \mathcal{M}_A^{\rm massive}=0.
\end{equation}

As a consistency check, we note that in the massless source limit, which amounts to taking $\rho \rightarrow 0$ in \eqref{eq:massive-dots}, 
\begin{equation}
    \lim_{\rho \rightarrow 0} \Delta F_{ru}^{(d - 2)} = \frac{Q_i}{2 \pi^{\frac{d-2}{2}}} \frac{\Gamma(d - 2)}{\Gamma(\frac{d - 2}{2})}\lim_{\rho \rightarrow 0} \left( \frac{\rho_i}{\rho_i^2 + |y - z_i|^2}\right)^{d - 2} = \frac{Q_i}{2}\delta^{(d - 2)}(y - z_i),
\end{equation}
where we used the identity (see appendix A of \cite{deGioia:2024yne} for a proof)
\begin{equation}
    \lim_{\rho \rightarrow 0} \left( \frac{\rho_i}{\rho_i^2 + |y - z_i|^2}\right)^{d - 2} = \pi^{\frac{d-2}{2}} \frac{\Gamma\left(\frac{d - 2}{2} \right)}{\Gamma\left(d-2 \right)} \delta^{(d - 2)}(y - z_i).
\end{equation}
We see that this result is in perfect agreement with the memory effect sourced by massless particles \eqref{eq:even-memory-observable}.

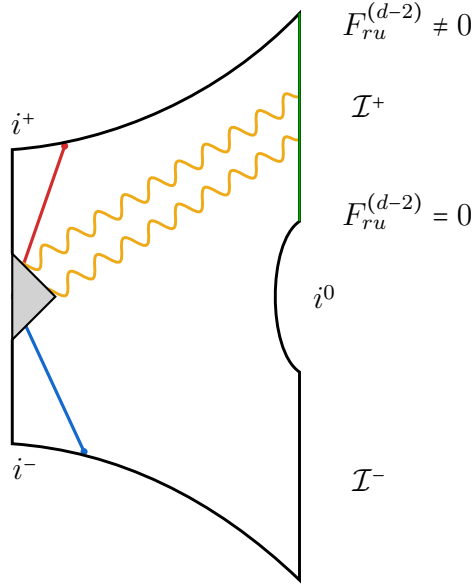
\begin{figure}
\centering
\begin{tikzpicture}[
  scale=0.95,
  every node/.style={font=\large},
  massive/.style={line width=1.2pt},
  radiation/.style={
    line width=1.1pt,
    draw=radiationyellow,
    decorate,
    decoration={snake, amplitude=1.2mm, segment length=4.2mm}
  }
]
\definecolor{particleblue}{RGB}{20,105,205}
\definecolor{particlered}{RGB}{210,45,45}
\definecolor{radiationyellow}{RGB}{238,170,24}
\def\L{0.3}

\draw[line width=1.1pt]
  (-1.55,  0.60) -- (-1.55,  2.05)
  .. controls (-0.15, 2.15) and (1.35, 2.80) .. (2.45, 3.95)
  -- (2.45, 1.05)
  .. controls (2.00, 0.75) and (2.00,-0.75) .. (2.45,-1.05)
  -- (2.45,-3.95)
  .. controls (1.25,-2.80) and (-0.10,-2.15) .. (-1.55,-2.05)
  -- (-1.55,-0.60);

\filldraw[fill=gray!35, draw=black, line width=0.8pt]
  (-1.55,  0.60) -- (-0.95,0) -- (-1.55,-0.60) -- cycle;

\begin{pgfonlayer}{particlelayer}
\draw[massive, particleblue] (-0.55,-2.16) -- (-1.55,0);
\fill[particleblue] (-0.55,-2.16) circle (1.4pt);

\draw[massive, particlered] (-1.55,0) -- (-0.82,2.10);
\fill[particlered] (-0.82,2.10) circle (1.4pt);
\draw[radiation] (-1.55 ,\L) -- (2.45,2.80);
\draw[radiation]  (-1.55 ,\L-2*\L) -- (2.45,2.80-2*\L);
\end{pgfonlayer}

\draw[thick, color = green!65!black] 
(2.45, 3.95)
  -- (2.45, 1.05);

\node[left] at (-1.07, 2.42) {$i^+$};
\node[left] at (-1.05,-2.40) {$i^-$};
\node at (2.8,0) {$i^0$};
\node[right] at (3.05,2.65) {$\mathcal I^+$};
\node[right] at (3.05,-2.55) {$\mathcal I^-$};
\node[right] at (2.9,3.85) {$F_{ru}^{(d-2)} \neq 0$};
\node[right] at (2.9,1.2) {$F_{ru}^{(d-2)} = 0$};
\end{tikzpicture}
\caption{In even spacetime dimension, the Coulombic order of the Maxwell constraint evaluated at $\mathcal{I}^{\pm}$ reveals a Coulombic memory effect sourced by the change in $F_{ru}^{(d - 2)}.$ }
\label{fig:massive-even}
\end{figure}

The boundary value of the radial electric field \eqref{eq:rdm2} carries information about the massive particle flux through the hyperbolic resolution of future timelike infinity and sources a non-vanishing memory effect just as in $d = 4$. We illustrate this in Figure \ref{fig:massive-even} on a Minkowski spacetime diagram including the resolution of $i^{\pm}$ and $i^0$ following \cite{Compere:2023qoa}. 

In the following section, we will show that just as in the massless source case, the Maxwell constraint is trivial at Coulombic order when evaluated at $\mathcal{I}^{\pm}$ and there is therefore no memory effect. The memory effect is nevertheless present on a surface at any finite (large) distance, such as a lightsheet at $r = r_0$. Again, the key observation is that the integral computing the memory effect does not commute with the $r_{0} \rightarrow \infty$ limit.

\subsection{Odd spacetime dimensions}
\label{sec:odd-massive}

Using the solution \eqref{eq:odd-dim-int} and accounting for the constraint imposed by the theta function by repeating the analysis in section \ref{sec:even-massive}, we obtain the odd-dimensional gauge field 
\begin{equation}
\begin{split}
    A^{\mu}(x) &= -\frac{(-1)^{\frac{d-1}{2}}}{2 \pi^{\frac{d}{2}}} \Gamma\left(\frac{d - 2}{2} \right) \sum_{i = 1}^{M}\Big[Q_i p_i^{\mu}\Theta(-x^2) \int_0^{\infty} d\tau \left[(\tau - \tau_-^i)(\tau - \tau_+^i) \right]^{-\frac{d-2}{2}} \\
    &+ Q_i p_i^{\mu}\Theta(x^2)\int_{\tau_+(p_i)}^{\infty} d\tau \left[(\tau - \tau_-^i)(\tau - \tau_+^i) \right]^{-\frac{d-2}{2}} \Big]\\ &- \frac{(-1)^{\frac{d-1}{2}}}{2 \pi^{\frac{d}{2}}} \Gamma\left(\frac{d - 2}{2} \right) \sum_{i = M+1}^{M+N}  Q_i p_i^{\mu}\Theta(-x^2) \int_0^{\tau_-(p_i)} d\tau \left[(\tau - \tau_-^i)(\tau - \tau_+^i) \right]^{-\frac{d-2}{2}}.
    \end{split}
\end{equation}
The three terms can be evaluated using the following result
\begin{equation}
\begin{split}
  I(\alpha) \equiv  \int_0^{\alpha} d \tau \left[(\tau - \tau_-)(\tau - \tau_+) \right]^{-\frac{d-2}{2}}&= \frac{2(\tau_- \tau_+)^{2 - \frac{d}{2}}{}_2F_1\left[1,4 - d;3 - \frac{d}{2}; \frac{\tau_-}{\tau_- - \tau_+} \right]}{(d - 4)(\tau_- - \tau_+)} \\
    &- 2 \frac{((\alpha - \tau_-)(\alpha - \tau_+))^{2 - \frac{d}{2}}{}_2F_1\left[1, 4 - d; 3 - \frac{d}{2}; \frac{\tau_- - \alpha}{\tau_- - \tau_+} \right]}{(d - 4)(\tau_- - \tau_+)}
    \end{split}
\end{equation}
and its limits as $\alpha \rightarrow \tau_{\pm}(p_i)$ and $\alpha \rightarrow \infty$. The result admits the equivalent representation 
\begin{equation}
    I(\alpha) = \Delta^{3 - d} \left[B_{z(\alpha)}(2 - \frac{d}{2},d - 3) - B_{z(0)}(2 - \frac{d}{2},d - 3) \right], \quad z(\alpha) = \frac{\alpha - \tau_+}{\alpha - \tau_-},
\end{equation}
where 
\begin{equation}
    B_{z}(a,b) \equiv \int_0^z dt t^{a - 1} (1 - t)^{b - 1}
\end{equation}
is the incomplete beta function.

The relevant limits are 
\begin{equation}
    \begin{split}
      I(\tau_-) &\equiv  \lim_{\alpha \rightarrow \tau_- + \epsilon} I[\alpha] = J - 2 \frac{\epsilon^{2 - \frac{d}{2}}(\tau_- - \tau_+)^{1 - \frac{d}{2}}}{d - 4},\\
   I(\tau_+) &\equiv      \lim_{\alpha \rightarrow \tau_+ + \epsilon} I[\alpha] = J + 2 \frac{(-\epsilon)^{2 - \frac{d}{2}}(\tau_- - \tau_+)^{1 - \frac{d}{2}}}{d - 4},\\
 I(\infty) &\equiv         \lim_{\alpha \rightarrow \infty} I[\alpha] = J +(\tau_- - \tau_+)^{3 - d} B\left(2 - \frac{d}{2}, d - 3 \right),
    \end{split}
\end{equation}
where 
\begin{equation}
\label{eq:J}
    J(\tau_+, \tau_-) \equiv  \frac{2(\tau_- \tau_+)^{2 - \frac{d}{2}}{}_2F_1\left[1,4 - d;3 - \frac{d}{2}; \frac{\tau_-}{\tau_- - \tau_+} \right]}{(d - 4)(\tau_- - \tau_+)} 
\end{equation}
and we used the identity
\begin{equation}
    {}_2F_1\left[1, 4 - d; 3 - \frac{d}{2}; 1\right] = -1, \quad d = 2k + 1, \quad k \in \mathbb{N}.
\end{equation}
Putting everything together, we find that the gauge field takes the form
\begin{equation}
\begin{split}
    A^{\mu}(x) &= -\frac{(-1)^{\frac{d - 1}{2}}}{2 \pi^{\frac{d}{2}}} \Gamma\left(\frac{d - 2}{2} \right) \sum_{i = 1}^M \left[Q_i p_i^{\mu} \left(I(\infty) - \Theta(x^2)I(\tau_+^i) \right) \right]\\
    &-\frac{(-1)^{\frac{d - 1}{2}}}{2 \pi^{\frac{d}{2}}} \Gamma\left(\frac{d - 2}{2} \right) \sum_{i = M+1}^{M+N} Q_i p_i^{\mu} \Theta(-x^2) I(\tau_-^i).
    \end{split}
\end{equation}
This expression diverges in the limit $\epsilon \rightarrow 0$, just like the gauge field sourced by massless particles in odd spacetime dimension (see Section \ref{sec:odd-dim-gf}). The divergences have the same origin as those in \eqref{eq:gf-odd}, and therefore we can employ the same regularization procedure described in Section \ref{sec:odd-dim-gf}. We therefore obtain the regularized gauge field
\begin{equation}
\label{eq:reg-gf-massive}
    \begin{split}
    A_R^{\mu}(x) &= -\frac{(-1)^{\frac{d - 1}{2}}}{2 \pi^{\frac{d}{2}}} \Gamma\left(\frac{d - 2}{2} \right) \sum_{i = 1}^M \left[Q_i p_i^{\mu} \left(I(\infty) - \Theta(x^2)J(\tau_+^i, \tau_-^i) \right) \right]\\
    &-\frac{(-1)^{\frac{d - 1}{2}}}{2 \pi^{\frac{d}{2}}} \Gamma\left(\frac{d - 2}{2} \right) \sum_{i = M+1}^{M+N} Q_i p_i^{\mu} \Theta(-x^2) J(\tau_+^i, \tau_-^i).
    \end{split}
\end{equation}

\subsubsection{Field strength}

The relevant formulas are
\begin{equation}
\label{eq:Jfs}
    p_{[\nu} \p_{\mu]} J(x,p) \equiv  (x_{\mu}p_{\nu} - x_{\nu} p_{\mu})\frac{2}{d - 4}\left[\left(4 - d\right) \frac{(-x^2)^{1 - \frac{d}{2}}}{\Delta} H(z) + 4\frac{(-x^2)^{2 - \frac{d}{2}}}{\Delta^3} H(z)+ 4\frac{(x\cdot p)(-x^2)^{2 - \frac{d}{2}}}{\Delta^4} H'(z)\right],
\end{equation}
where $\Delta$ was defined in \eqref{eq:def-fd},
\begin{equation}
\label{eq:H}
    H(z) =  {}_2F_1\left[1, 4 - d; 3 - \frac{d}{2}; z\right], \quad z = \frac{2x\cdot p + \Delta}{2\Delta}
\end{equation}
and
\begin{equation}
     p_{[\nu} \p_{\mu]} (\tau_- - \tau_+)^{a} = 4 a (\tau_- - \tau_+)^{a - 2} (x_{\mu}p_{\nu} - x_{\nu} p_{\mu}).
\end{equation}
Using the identity 
\begin{equation}
    \left[ 2 - \frac{d}{2} + 2 z(z - 1) + (2z - 1) z (z - 1) \p_z \right] {}_2 F_1\left[1, 4 - d; 3 - \frac{d}{2}; z \right] = (2 - \frac{d}{2}) {}_2 F_1\left[ 1, 2 - d; 2 - \frac{d}{2}; z\right],
\end{equation}
\eqref{eq:Jfs} simplifies to 
\begin{equation}
      p_{[\nu} \p_{\mu]} J(x,p) = 2(x_{\mu}p_{\nu} - x_{\nu} p_{\mu}) \frac{(\tau_+ \tau_-)^{1 - \frac{d}{2}}}{\tau_- - \tau_+} {}_2 F_1\left[1, 2 - d; 2 - \frac{d}{2}; \frac{\tau_-}{\tau_- - \tau_+} \right].
\end{equation}
We prove this identity in Appendix \ref{eq:idoddm}.

Away from $x^2 = 0$, the field strength associated with the regularized gauge field \eqref{eq:reg-gf-massive} is then
\begin{equation}
\label{eq:reg-mass-odd}
    \begin{split}
    F_{\mu\nu} &= -\frac{(-1)^{\frac{d - 1}{2}}}{2 \pi^{\frac{d}{2}}}\Gamma\left( \frac{d - 2}{2} \right)\Big[ \sum_{i = 1}^{M+N} \Theta(-x^2) 2 Q_i \frac{(\tau_+^i \tau_-^i)^{1 - \frac{d}{2}}}{\tau^i_- - \tau^i_+} {}_2 F_1\left[1, 2 - d; 2 - \frac{d}{2}; \frac{\tau_-^i}{\tau_-^i - \tau_+^i} \right](x_{\mu}p_{i,\nu} - x_{\nu} p_{i,\mu}) \\
    &+ \sum_{i = 1}^M 2(2 - d) Q_i  B(2 - \frac{d}{2}, d - 2) (\tau_-^i - \tau_+^i)^{1 - d}  (x_{\mu}p_{i,\nu} - x_{\nu} p_{i,\mu}) \Big] \equiv \sum_{i = 1}^{M+N} (x_{\mu}p_{i,\nu} - x_{\nu} p_{i,\mu})\Phi_i.
    \end{split}
\end{equation}
It is straightforward to show that the regularized field strength obeys the Maxwell equations, however we include a proof of this fact, for completeness, in Appendix \ref{sec:sol-odd}. 
Using the null frame defined in \eqref{eq:null-frame}, we obtain
\begin{equation}
\label{eq:frame-massive-odd}
\begin{split}
    F_{ru} &= \sum_{i = 1}^{M+N}\frac{1}{2}(r q\cdot p_i - u n \cdot p_i) \Phi_i, \\
    F_{Au} &= \sum_{i = 1}^{M+N}\frac{r^2}{2} \p_A(q\cdot p_i) \Phi_i,
    \end{split}
\end{equation}
where 
\begin{equation}
\label{eq:phii}
\begin{split}
    \Phi_i &= -\frac{(-1)^{\frac{d - 1}{2}}}{2 \pi^{\frac{d}{2}}}\Gamma\left( \frac{d - 2}{2} \right) \Big[ \Theta(-x^2) 2 Q_i \frac{(\tau_+^i \tau_-^i)^{1 - \frac{d}{2}}}{\tau_-(p_i) - \tau_+(p_i)} {}_2 F_1\left[1, 2 - d; 2 - \frac{d}{2}; \frac{\tau_-^i}{\tau_-^i - \tau_+^i} \right]\\
    &+ \delta_{i \leq M} 2(2 - d) Q_i  B(2 - \frac{d}{2}, d - 2) (\tau_-^i - \tau_+^i)^{1 - d}  \Big].
    \end{split}
\end{equation}

\subsubsection{Maxwell constraint}

Generalizing the analysis of the Coulombic memory sourced by massless particles in Section \ref{sec:odd-mem}, in this section we show that massive particles also give rise to a non-vanishing memory effect at Coulombic order in odd spacetime dimensions. The key insight is again that the memory effect should be measured at finite distance from the sources and follows from the Maxwell constraint evaluated on a fixed $r = r_0$ hypersurface. 

We consider the pullback of Maxwell's equations \eqref{eq:Maxwell-0} to the $r = r_0$ hypersurface and perform the integral on $u$ to find 
\begin{equation}
\label{eq:Maxwell-constraint-fixed-r-fin}
  2 r_0^{d - 2} \int_{-\infty}^{\infty} du \p_u F_{ru} - r_0^{d - 4}  \int_{-\infty}^{\infty} du D^AF_{Au} = - r_0^{d - 2} \int_{-\infty}^{\infty} du j_u .
\end{equation}
It is clear from the solution \eqref{eq:reg-mass-odd}, \eqref{eq:frame-massive-odd} that the first term is the difference of boundary terms at $|u| \rightarrow \infty$ which vanishes, just as in the massless source case. We next compute the memory integral
\begin{equation}
    \mathcal{M}(y) \equiv  D^A \int_{-\infty}^{\infty} du F_{Au}
\end{equation}
using the solution \eqref{eq:frame-massive-odd}. This consists of two components, involving the integrals
\begin{equation}
    J_1 = \int_0^{\infty} du \frac{(\tau_+ \tau_-)^{1 - \frac{d}{2}}}{\tau_- - \tau_+} {}_2 F_1\left[1, 2 - d; 2 - \frac{d}{2}; \frac{\tau_-}{\tau_- - \tau_+} \right]
\end{equation}
and 
\begin{equation}
    J_2 = \int_{-\infty}^{\infty} du (\tau_- - \tau_+)^{1 - d}.
\end{equation}

The evaluation of these integral may be simplified by first defining
\begin{equation}
    \begin{split}
        \lambda &\equiv \frac{|y - y_p|^2}{\rho^2}, \\
        t & \equiv \frac{u}{r \rho^2}
    \end{split}
\end{equation}
 and using the parameterization in \eqref{eq:massive-dots}, 
\begin{equation}
    \begin{split}
        q\cdot p = \eta \frac{\rho^2 + |y - y_p|^2}{2\rho}, \quad n \cdot p = \frac{\eta}{2\rho}, \quad x^2 = -ru,
    \end{split}
\end{equation}
where $\eta = 1$ ($\eta = -1$) for incoming (outgoing) particles. 
Therefore
\begin{equation}
    \tau_{\pm}(p) = \frac{r \rho}{2}\left[-\eta (1 + \lambda + t) \pm \sqrt{(t + 1 + \lambda)^2 - 4t} \right]
\end{equation}
and
\begin{equation}
\begin{split}
\tau_- - \tau_+ &= -r \rho \sqrt{(t - 1 + \lambda)^2 + 4 \lambda},\\
    \frac{\tau_-}{\tau_- - \tau_+} &= \frac{1}{2} + \frac{\eta}{2} \frac{1 + \lambda + t}{\sqrt{(t - 1 + \lambda)^2 + 4 \lambda}}.
    \end{split}
\end{equation}
Then 
\begin{equation}
\label{eq:int-odd}
    J_1(\eta) = -\int_0^{\infty} dt r^{2 - d} \rho^{3 - d} \frac{t^{1-  \frac{d}{2}}}{\sqrt{(t - 1 + \lambda)^2 + 4 \lambda}}{}_2 F_1\left[1, 2 - d; 2 - \frac{d}{2}; \frac{1}{2} + \frac{\eta}{2} \frac{1 + \lambda + t}{\sqrt{(t - 1 + \lambda)^2 + 4 \lambda}} \right].
\end{equation}

This integral may be evaluated by first observing that 
\begin{equation}
    J_1(\eta) = -J_1(-\eta).
\end{equation}
Let
\begin{equation}
    J_1(1) = J_+.
\end{equation}
One can check that the integrand for $\eta = 1$ is proportional to
\begin{equation}
    f_d(t, \lambda) \equiv \frac{t^{1-  \frac{d}{2}}}{\sqrt{(t - 1 + \lambda)^2 + 4 \lambda}}{}_2 F_1\left[1, 2 - d; 2 - \frac{d}{2}; \frac{1}{2} + \frac{1}{2} \frac{1 + \lambda + t}{\sqrt{(t - 1 + \lambda)^2 + 4 \lambda}} \right],
\end{equation}
which obeys the following differential identity
\begin{equation}
    f_{d+2}(t, \lambda) = \frac{2}{d-2}\left(\p_{\lambda} - \p_t \right) f_d(t,\lambda).
\end{equation}
It follows that $f_d$ for all odd $d$ can be related to $f_3$ by repeatedly acting with the following differential operator
\begin{equation}
\label{eq:main-id}
    f_{d}(t,\lambda) = \left(\prod_{j = 1}^k \frac{2}{2j - 1} \right) (\p_{\lambda} - \p_t)^{k-1} f_3, \quad d = 2k + 1.
\end{equation}
We therefore find that 
\begin{equation}
    J_+ = - r^{2 - d} \rho^{3 - d} \frac{2^{\frac{d - 1}{2}}}{(d-2)!!} \p_{\lambda}^{\frac{d - 3}{2}} \int_0^{\infty}dt f_3(t, \lambda) + {\rm bdry.~terms},
\end{equation}
where the boundary terms are the contribution to the integral from the total derivatives in $t$ in \eqref{eq:main-id}. It is clear that all boundary terms at $t = \infty$ vanish for $d \geq 3$, while the $t = 0$ boundary terms yield a divergent power series, whose finite part vanishes. Upon subtracting these divergences, or equivalently, analytically continuing the result of the integral \eqref{eq:int-odd} to odd integer $d$ outside $2<d<4$, we obtain
\begin{equation}
    J_+ = (-)^{\frac{d - 3}{2}} r^{2 - d} \rho^{3 - d} \eta \pi \lambda^{1 - \frac{d}{2}},
\end{equation}
where we used that 
\begin{equation}
    \int_0^{\infty} dt f_3(t,\lambda) = -\frac{\pi}{\sqrt{\lambda}}.
\end{equation}

On the other hand, for odd $d$, we have
\begin{equation}
    J_2 = r^{2 - d} \rho^{3 - d} \int_{-\infty}^{\infty} dt \sqrt{(t - 1 + \lambda)^2 + 4 \lambda} = r^{2 - d} \rho^{3 - d}\sqrt{\pi} \frac{\Gamma(\frac{d - 2}{2})}{\Gamma(\frac{d - 1}{2})} (2 \sqrt{\lambda})^{2-d},
\end{equation}
where the integral reduces to a standard Beta function integral after changing variables $ t - 1 + \lambda \rightarrow t$ followed by $1/t^2 \rightarrow t$.  Furthermore,
\begin{equation}
    (2 - d) B(2 - \frac{d}{2}, d - 2) J_2 = -(-1)^{\frac{d-3}{2}} \pi r^{2 -  d} \rho^{3 - d} \lambda^{1 - \frac{d}{2}}.
\end{equation}
We hence have
\begin{equation}
\begin{split}
    \int_{-\infty}^{\infty} du \Phi_i &= -\frac{(-1)^{\frac{d - 1}{2}}}{2 \pi^{d/2}} \Gamma\left(\frac{d - 2}{2} \right) 2 Q_i (-\eta_i + \delta_{i\leq M}) \pi (-1)^{\frac{d + 3}{2}} r^{2 - d} \rho_i^{3 - d}\lambda_i^{1 - \frac{d}{2}}\\
    &=-\frac{1}{2 \pi^{d/2}} \Gamma\left(\frac{d - 2}{2} \right) 2 Q_i (-\eta_i + \delta_{i\leq M}) \pi r^{2 - d} \rho_i^{3 - d}\lambda_i^{1 - \frac{d}{2}}.
    \end{split}
\end{equation}
We see that the incoming contribution cancels (recall $\eta = 1$ for incoming particles), while the outgoing contribution gives a nontrivial memory integral
\begin{equation}
\label{eq:memory-massive-odd}
\begin{split}
    \mathcal{M}(y) &= \int_{-\infty}^{\infty}du D^A F_{Au} =  \sum_{i = M+1}^{M+N} D_A\left(\frac{(y - y_i)^A}{|y - y_i|^{d - 2}} \right) \frac{1}{2 \pi^{d/2-1}} \Gamma\left(\frac{d - 2}{2} \right) Q_i \\
    &= \sum_{i = M+1}^{M+N} Q_i \delta^{(d-2)}(y - y_i).
    \end{split}
\end{equation}

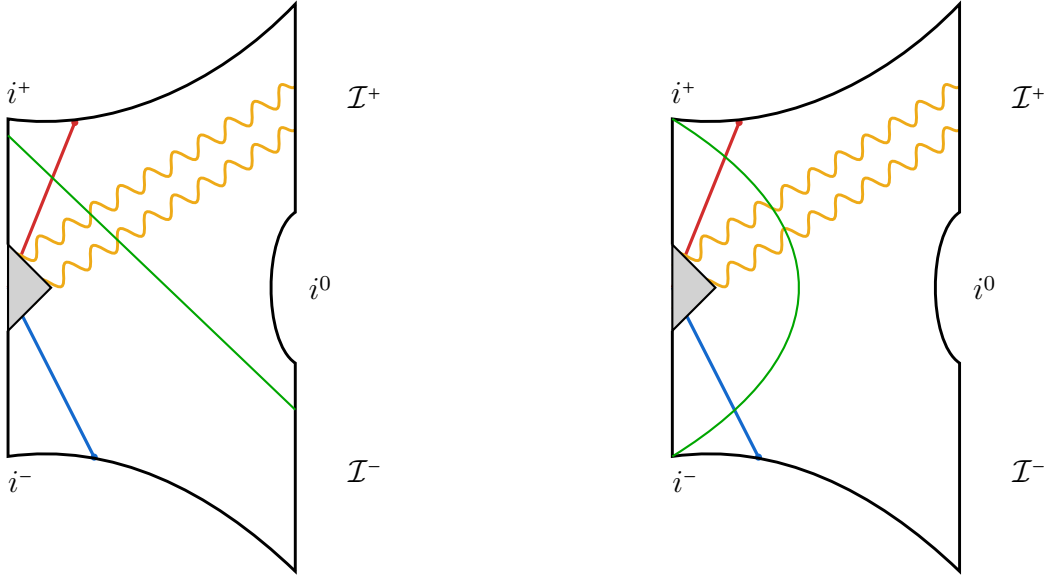
\begin{figure}[h]
    \begin{subfigure}[t]{0.48\textwidth}
        \centering
        \begin{tikzpicture}[
  scale=0.95,
  every node/.style={font=\large},
  massive/.style={line width=1.2pt},
  radiation/.style={
    line width=1.1pt,
    draw=radiationyellow,
    decorate,
    decoration={snake, amplitude=1.2mm, segment length=4.2mm}
  }
]
\definecolor{particleblue}{RGB}{20,105,205}
\definecolor{particlered}{RGB}{210,45,45}
\definecolor{radiationyellow}{RGB}{238,170,24}
\def\L{0.3}
\def\offs{0.8}

\draw[line width=1.1pt]
  (-1.55,  0.60) -- (-1.55,  2.35)
  .. controls (-0.15, 2.15) and (1.35, 2.80) .. (2.45, 3.95)
  -- (2.45, 1.05)
  .. controls (2.00, 0.75) and (2.00,-0.75) .. (2.45,-1.05)
  -- (2.45,-3.95)
  .. controls (1.25,-2.80) and (-0.10,-2.15) .. (-1.55,-2.35)
  -- (-1.55,-0.60);

\filldraw[fill=gray!35, draw=black, line width=0.8pt]
  (-1.55,  0.60) -- (-0.95,0) -- (-1.55,-0.60) -- cycle;

  \draw[line width=0.8pt,color=green!65!black]
 (-1.55,1.32+\offs) -- (2.45,-2.5+\offs);

\begin{pgfonlayer}{particlelayer}
\draw[massive, particleblue] (-0.35,-2.36) -- (-1.55,0);
\fill[particleblue] (-0.35,-2.36) circle (1.4pt);

\draw[massive, particlered] (-1.55,0) -- (-0.62,2.30);
\fill[particlered] (-0.62,2.30) circle (1.4pt);
\draw[radiation] (-1.55 ,\L) -- (2.45,2.80);
\draw[radiation]  (-1.55 ,\L-2*\L) -- (2.45,2.80-2*\L);
\end{pgfonlayer}

\node[left] at (-1.07, 2.72) {$i^+$};
\node[left] at (-1.05,-2.70) {$i^-$};
\node at (2.8,0) {$i^0$};
\node[right] at (3.05,2.65) {$\mathcal I^+$};
\node[right] at (3.05,-2.55) {$\mathcal I^-$};
\end{tikzpicture}
        \caption{The Coulombic memory effect on a lightsheet at large distance receives constributions from outgoing particles only.}
        \label{fig:null}
    \end{subfigure}
    \hfill
    \begin{subfigure}[t]{0.48\textwidth}
        \centering
          \centering
        \begin{tikzpicture}[
  scale=0.95,
  every node/.style={font=\large},
  massive/.style={line width=1.2pt},
  radiation/.style={
    line width=1.1pt,
    draw=radiationyellow,
    decorate,
    decoration={snake, amplitude=1.2mm, segment length=4.2mm}
  }
]
\definecolor{particleblue}{RGB}{20,105,205}
\definecolor{particlered}{RGB}{210,45,45}
\definecolor{radiationyellow}{RGB}{238,170,24}
\def\L{0.3}
\def\offs{0.8}

\draw[line width=1.1pt]
  (-1.55,  0.60) -- (-1.55,  2.35)
  .. controls (-0.15, 2.15) and (1.35, 2.80) .. (2.45, 3.95)
  -- (2.45, 1.05)
  .. controls (2.00, 0.75) and (2.00,-0.75) .. (2.45,-1.05)
  -- (2.45,-3.95)
  .. controls (1.25,-2.80) and (-0.10,-2.15) .. (-1.55,-2.35)
  -- (-1.55,-0.60);

\filldraw[fill=gray!35, draw=black, line width=0.8pt]
  (-1.55,  0.60) -- (-0.95,0) -- (-1.55,-0.60) -- cycle;

  \draw[line  width=0.8pt,color=green!65!black]
(-1.55,  2.35) .. controls (0.8, 0.9) and (0.8, -0.9) ..  (-1.55,-2.35);

\begin{pgfonlayer}{particlelayer}
\draw[massive, particleblue] (-0.35,-2.36) -- (-1.55,0);
\fill[particleblue] (-0.35,-2.36) circle (1.4pt);

\draw[massive, particlered] (-1.55,0) -- (-0.62,2.30);
\fill[particlered] (-0.62,2.30) circle (1.4pt);
\draw[radiation] (-1.55 ,\L) -- (2.45,2.80);
\draw[radiation]  (-1.55 ,\L-2*\L) -- (2.45,2.80-2*\L);
\end{pgfonlayer}

\node[left] at (-1.07, 2.72) {$i^+$};
\node[left] at (-1.05,-2.70) {$i^-$};
\node at (2.8,0) {$i^0$};
\node[right] at (3.05,2.65) {$\mathcal I^+$};
\node[right] at (3.05,-2.55) {$\mathcal I^-$};
\end{tikzpicture}
        \caption{The Coulombic memory effect evaluated on a timelike surface at large distance receives contributions from both incoming at outgoing particles. }
        \label{fig:timelike}
    \end{subfigure}
    \caption{The Maxwell constraint evaluated on asymptotic null (left) and timelike (right) surfaces (green) implies that the Coulombic memory effect is non-vanishing. }
    \label{fig:massive-odd}
\end{figure}

This contribution is precisely what one would expect from the Maxwell constraint \eqref{eq:Maxwell-constraint-fixed-r-fin}. Indeed, unlike at $\mathcal{I}^+$, the change in $F_{ru}$ between $u = \pm \infty$ vanishes on the $r = r_0$ hypersurface. On the other hand, the flux through this finite-distance hypersurface is non-vanishing, due to outgoing particles crossing the lightsheet. It is not hard to see that this outgoing flux is precisely equal to \eqref{eq:memory-massive-odd}. Note that there is no flux due to the incoming particles, since these particles can never cross the light-sheet at $r = r_0$. We illustrate this in Figure \ref{fig:null}. The situation would be similar if we considered a timelike cylinder at $r = r_0$, except for in this case, both the memory integral and the current receive contributions from both incoming and outgoing particles. This is illustrated in Figure \ref{fig:timelike}. We conclude by observing that the shockwaves discussed in Section \ref{sec:shockwaves} are absent for massive particles. The massless limit of \eqref{eq:memory-massive-odd} is hence directly compatible with the results in Section \ref{sec:Coulomb-mem}, but the memory alone cannot resolve the appearance of the shockwaves in the ultra-boosted limit. It would be interesting to study the massless limit at the level of other observables, such as the field-strength itself. 

\subsection{Soft theorem}
\label{sec:soft-thm-massive}

The relation between memory effects and soft theorems for massive sources carries through in the same way as for massless ones. We conclude this section by sketching the derivation, outlining the differences with respect to the discussion in Section \ref{eq:mc-st}. 

As before, the bridge between classical gauge fields and scattering amplitudes is given by the Ward identity \eqref{eq:Ward-id}. For point-particle sources moving along straight line trajectories, the LHS is again
\begin{equation}
\label{eq:soft-fact-comm}
\begin{split}
     \langle {\rm out}| A^{\rm out}_{\mu}(x) S  - S A^{\rm in}_{\mu}(x)|{\rm out}\rangle &= \sum_{i = 1}^n \int \frac{d^{d-1}q}{2 q^0 (2\pi)^{d-1}} \left(e^{iq\cdot x}\frac{ Q_i p_{i,\mu}}{p_i \cdot q - i\epsilon} + e^{-iq \cdot x}\frac{Q_i p_{i,\mu}}{p_i \cdot q + i\epsilon} \right)  \langle {\rm out}|S|{\rm in}\rangle,
     \end{split}
\end{equation}
where $p_i$ are now massive momenta. The computation of the Fourier transforms proceeds as in Appendix \ref{app:sf} with \eqref{eq:d-ft} unchanged. As a result, we immediately find
\begin{equation}
\label{eq:AcS}
    \begin{split}
         \frac{\langle {\rm out}| A^{\rm out}_{\mu}(x) S  - S A^{\rm in}_{\mu}(x)|{\rm out}\rangle}{\langle {\rm out}| S |{\rm out}\rangle} &= \Big[-i\Gamma\left(\frac{d}{2}-1\right)\sum_{i = 1}^n \frac{Q_ip_{i,\mu}}{2\pi^{\frac{d}{2}}} \int_0^{\infty} ds (x^2 - 2 s x\cdot p_i - s^2 + i\epsilon(x^0 - s p_i^0))^{1 - \frac{d}{2}}\\
         &+ i\Gamma\left(\frac{d}{2}-1\right)\sum_{i = 1}^n \frac{Q_ip_{i,\mu}}{2\pi^{\frac{d}{2}}} \int_0^{\infty} ds (x^2 - 2 s x\cdot p_i - s^2 - i\epsilon(x^0 - s p_i^0))^{1 - \frac{d}{2}} \Big] .
    \end{split}
\end{equation}
We now recognize that the integrand can be rewritten in terms of
\begin{equation}
    [-(s - s_+)(s - s_-)]^{1 - \frac{d}{2}},
\end{equation}
where 
\begin{equation}
s_{\pm} = \tau_{\pm}(-p_i). 
\end{equation}
We therefore recover the gauge field computed in Section \ref{sec:em-gf} up to a $\Theta(x^0 - s p_i^0)$ in the integrand. In other words, \eqref{eq:AcS} computes a linear combination of retarded and advanced gauge fields. One can trace back this mismatch to the use of Feynman propagators in the the computation of S-matrix elements, which are momentum space counterparts of time-ordered as opposed to retarded/advanced Green's functions.  The matrix element of the commutator then ensures that the object computed in \eqref{eq:AcS} is a linear combination of advanced and retarded gauge fields.  

\section{Discussion}
\label{sec:disc}

In this section, we conclude with some comments on the relation between our results and flat space holography, as well as the generalization to gravity.

\subsection{Conformal primary wavefunctions in higher dimensions}

By construction, all gauge fields considered in this work transform covariantly under Poincar\'e transformations. In the coordinates \eqref{eq:Mink-lightcone} these are given by
\begin{equation}
    \begin{split}
        P_- &= \p_u, \quad P_+ = \p_r  + y^2 \p_u - \frac{y^A}{r}\p_A, \quad
        P_A = -2 y_A \p_u + \frac{1}{r} \p_A, \\
        M_{AB} &= y_A \p_B - y_B \p_A, \quad M_{+-} = r \p_r - u \p_u -y^A \p_A, \\
        M_{-A} &= \p_A, \quad
        M_{+A} = 2 r y_A \p_r - 2u y_A \p_u + \left(\frac{u}{r} + y^2 \right) \p_A - 2 y_A y^B \p_B,
    \end{split}
\end{equation}
where the first line contains the translation generators and the rest are Lorentz generators. 

Under Poincar\'e transformations, the gauge field transforms covariantly as a one-form, namely
\begin{equation}
  \mathcal{L}_{\xi} A_{\mu} =  \xi^{\nu} \p_{\nu} A_{\mu} +  \p_{\mu} \xi^{\nu} A_{\nu}.
\end{equation}
In  particular, under the Lorentz subalgebra, we find that the gauge field sourced by a particle of momentum $p$ transforms as
\begin{equation}
\label{eq:LTgf}
    \mathcal{L}_{M_{\rho\sigma}} A_{\mu}(x;p) = -\left( p_{\rho} \frac{\p}{\p p^{\sigma}} -  p_{\sigma} \frac{\p}{\p p^{\rho}} \right) A_{\mu}(x;p).
\end{equation}
This formula is valid for both massless and massive source momenta. To derive \eqref{eq:LTgf}, we simply use the representations of the gauge field as a Fourier transform of the soft factor, which was derived in sections \eqref{eq:mc-st} and \eqref{sec:soft-thm-massive}. 

Equation \eqref{eq:LTgf} is the infinitesimal version of the defining property of a conformal primary wavefunction introduced in \cite{Pasterski:2016qvg,Pasterski:2017kqt}.
For massless sources, and away from the source trajectories, the solutions are therefore nothing but $d$-dimensional conformal primary wavefunctions of dimension, or Lorentz boost weight $\Delta = 1$. For massive sources, we obtain a cousin of the conformal primary wavefunctions, namely spin-1 Minkowski propagators parameterized by a spacetime point $x$ and a point on the hyperbolic resolution of $i^{\pm}$,
\begin{equation}
\label{eq:btb-prop}
  K_{\Delta}^{\pm}(x;p) \propto  \int d^{d - 1}q  \frac{e^{\pm iq \cdot x}}{(-q \cdot p\pm i\epsilon)^{\Delta}}.
\end{equation}
These observations lead to a novel class of (sourced) conformal primary wavefunctions defined in terms of the propagators, namely
\begin{equation}
\label{eq:primaries-gf}
    \Psi_{\delta; \Delta}(x;\hat{q}) \equiv  \int d^{d-1} p \frac{K_{\Delta}^{\pm}(x;p)}{(-\hat{q}\cdot p \pm i\epsilon)^{\delta}} .
\end{equation}
Naively, these wavefunctions obey the massless wave equations since $q^2 = 0$. However, as we have seen before, this is only the case away from the source trajectories. In general, \eqref{eq:primaries-gf} are in fact Green's functions for the massless wave equation that transform as primaries with respect to the Lorentz subalgebra of Poincar\'e.

These objects appear to be the natural flat space counterparts of the AdS bulk-to-boundary propagators for gauge fields sourced by massive particles.   We note that they are neither massless nor massive conformal primary wavefunctions. In contrast to \eqref{eq:primaries-gf}, the massive conformal primary wavefunctions are obtained by a Fourier transform of the soft factor with respect to the massive momentum $p$. 
For $\Delta = 1$ and
\begin{equation}
    K_{\Delta}(x;p) = K_{\Delta}^+(x;p) + K_{\Delta}^-(x;p),
\end{equation}
the Fourier transform \eqref{eq:btb-prop} gives rise to the gauge fields explicitly derived for even and odd $d$ in Section \ref{sec:em-gf}. Analogous wavefunctions associated with solutions to the Klein Gordon equation in AdS$_4$ were constructed in \cite{Navarro:2025xln,Navarro:2026rna}, which should generalize to gauge fields and gravity and reduce to these structures in a flat space limit.  

\subsection{Gravitational memory}
\label{sec:gravity}

We conclude by remarking that the generalization to linearized gravity is straightforward. The simplest way to see this is via the all-order in $r$ relation between the position and momentum space fields sourced by particles moving at constant velocity \eqref{eq:soft-fact0} which in gravity will take the form 
    \begin{equation}
\label{eq:soft-fact-grav}
\begin{split}
     \langle {\rm out}| h^{\rm out}_{\mu\nu}(x) S  - S h^{\rm in}_{\mu\nu}(x)|{\rm out}\rangle &\simeq \frac{\kappa}{2} \sum_{i = 1}^n \int \frac{d^{d-1}q}{2 q^0 (2\pi)^{d-1}} \left(e^{iq\cdot x}\frac{ p_{i,\mu} p_{i,\nu}}{p_i \cdot q + i\epsilon} + e^{-iq \cdot x}\frac{p_{i,\mu} p_{i,\nu}}{p_i \cdot q - i\epsilon} \right)  \langle {\rm out}|S|{\rm in}\rangle,
     \end{split}
\end{equation}
where $\kappa = \sqrt{32\pi G_N}$ is the gravitational coupling. We see that the only change with respect to the electromagnetic analysis is a formal replacement of the couplings $Q_i \rightarrow \kappa p_{i\nu}$ which does not affect any of the calculations and results derived in this paper. Consequently, we conclude that the soft graviton theorem in all higher-dimensions implies the existence of a Coulombic order memory effect measurable by an asymptotic observer at large, but finite distance in a parametrically larger, finite time.

\section*{Acknowledgements}

It is a pleasure to thank Martin Ammon, Federico Capone, Laurent Freidel, Temple He, Prahar Mitra, Monica Pate, and Andy Strominger for discussions related to this problem over the last decade, and especially Biswajit Sahoo for initial collaboration, many insightful discussions and extremely valuable comments on a draft. We acknowledge ChatGPT Plus which in conjunction with Mathematica allowed us complete the generalization of our results to massive sources in a finite amount of time.

\appendix

\section{Large-$r$ expansions of Maxwell's equations}
\label{app-large-r-exp}

In this appendix we review the large-$r$ expansions of Maxwell's equations in higher-dimensional spacetimes. We will work in the light-cone coordinates \eqref{eq:coords} in which the Minkowski metric takes the form \eqref{eq:Mink0}. Using the formula for the divergence of an antisymmetric tensor
\begin{equation}
    \nabla_{\mu} F^{\mu\nu} = \frac{1}{\sqrt{\det g}}\p_{\mu}\left(\sqrt{\det g} F^{\mu\nu}\right),
\end{equation}
Maxwell's equations 
\begin{equation}
    \nabla^{\mu}F_{\mu\nu} = e^2j_{\nu}
\end{equation}
become
\begin{equation}
    \begin{split}
        &r^{-2m}\partial_r(r^{2m} F^{ru}) + D_A F^{Au} = e^2 j^u,\\
&\partial_u F^{ur} + D_A F^{Ar} = e^2 j^r,\\
&\partial_u F^{uA} + r^{-2m}\partial_r(r^{2m} F^{rA}) + D_B F^{BA} = e^2 j^A.\\
    \end{split}
\end{equation}
Equivalently with lowered indices
\begin{equation}
\label{eq:Maxwell}
    \begin{split}
        &2 r^{-2m}\partial_r(r^{2m} F_{ur}) - D_A F^{A}_{~r} = -e^2 j_r,\\
&2\partial_u F_{ru} - D_A F^{A}_{~u} = -e^2 j_u,\\
&-2\partial_u F_{rA} - 2r^{2-2m}\partial_r(r^{2m-2} F_{uA}) + D_B F^{B}_{~A} = e^2 j_A.\\
    \end{split}
\end{equation}
These equations are precisely eq. (2.5) in \cite{He:2019jjk}.

Following \cite{He:2019jjk}, we further assume the following fall-offs of the field strength
\begin{equation}
\label{eq:expf}
    \begin{split}
        F_{ur} &= \sum_{n = 0}^{\infty} \frac{F_{ur}^{(m+1+n)}}{r^{m+n+1}},\quad F_{rA}= \sum_{n = 0}^{\infty} \frac{F_{rA}^{(n+m)}}{r^{m+n}}, \\
        F_{uA} &= \sum_{n = 0}^{\infty} \frac{F_{uA}^{(m+n-1)}}{r^{m - 1 + n}}, \quad
        F_{AB} = \sum_{n = 0}^{\infty}\frac{F_{AB}^{(m-1+n)}}{r^{m-1+n}}, \quad n \in \frac{1}{2}\mathbb{Z}
    \end{split}
\end{equation}
and the sources 
\begin{equation}
    \begin{split}
        j_u \sim \mathcal{O}(r^{-2m}), \quad j_r \sim \mathcal{O}(r^{-2m - 2}), \quad j_A \sim \mathcal{O}(r^{-2m}).
    \end{split}
\end{equation}
In general, $n$ is allowed to run over both integer and half integer values. In even spacetime dimension, ie. in the case where $m \in \mathbb{Z}$, it is consistent to restrict to $n \in \mathbb{Z}$. In odd spacetime dimensions, the radiative fields contain an integer and a half-integer series in $r^{-1}$, cf. eq. (2.27) in \cite{He:2019jjk}, while the Coulombic fields only contain an integer series. 

Let us further assume that there is no magnetic field, namely
\begin{equation}
    F_{AB} = 0.
\end{equation}
Substituting the expansions \eqref{eq:expf} into the Maxwell equations \eqref{eq:Maxwell}, we find
\begin{equation}
\label{eq:Maxwell-large-r}
\begin{split}
    &2(m-n-1) F_{ur}^{(m+n+1)} - D^AF_{Ar}^{(m + n)} = 0, \quad n \leq m -1,\\
    &2 \p_u F_{ru}^{(n+m+1)} - D^A F_{Au}^{(m+n-1)} = 0, \quad n \leq m-2,\\
    &\p_u F_{rA}^{(m+n)} + (m - n -1) F_{uA}^{(m+n-1)} = 0, \quad n \leq m-1.
    \end{split}
\end{equation}
The last equation determines $F_{Ar}$ in terms of $F_{Au}$ (for $n \leq m - 2$), while the first and second equations can be seen to be equivalent after accounting for this relation. 

Also useful will be the leading order sourced Maxwell constraint
\begin{equation}
    2 \p_u F_{ru}^{(2m)} - D^A F_{Au}^{(2m-2)} = -e^2 j_u^{(2m)}
\end{equation}
together with the Bianchi identities
\begin{equation}
\label{eq:Bianchi}
    \nabla_{\mu} F_{\nu \rho} + \nabla_{\nu} F_{\rho \mu} + \nabla_{\rho} F_{\mu\nu} = 0.
\end{equation}
For $\mu = A, \nu = r, \rho = u$, \eqref{eq:Bianchi} reduce to
\begin{equation}
    \p_A F_{ru} + \p_{r} F_{uA} + \p_u F_{Ar} =  0,
\end{equation}
which at large $r$ become
\begin{equation}
\begin{split}
    &\p_A F_{ru}^{(m+n+1)} - (m+n) F_{uA}^{(m+n)} + \p_{u}F_{Ar}^{(m+n+1)} = 0 \implies\\
     &\p_A F_{ru}^{(m+n+1)} - (m+n) F_{uA}^{(m+n)} + (m- n - 2) F_{uA}^{(m+n)} = 0 \implies \\
     &\p_A F_{ru}^{(m+n+1)} - 2(n+1) F_{uA}^{(m+n)} = 0.
    \end{split}
\end{equation}

Substituting this into the second equation of \eqref{eq:Maxwell-large-r}, we see that the subleading components of $F_{ru}$ are recursively determined in terms of the leading radiative one, namely
\begin{equation}
    2 \p_u F_{ru}^{(n+m+1)} + \frac{1}{2 n} D^A \p_A F_{ru}^{(m+ n)} = 0, \quad   2m - 2 \geq n > 0.
\end{equation}

\section{Retarded Green's functions}
\label{app:Greensf}

The retarded Green's function can be expressed in terms of a Fourier transform
\be\label{eq:GR_mom_space}
G_R(x-y)=\int\f{d^d\ell}{(2\pi)^d}\f{e^{i\ell \cdot (x-y)}}{(\ell^0+i\epsilon)^2-\vec{\ell}^2}, 
\ee
where the choice of $i\epsilon$ prescription ensures that the result vanishes  for $x^0<y^0$. Indeed, in this case, the $\ell^0$ contour can be closed in the upper half complex $\ell^0$ plane where there are no poles.
For $x^0>y^0$ the $\ell^0$ integral can be evaluated by closing the contour in the  lower half plane and picking up the  contributions from the poles at $\ell^0=\pm|\vec{\ell}|-i\epsilon$. The result of the integration is
\be
\label{eq:GR_deneral_d}
G_R(x-y)=-\f{i}{2}e^{-\epsilon(x^0-y^0)}\Theta(x^0-y^0)\int\f{d^{d-1}\vec{\ell}}{(2\pi)^{d-1}}\f{1}{|\vec{\ell}|}\, e^{i\vec\ell \cdot (\vec x-\vec y)}\left[e^{-i|\vec \ell|(x^0-y^0)}-e^{i|\vec \ell|(x^0-y^0)}\right]. 
\ee

The remaining integral can be performed by parameterizing $\vec{\ell}$ with spherical coordinates in which $d^{d-1}\vec\ell=|\vec \ell|^{d-2}d\theta (\sin\theta)^{d-3}d|\vec\ell|d\Omega_{d-3}$, where $\theta$ is the angle between $\vec\ell$ and $\vec x-\vec y$ and $d\Omega_{d-3}$ is the measure on the $(d-3)$-dimensional sphere $S^{d-3}$. Then \eqref{eq:GR_deneral_d} simplifies to 
\begin{equation}
\begin{split}
G_R(x-y)&=-\f{i}{2(2\pi)^{d-1}}e^{-\epsilon(x^0-y^0)}\Theta(x^0-y^0)\text{Vol}(S_{d-3}) \\
&\times \int_0^\infty d|\vec{\ell}| |\vec{\ell}|^{d-3} \left[e^{-i|\vec \ell|(x^0-y^0)}-e^{i|\vec \ell|(x^0-y^0)}\right]\int_0^\pi d\theta \left(\sin\theta\right)^{d-3} e^{i|\vec\ell||\vec x-\vec y|\cos\theta}, 
\end{split}
\end{equation}
where 
\begin{equation}
    \text{Vol}(S_{d-3})=\f{2(\pi)^{\f{d}{2}-1}}{\Gamma\left(\f{d}{2}-1\right)}. 
    \end{equation}
The integration over $\theta$ can be performed using the integral representation of Bessel function of first kind $J_\nu(z)$ 
\begin{equation}
    J_{\nu}(z) = \frac{\left(\frac{z}{2} \right)^{\nu}}{\sqrt{\pi} \Gamma(\nu + \frac{1}{2})} \int_{0}^{\pi} d\theta (\sin\theta)^{2\nu} e^{i z \cos \theta} 
\end{equation}
which yields
\begin{equation}
\label{eq:Gr-int}
\begin{split}
G_R(x-y)&= -\f{i}{2(2\pi)^{d-1}}\f{2^{\f{d-3}{2}}\sqrt{\pi}\, \Gamma\left(\f{d-2}{2}\right)}{|\vec x-\vec y|^\f{d-3}{2}}e^{-\epsilon(x^0-y^0)}\Theta(x^0-y^0)\text{Vol}(S_{d-3})\\
&\times \int_0^\infty d|\vec{\ell}| |\vec{\ell}|^{\f{d-3}{2}} \left[e^{-i|\vec \ell|(x^0-y^0)}-e^{i|\vec \ell|(x^0-y^0)}\right]J_{\f{d-3}{2}}\left(|\vec\ell|\, |\vec x-\vec y|\right). 
\end{split}
\end{equation}
This integral can be evaluated using the identity 
\bea
\label{eq:Mellin-J}
\int_0^\infty dxx^\nu e^{-\alpha x}J_\nu (\beta x)=\frac{(2\beta)^\nu \Gamma\left(\nu+\f{1}{2}\right)}{\sqrt{\pi}(\alpha^2+\beta^2)^{\nu+\f{1}{2}}}, 
\eea
which holds for $\text{Re}(\alpha)>\text{Im}(\beta)$. To apply this result, we shift $\alpha = i(x^0 - y^0) \rightarrow i(x^0 - y^0) + \varepsilon$ in the first term and $x^0 - y^0 \rightarrow x^0 - y^0 + i\varepsilon$ in the second term of \eqref{eq:Gr-int}, and integrate over $|\vec\ell|$ to obtain
\begin{equation}
\label{eq:finalG}
\begin{split}
G_R(x-y)&= -\f{i}{4\pi^\f{d}{2}} \Gamma\left(\f{d-2}{2}\right)e^{-\epsilon(x^0-y^0)}\Theta(x^0-y^0)\\
&\times \lim_{\varepsilon\rightarrow 0}\left[\f{1}{\left[-(x^0-y^0-i\varepsilon)^2+|\vec x-\vec y|^2\right]^\f{d-2}{2}}-\f{1}{\left[-(x^0-y^0+i\varepsilon)^2+|\vec x-\vec y|^2\right]^\f{d-2}{2}}\right].
\end{split}
\end{equation}
Note that in presence of $\Theta(x^0-y^0)$ the factor 
\be
(x^0-y^0\pm i\varepsilon)^2-|\vec x-\vec y|^2=(x^0-y^0)^2-|\vec x-\vec y|^2 \pm i\varepsilon,
\ee
behaves differently depending on whether its power is even (pole singularity) or odd (branch cut singularity). As a result, the limit $\varepsilon\rightarrow 0$ has to be treated separately for even and odd values of $d$.

\paragraph{Even spacetime dimensions:} Introducing $A\equiv (x^0-y^0)^2-|\vec x-\vec y|^2$, the expression in \eqref{eq:GR_deneral_d} can be rewritten as
\begin{equation}
\begin{split}
G_R(x-y) &\overset{d\in\text{even}}{=} (-1)^\f{d}{2}\f{i}{4(\pi)^\f{d}{2}} \Gamma\left(\f{d-2}{2}\right)e^{-\epsilon(x^0-y^0)}\Theta(x^0-y^0)\\
&\times  \f{(-1)^\f{d-4}{2}}{\Gamma\left(\f{d-2}{2}\right)}\left(\f{d}{dA}\right)^\f{d-4}{2}\left[\f{1}{A-i\varepsilon}-\f{1}{A+i\varepsilon}\right]. 
\end{split}
\end{equation}
Using the identity $\f{1}{A\pm i0}=\text{PV}\f{1}{A}\mp i\pi\delta(A)$ and explicitly substituting the value of $A$, the final form of even dimensional Green's function reads 
\be\label{eq:GR_even_d}
\boxed{G_R(x-y)
\overset{d\in\text{even}}{=}-\f{1}{2\pi^\f{d-2}{2}} e^{-\epsilon(x^0 - y^0)}\Theta(x^0-y^0) \delta^{\left(\f{d-4}{2}\right)}(-(x-y)^2)}. 
\ee
 This result agrees, up to normalization, with that stated in  \cite{Garfinkle:2017fre}.

\paragraph{Odd spacetime dimensions:} For integer  $\nu$ values, the function $(a\pm i \varepsilon)^
{\f{1}{2}+\nu}$ has branch cuts for negative values of $a$ with branch points $\mp i \varepsilon$. This function is then given by $|a|^{\nu+\f{1}{2}} e^{\pm i\pi \left(\nu+\f{1}{2}\right)}$ for $a<0$. Therefore we get 
\be
\label{eq:theta-id}
\lim_{\varepsilon\rightarrow 0}\left[(a+ i \varepsilon)^
{\f{1}{2}+\nu}-(a- i \varepsilon)^
{\f{1}{2}+\nu}\right]=2i \sin\left(\nu\pi+\f{\pi}{2}\right) (-a)^{\nu+\f{1}{2}}\Theta(-a)=2i(-1)^\nu  (-a)^{\nu+\f{1}{2}}\Theta(-a), \quad \nu > 0,
\ee
 while for $\nu < 0$ there is an additional delta function series contribution due to the fact that we can write
\begin{equation}
    (a+i\epsilon)^{\frac{1}{2} + \nu} = (-1)^{-\nu - 1} \frac{\sqrt{\pi}}{\Gamma\left( - \nu - \frac{1}{2} \right)} \left(\frac{d}{da} \right)^{-\nu - 1} \frac{1}{\sqrt{a + i\epsilon}}
\end{equation}
and therefore, applying \eqref{eq:theta-id} with $\nu = -1$,
\begin{equation}
\label{eq:id-ds}
    \lim_{\varepsilon\rightarrow 0}\left[(a+ i \varepsilon)^
{\f{1}{2}+\nu}-(a- i \varepsilon)^
{\f{1}{2}+\nu}\right] = 2i (-1)^{-\nu} \frac{\sqrt{\pi}}{\Gamma\left( - \nu - \frac{1}{2} \right)} \left(\frac{d}{da} \right)^{-\nu - 1} \left[(-a)^{-\frac{1}{2}} \Theta(-a) \right], \quad \nu < 0.
\end{equation}
Comparing these identities with  \eqref{eq:finalG}, we identify $a=(x-y)^2$ and $\nu=-\f{d-1}{2}$, which is a negative integer for odd values of $d \geq 3$. This identity allows us to express the finite part of the odd dimensional Green's function as
\be\label{eq:GR_odd_d}
\boxed{G_R(x-y)\overset{d\in\text{odd}}{=}(-1)^\f{d-1}{2}\f{1}{2\pi^\f{d}{2}} \Gamma\left(\f{d-2}{2}\right)e^{-\epsilon(x^0-y^0)} \left(-(x-y)^2\right)^{-\f{d-2}{2}} \Theta(x^0-y^0-|\vec x-\vec y|)},
\ee
where we used the fact that $\Theta(x^0-y^0)\Theta(-(x-y)^2)=\Theta(x^0-y^0-|\vec x-\vec y|)$. For odd values of $d\geq 5$, we can set $\epsilon=0$. When $d = 3$, $\epsilon$ plays the role of an IR cut-off \cite{Shimizu:2025hfl}. The divergent series is accounted for by using the identity \eqref{eq:id-ds} which yields \cite{Satishchandran:2017pek}
\begin{equation}
\label{eq:full-gf-odd}
     G_R(x-y) \overset{d\in \text{odd}}{=}  \frac{-1}{2\pi^{\frac{d-1}{2}}} \Theta(x^0-y^0) \left(\frac{d}{d\sigma} \right)^{\frac{d - 3}{2}}\left[ \sigma^{-\frac{1}{2}} \Theta(\sigma) \right],\quad 
    \sigma = -(x-y)^2.
\end{equation}

 \section{Divergences in the field strength in odd $d$}
 \label{sec:div-odd}

 In this appendix, we examine the divergent contributions to the field strength \eqref{eq:odd-components} and show that they do not contribute to the Coulombic order constraints. It is straightforward to show that the divergent terms are given by
 \begin{equation}
\label{eq:odd-components-div}
\begin{split}
        F_{ru}^{{\rm div},i}(x)
        &=  2 N Q_i\,\delta(-x^2)(q\cdot n)
        \frac{u(n\cdot p_i)-r(q\cdot p_i)}{(d-4)x\cdot p_i}
        (-x^2)^{2-\frac d2},\\
        F^{{\rm div},i}_{Au}(x)
        &= 2 N Q_i \,\delta(-x^2)
        \frac{r^2(n\cdot q)(\p_Aq\cdot p_i)}{(d-4)x\cdot p_i}
        (-x^2)^{2-\frac d2}.
    \end{split}
\end{equation}
Since $-x^2 = ur$, they are localized at $u = 0$ and therefore, it is clear that the change of the Coulomb field between $u = \pm \infty$ vanishes, 
\begin{equation}
\Delta F_{ru}^{\rm div, i} = 0.
\end{equation}
As for $F_{Au}^{{\rm div}, i}$, it is clear the order by order in the large-$r$ expansion there are no Coulombic order contributions. On the other hand, at finite $r = r_0$, we find
\begin{equation}
   \int du F_{Au}^{{\rm div},i} = 2 N Q_i \, r_0^{4 - d} \lim_{u \rightarrow 0}
        \frac{(n\cdot q)(\p_Aq\cdot p_i)}{(d-4)(\frac{u}{r_0} n\cdot p_i + q \cdot p_i)}
        \left(\frac{u}{r_0}\right)^{2-\frac d2} .
\end{equation}
Then setting 
\begin{equation}
    \epsilon^2 = \frac{u}{r_0}
\end{equation}
and using the delta function identity (see Section \ref{sec:rel-previous})
\begin{equation}
  \lim_{\epsilon \rightarrow 0}  \frac{\epsilon^{4 - d}}{\epsilon^2 + |y - z_i|^2} = \pi^{\frac{d - 2}{2}}\Gamma\left( 2 - \frac{d}{2} \right) \delta^{(d  - 2)}(y - z_i)
\end{equation}
we find
\begin{equation}
     \int du F_{Au}^{{\rm div},i}(x) = 2 \pi^{\frac{d - 2}{2}} \Gamma\left(2 - \frac{d}{2} \right)N Q_i \, r_0^{4 - d} 
        \frac{(\p_Aq\cdot p_i)}{(d-4)} \delta^{(d - 2)}(y - z_i) = 0,
\end{equation}
where the last equality follows from
\begin{equation}
    \p_Aq \cdot p_i \propto (y - z_i)_A.
\end{equation}

We conclude by noting that these terms would either way cancel against the terms proportional to $\delta(-x^2)$ appearing in the divergent gauge field in \eqref{eq:gf-odd}, so we can regard them as artifacts of the regularization procedure.

\section{Soft theorem and eikonal currents}

In this appendix, we give details of the computations in Section \eqref{eq:mc-st} which establish a relation between soft photon Ward identities, gauge fields sourced by particles of constant momenta and the Maxwell constraint to all orders in a $r^{-1}$ expansion. 

\subsection{Fourier transform of soft factor to all orders in $r$}
\label{app:sf}

In this appendix, we prove \eqref{eq:FT-soft-factor}. We first express the soft factor as a Mellin integral
\begin{equation}
\int \frac{d^{d-1} q}{2q^0} e^{iq \cdot x} \frac{p_i^{\mu}}{p_i \cdot q - i\epsilon} = i p_i^{\mu}\int \frac{d^{d - 1} q}{2q^0} e^{iq\cdot x} \int_0^{\infty} ds e^{-is (p_i \cdot q - i\epsilon)}.
\end{equation}
The $q$ integral can then be evaluated using the integral transform
\begin{equation}
\label{eq:d-ft}
    \int \frac{d^{d-1} q}{2q^0} e^{iq\cdot(y -i\epsilon t)} = 2^{d - 3} \pi^{d/2 - 1} \Gamma\left(\frac{d}{2} - 1 \right)(y^2 + i\epsilon y^0)^{1 - d/2},
\end{equation}
where $t$ is a timelike vector, $q^0 >0$ and 
\begin{equation}
    y = x - s p_i \implies y^2 = x^2 - 2 s x \cdot p_i.
\end{equation}
Eq. \eqref{eq:d-ft} can be derived by setting $q = \omega \hat{q}$, observing that 
\begin{equation}
    \frac{d^{d-1} q}{2 q^0} = \frac{\omega^{d - 2} d\omega d^{d-2}\hat{q}}{2}
\end{equation}
and that
\begin{equation}
  \hat{q}\cdot y = - \frac{1}{2}(\hat{q} - y)^2 +  \frac{y^2}{2}.
\end{equation}
The $i\epsilon$ prescription is inherited from the soft factor and follows from the identity
\begin{equation}
    (y - i\epsilon t)^2 \sim y^2 + i\epsilon y^0.
\end{equation}
The Gaussian integral in $\mathbb{R}^{d - 2}$ can be performed straightforwardly using the formula
\begin{equation}
    \int d^{d - 2} \hat{q} e^{-i \frac{\omega}{2}(\hat{q} - y)^2} = \left(\frac{2\pi}{i\omega} \right)^{\frac{d - 2}{2}},
\end{equation}
followed by the Mellin transform in $\omega$,
\begin{equation}
\begin{split}
    \frac{1}{2}\int_0^{\infty}d \omega \omega^{d-3} \left(\frac{2\pi}{i\omega} \right)^{\frac{d - 2}{2}} e^{\frac{i\omega}{2}(y^2 + i\epsilon y^0)} &= \frac{1}{2} \Gamma\Big(\frac{d - 2}{2}\Big) (-2\pi i)^{\frac{d - 2}{2}} \Big(-\frac{2}{i(y^2 + i\epsilon y^0)}\Big)^{\frac{d-2}{2}}\\
    &=2^{d - 3} \pi^{\frac{d-2}{2}}\Gamma\Big(\frac{d - 2}{2}\Big)(y^2 + i\epsilon y^0)^{1 - \frac{d}{2}}.
    \end{split}
\end{equation}
In the final step we need to evaluate the integral on $s$. Noting that
\begin{equation}
    (x^2 - 2 s x \cdot p_i + i\epsilon (x^0 - sp_i^0))^{1 - d/2} = -\frac{1}{(4 - d) x\cdot p}  \frac{d}{ds}(x^2 - 2 s x \cdot p_i + i\epsilon (x^0 - sp_i^0))^{2 - d/2},
\end{equation}
we immediately find that for $d > 4$
\begin{equation}
\label{eq:FT}
    \int_0^{\infty} ds (x^2 - 2 s x \cdot p_i + i\epsilon (x^0 - sp_i^0))^{1 - d/2} = -\frac{1}{(d - 4) x\cdot p_i} (x^2 + i\epsilon x^0)^{2 - d/2}.
\end{equation}

The result involving the complex conjugate of the Fourier kernel is related to \eqref{eq:FT} by complex conjugation, or equivalently setting $x \rightarrow - x$, namely
\begin{equation}
    \int \frac{d^{d-1} q}{2q^0} e^{-iq \cdot x} \frac{p_i^{\mu}}{p_i \cdot q +i\epsilon} = ip_i^{\mu} 2^{d - 3} \pi^{d/2 - 1}\Gamma\left(\frac{d}{2} - 1\right)\frac{(x^2 - i\epsilon x^0)^{2 - \frac{d}{2}}}{(d - 4) x \cdot p_i}.
\end{equation}
The phase multiplying $x^2$ will be important for odd $d.$ 

\subsection{Eikonal current action on asymptotic states}
\label{app:current}

We consider for concreteness the current sourced by a massless scalar field
\begin{equation}
    J_{\mu}(x) = -i Q : \phi \overleftrightarrow{\partial} \phi:(x) \supset Q \int \frac{d^{d-1}k}{(2\pi)^{d-1} 2k^0}\int \frac{d^{d-1}p}{(2\pi)^{d-1} 2p^0} (k+p)^{\mu} a(\vec{k})^{\dagger} a(\vec{p}) e^{i(p - k)\cdot x}  + \cdots,
\end{equation}
where $\cdots$ contain two-particle creation/annihilation terms. Using the canonical commutation relations, we obtain the action of the current on an incoming momentum eigenstate 
\begin{equation}
\label{eq:current-state}
    J_{\mu}(x)|p\rangle = Q \int \frac{d^{d-1}k}{(2\pi)^{d-1} 2k^0} (k+p)_{\mu} e^{i(p - k)\cdot x}|k\rangle + \cdots .
\end{equation}

We now consider the eikonal approximation \cite{Bloch:1937pw}
\begin{equation}
    k = p + \ell, \quad \ell^0 \ll p^0.
\end{equation}
Then 
\begin{equation}
\label{eq:eik-u}
    0= k^2 = 2 p \cdot \ell + \ell^2 \simeq 2p\cdot \ell
\end{equation}
where the last identity is a result of the approximation. In this limit, \eqref{eq:current-state}  becomes
\begin{equation}
\begin{split}
     J_{\mu}(x)|p\rangle &= Q \int \frac{d^{d-1}\ell}{(2\pi)^{d-1}} \left( \frac{p^{\mu}}{p^0} e^{-i\ell\cdot x}|p\rangle + \mathcal{O}(\frac{\ell^0}{p^0}) \right)\\
     &= \frac{p^{\mu}}{p^0} \delta^{(d-1)}\Big(\vec{x} - \frac{\vec{p}}{p^0}x^0\Big)|p\rangle,
     \end{split}
\end{equation}
where $\cdots$ disappear since they are multiplied by $e^{\pm 2ip\cdot x}$ which are highly suppressed. 
We used 
\begin{equation}
    \ell \cdot x = -\ell^0 x^0 + \vec{\ell}\cdot \vec{x} = \vec{\ell}\cdot(\vec{x} - \frac{\vec{p}}{p^0}x^0),
\end{equation}
which follows from \eqref{eq:eik-u}. 

As a result,
\begin{equation}
    \langle {\rm out}| S J^{\mu}(x)|{\rm in}\rangle = \sum_{i = 1}^n \Theta(-x^0)Q_i \frac{p_i^{\mu}}{p_i^0} \delta^{(d-1)}\Big(\vec{x} - \frac{\vec{p}_i}{p_i^0}x^0\Big)\langle {\rm out}| S |{\rm in}\rangle 
\end{equation}
while
\begin{equation}
     \langle {\rm out}| J_{\mu}(x) S |{\rm in}\rangle = \sum_{i = 1}^n \Theta(x^0)Q_i\frac{p_i^{\mu}}{p_i^0} \delta^{(d-1)}\Big(\vec{x} - \frac{\vec{p}_i}{p_i^0}x^0\Big) \langle {\rm out}| S |{\rm in}\rangle,
\end{equation}
where the theta functions appear since the current is inserted before and after the interaction assumed to be localized at $x^0 = 0$. 
We now use the identities
\begin{equation}
   \frac{1}{p^0} \Theta(x^0) = \int_0^{\infty} d\tau \delta(x^0 - p^0 \tau), \quad p^0 > 0
\end{equation}
and
\begin{equation}
   \frac{1}{p^0} \Theta(-x^0) = -\int_0^{\infty} d\tau \delta(-x^0 + p^0 \tau), \quad p^0 < 0,
\end{equation}
where we recall that we are working in the all-outgoing convention. Since $\delta(x) = \delta(-x)$, we obtain
\begin{equation}
      \langle {\rm out}| J^{\mu}(x) S |{\rm in}\rangle- \langle {\rm out}| S J^{\mu}(x)|{\rm in}\rangle = \sum_{i = 1}^n Q_i  p^{\mu}_i \int_0^{\infty} d\tau \delta^{(d)}(x - p_i \tau) \langle {\rm out}|  S |{\rm in}\rangle
\end{equation}
which confirms \eqref{eq:current}.

\section{Massive sources in even dimensions}

In this appendix, we collect the derivations related to the results presented in Section \ref{sec:even-massive}.

\subsection{Proof of eq. \eqref{eq:df-id}}
\label{app:delta-fid}

The proof of \eqref{eq:df-id} proceeds by writing
\begin{equation}
    \int_{I} d\tau g(\tau) \delta^{(n)}(f) = \p_\lambda^n \int_{I} d\tau g(\tau) \delta(f + \lambda)\left. \right|_{\lambda = 0}
\end{equation}
and observing that the roots of the new, $\lambda$-deformed equation
\begin{equation}
    (\tau - \tau_+)(\tau - \tau_-) + \lambda = 0
\end{equation}
are
\begin{equation}
    \tau_{\pm}(\lambda) = \frac{\tau_+ + \tau_-}{2} \pm \frac{1}{2} \sqrt{(\tau_+ - \tau_-)^2 - 4 \lambda},
\end{equation}
namely precisely \eqref{eq:tau-lpm}. Using the standard identity
\begin{equation}
    \delta\Big((\tau - \tau_+(\lambda))(\tau - \tau_-(\lambda))\Big) = \frac{1}{|\tau_+(\lambda) - \tau_-(\lambda)|} \left(\delta(\tau - \tau_+(\lambda)) + \delta(\tau - \tau_-(\lambda))\right),
\end{equation}
we arrive immediately at \eqref{eq:df-id} since
\begin{equation}
    \tau_+(\lambda) - \tau_-(\lambda) = \sqrt{\Delta^2 - 4\lambda}.
\end{equation}

\subsection{Proof of eq. \eqref{eq:theta-der-m}}
\label{sec:theta-der}

In this Appendix we prove the identity
\begin{equation}
    \p_{\lambda}^m \Theta(\tau_{\pm}(\lambda)) = \sum_{j = 0}^{m - 1} (\mp)^{j+1} \frac{(2m-j-2)!}{j! (m - j - 1)!} \frac{\delta^{(j)}(\tau_{\pm}(\lambda))}{(\sqrt{\Delta^2 - 4\lambda})^{2m - j - 1}}, \quad m \geq 1.
\end{equation}
We first notice that
\begin{equation}
    \p_{\lambda}\tau_{\pm}(\lambda) =  \mp \frac{1}{\sqrt{\Delta^2 - 4 \lambda}}.
\end{equation}
Then
\begin{equation}
\begin{split}
    \p_{\lambda}^m \Theta(\tau_{\pm}(\lambda)) &= \mp \sum_{j = 0}^{m - 1} \left(\begin{matrix}
        m - 1\\
        j
    \end{matrix} \right) \p_{\lambda}^{j}\delta(\tau_{\pm}) \frac{(2(m -1 - j))!}{(m - j - 1)!} \frac{1}{(\sqrt{\Delta^2 - 4\lambda})^{2(m - 1 - j)+1}}\\
    &= \sum_{j = 0}^{m - 1} (\mp)^{j+1} A_{m;j}\delta^{(j)}(\tau_{\pm}) \frac{1}{(\sqrt{\Delta^2 - 4\lambda})^{2m - 1 - j}}
    \end{split}
\end{equation}
where in the last line $A_{m;j}$ are the result of resummation when applying the chain rule for the derivative of the delta function. It can be computed more easily by observing that the coefficients $A_{m;j}$ obey the recursion relation
\begin{equation}
\label{eq:rr}
    A_{m+1;j} = A_{m;j-1} +  2(2m - j - 1) A_{m;j},
\end{equation}
which follows from the identity
\begin{equation}
    \p_{\lambda}^{m+1}\Theta(\tau_{\pm}) = \sum_{j = 0}^{m - 1} (\mp 1)^{j+1}\left[ A_{m;j} (\mp) \frac{\delta^{(j+1)}(\tau_{\pm})}{(\sqrt{\Delta^2 - 4 \lambda})^{2m-j}} + 2(2m - j - 1) A_{m;j} \frac{\delta^{(j)}(\tau_{\pm})}{(\sqrt{\Delta^2 - 4 \lambda})^{2m - j + 1}} \right].
\end{equation}
The solution to \eqref{eq:rr} is easily found to be 
\begin{equation}
    A_{m;j} = \frac{(2m - j - 2)!}{j!(m - j - 1)!}.
\end{equation}

\subsection{Solutions to Maxwell's equations}
\label{sec:Maxwell-sol-mass}

In this appendix, we verify explicitly that the field strength 
\begin{equation}
\label{eq:fmunumassive-app}
\begin{split}
    F_{\mu\nu}(p_i) 
    &= \frac{Q_ip_{[\mu} x_{\nu]}}{\pi^{\frac{d-2}{2}}} \sum_{j = -1}^n a_j \frac{\delta^{(j)}(\tau_{\pm})}{\Delta^{2n - j + 2}} \equiv p_{[\mu} x_{\nu]} f(\tau_{\pm}, \Delta), \quad a_j = (\mp)^{j-1}\frac{(2n-j+1)!}{(j+1)!(n-j)!}
    \end{split}
\end{equation}
obeys the source-free Maxwell constraint order by order in the large-$r$ expansion, up to delta function contact terms that appear at the Coulombic order. 
We have 
\begin{equation}
    \p^{\mu} F_{\mu\nu} = \p^{\mu}\left[(p_{\mu} x_{\nu}- x_{\mu} p_{\nu}) f(\tau_{\pm}, \Delta) \right] = (1 - d) p_{\nu} f + x_{\nu} p \cdot \p f - p_{\nu} x\cdot \p f.
\end{equation}
Then using the identities
\begin{equation}
\label{eq:identities-del-tpm}
\begin{split}
    p \cdot \p \tau_{\pm} &= 1, \quad ~~ p \cdot \p \Delta = 0, \\
    x\cdot \p \tau_{\pm} &= \tau_{\pm}, \quad x \cdot \p \Delta = \Delta
  \end{split}  
\end{equation}
we obtain
\begin{equation}
     \p^{\mu} F_{\mu\nu} = (1 - d) p_{\nu} f + x_{\nu} \p_{\tau_{\pm}} f - p_{\nu}\left( \tau_{\pm} \p_{\tau_{\pm}} + \Delta \p_{\Delta} \right) f.
\end{equation}
Setting $\nu = u$, collecting the terms proportional to $p_u$ and recalling that $d = 2n +4$, we find
\begin{equation}
    \left((1 - d) - \Delta \p_{\Delta} \right) \frac{\delta^{(j)}(\tau_{\pm})}{\Delta^{2n - j + 2}} = (1 - d+ d - 4 - j + 2) \frac{\delta^{(j)}(\tau_{\pm})}{\Delta^{2n - j + 2}} = -(j+1)\frac{\delta^{(j)}(\tau_{\pm})}{\Delta^{2n - j + 2}},
\end{equation}
while 
\begin{equation}
   - \tau_{\pm} \p_{\tau_{\pm}}\frac{\delta^{(j)}(\tau_{\pm})}{\Delta^{2n - j + 2}}= (j + 1)\frac{\delta^{(j)}(\tau_{\pm})}{\Delta^{2n - j + 2}}
\end{equation}
so the terms proportional to $p_{u}$ cancel. Then
\begin{equation}
    \p^{\mu} F_{\mu u} = x_u \p_{\tau_{\pm}} f = \frac{r}{2} \p_{\tau_{\pm}} f =  \frac{Q_i r}{2\pi^{\frac{d - 2}{2}}} \sum_{j = 0}^{n + 1}(\mp)^{j- 2} \frac{(2n - j + 2)!}{j! (n - j + 1)!} \frac{\delta^{(j)}(\tau_{\pm})}{\Delta^{2n - j + 3}}.
\end{equation}
Finally, one can show that the RHS is in fact $\propto \delta(x^2)$ near $\tau_- = 0.$ To see this, note that on the one hand, we have
\begin{equation}
\label{eq:delta-1}
    \delta(x^2 - \lambda) = \sum_{n = 0}^{\infty} \frac{(-\lambda)^n}{n!} \delta^{(n)}(x^2)
\end{equation}
while on the other hand, near $\tau_- = 0$
\begin{equation}
    x^2 = - \tau_- \tau_+ = - \tau_-^2 - \tau_- \Delta
\end{equation}
so
\begin{equation}
    x^2 - \lambda = - \tau_-^2 - \tau_- \Delta-\lambda = 0 \implies \tau_- = \frac{\Delta - \sqrt{\Delta^2 - 4\lambda}}{-2} \equiv -w(\lambda).
\end{equation}
Then
\begin{equation}
    \delta(x^2 - \lambda) =  \frac{\delta(\tau_- + w(\lambda))}{\Delta - 2 w(\lambda)} = \sum_{n = 0}^{\infty} \frac{w^n}{n!} \frac{1}{\Delta - 2 w} \delta^{(n)}(\tau_-).
\end{equation}
We would like to rewrite the RHS as a power series in $\lambda$. To extract the coefficient of $\lambda^N$, it is sufficient to compute the residue 
\begin{equation}
   \oint_{\lambda} d\lambda \frac{w(\lambda)^n}{\Delta - 2 w(\lambda)} \lambda^{-N - 1} = \oint_{w} dw \frac{w^n}{w^{N+1}(\Delta - w)^{N+1}},
\end{equation}
where we used the identities
\begin{equation}
    \lambda = w(\Delta - w), \quad d\lambda = (\Delta - 2 w) dw.
\end{equation}
Furthermore,
\begin{equation}
    \oint_{w} dw \frac{w^n}{w^{N+1}(\Delta - w)^{N+1}} = \sum_{j = 0}^{\infty}  \oint_{w} dw  w^{n - N - 1} (-w)^j \Delta^{-N-1-j} \left(\begin{matrix}
        -N - 1\\
        j
    \end{matrix} \right) = \frac{(2N-n)!}{N! (N - n)!} \frac{1}{\Delta^{2N - n +1}}.
\end{equation}
It follows that 
\begin{equation}
\label{eq:delta-2}
    \delta(x^2 - \lambda) = \sum_{N = 0}^{\infty} \frac{\lambda^N}{N!} \sum_{n = 0}^N \frac{(2N - n)!}{n! (N - n)!} \frac{\delta^{(n)}(\tau_-)}{\Delta^{2N - n + 1}}.
\end{equation}
Equating \eqref{eq:delta-1} and \eqref{eq:delta-2}, we obtain the identity
\begin{equation}
    \sum_{k = 0}^N \frac{(2N - k)!}{k! (N - k)!} \frac{\delta^{(k)}(\tau_-)}{\Delta^{2N - k + 1}} = (-)^N \delta^{(N)}(x^2).
\end{equation}
Then setting $N = n+1$, we conclude that, near $\tau_- = 0,$
\begin{equation}
\label{eq:fs-con}
    \p^{\mu} F_{\mu u} = \frac{Q_i r}{2\pi^{\frac{d - 2}{2}}} (-1)^{n+1} \delta^{(n+1)}(x^2), \quad \tau_- = 0.
\end{equation}
The analysis can be repeated near $\tau_+ = 0$ in which case we again find \eqref{eq:fs-con}. Since $\p^{\mu} F_{\mu u}$ is independent on the momenta, this establishes that the constraint is obeyed for charge-conserving field configurations, up to delta function contact terms. 

\subsection{Large-r expansions}
\label{sec:large-r}

We would like to extract the $\mathcal{O}(r^{-2n-2})$ of $F_{ru}$.  We start by expressing $u$ in terms of the roots $\tau_{\pm}$. We have
\begin{equation}
    (x - p \tau)^2 = 0 \implies -ru - 2\tau (p\cdot q r + p\cdot n u) - \tau^2 = 0
\end{equation}
so 
\begin{equation}
    u = \frac{2 \tau r p\cdot q + \tau^2}{-r - 2 \tau p\cdot n} = -\frac{\tau(\tau + 2r p\cdot q)}{r + 2 p\cdot n \tau}.
\end{equation}
This holds for $\tau = \tau_{\pm}$. Furthermore,
\begin{equation}
    \tau_{\pm} = -p\cdot x \pm \frac{\Delta}{2}
\end{equation}
so 
\begin{equation}
\begin{split}
   \pm \Delta &= 2 (\tau_{\pm} + r p \cdot q + u p \cdot n) = 2 \big(\tau_{\pm} + r p \cdot q -\frac{\tau_{\pm}(\tau_{\pm} + 2r p\cdot q)}{r + 2 p\cdot n \tau_{\pm}} p \cdot n\big)\\
   &= 2 \frac{\tau_{\pm}r + p \cdot n \tau_{\pm}^2 + r^2 p\cdot q}{r + 2 p\cdot n \tau_{\pm}}.
   \end{split}
\end{equation}
We also note that 
\begin{equation}
    \frac{du}{d\tau} = -\frac{2r \tau + 2 p\cdot n \tau^2 + 2 r^2 p\cdot q}{(r + 2p\cdot n \tau)^2}.
\end{equation}

With this preparation, we next show that 
\begin{equation}
\label{eq:delta-exp}
    \delta^{(j)}(\tau_{\pm}(u,r)) = \sum_{k = 0}^j r^{-j+k} a_{jk} \delta^{(k)}(u),
\end{equation}
where $a_{jk}$ are independent on $u, r$. We first observe that
\begin{equation}
    u = \tau_{\pm} T(\epsilon \tau_{\pm}) \equiv g(\tau_{\pm}), \quad \epsilon \equiv \frac{1}{r}
\end{equation}
with 
\begin{equation}
    T(0) \neq 0. 
\end{equation}
Then integrating the delta function against a test function
\begin{equation}
\label{eq:test-fct}
    \int du f(u) \delta^{(j)}(\tau) = \int d\tau g'(\tau) f(g(\tau)) \delta^{(j)}(\tau) = (-1)^j\p_{\tau}^j\left( g'(\tau) f(g(\tau))\right) \left. \right|_{\tau = 0}.
\end{equation}
From here, we see that that the expansion of $\delta^{(j)}(\tau)$ in terms of $\delta^{(k)}(u)$ can only contain term with $k \leq j$. Would-be higher powers contribute terms $\propto \tau^{k - j}$ to \eqref{eq:test-fct} and vanish at $\tau = 0.$ Furthermore, the coefficient of the $k$-th term in the expansion is given by
\begin{equation}
    (-1)^j \p_{\tau}^j \left( g'(\tau) \frac{g(\tau)^k}{k!}\right) \left. \right|_{\tau = 0} = (-1)^j \p_{\tau}^j\left( \sum_{m} \frac{c_m}{k!}(\epsilon \tau)^m \tau^k \right)\left. \right|_{\tau = 0} = (-1)^j \frac{c_{j-k}}{k!} \epsilon^{j - k}, 
\end{equation}
where we used that $g'(\tau) = 1$ at $\tau = 0$. This identity confirms \eqref{eq:delta-exp}. 

Finally, we show that this identity implies that the delta function contact term in $F_{ru}$ cannot contribute at $\mathcal{O}(r^{-2n-2})$. We have 
\begin{equation}
    F_{ru} \propto (-p\cdot q r + u p\cdot n) \sum_{j = 0}^{n} a_j \frac{\delta^{(j)}(\tau_{\pm})}{\Delta^{2n - j + 2}} \propto (-p\cdot q r + u p\cdot n) \sum_{k = 0}^nA_k \frac{\delta^{(k)}(u)}{r^{2n - k + 2} },
\end{equation}
where we have used that 
\begin{equation}
    \Delta = \pm 2r p\cdot q, \quad {\rm at} \quad \tau_{\pm} = 0.
\end{equation}
It is then clear that the first term gives all overleading contributions at large $r$, while the only contribution from the second term is 
$\propto u \delta(u) = 0$. We conclude that all delta function contact terms in $F_{ru}$ appear at orders overleading to $r^{-2n-2}$, thereby confirming the first equation in \eqref{eq:rdm2}.

In the large-$r$ expansion of $F_{Au}$ we are interested in the $\mathcal{O}(r^{-2n})$ term. We immediately see that the $\Theta$ function term is subleading by $r^{-1}$, while the full tower of contact terms can contribute at this order. To extract the coefficient, we integrate the distribution against a test function
\begin{equation}
 \sum_{j = 0}^n a_j \int du \frac{\delta^{(j)}(\tau(u))}{\Delta(\tau(u))^{2n - j + 2}} f(u)=    \sum_{j = 0}^n a_j \int d\tau \frac{\delta^{(j)}(\tau)}{\Delta^{2n - j + 2}} \frac{du}{d\tau} f(u(\tau)) = \sum_{j = 0}^n a_j \p_{\tau}^j\left(\frac{(-1)^j}{\Delta^{2n - j + 2}} \frac{du}{d\tau} f(u(\tau)) \right)\left. \right|_{\tau = 0}.
\end{equation}
It clear from the previous discussion (cf. eq. \eqref{eq:delta-exp}), that the coefficient of the $\mathcal{O}(r^{-2n})$ term in $F_{uA}$ will multiply a $\delta(u)$, so we just need to compute
\begin{equation}
    \sum_{j = 0}^n a_j j! \oint_{\tau} d\tau \tau^{-j-1} \left(\frac{(-1)^j}{\Delta^{2n - j + 2}} \frac{du}{d\tau} \right) =  \sum_{j = 0}^n a_j j! \oint_{\tau} d\tau \tau^{-j-1} (-1)^j (2\tau r + 2p\cdot n \tau^2 + 2r^2 p\cdot q)^{-2n+j-1} (r + 2p\cdot n \tau)^{2n -j}.
\end{equation}
Changing variables
\begin{equation}
    y = \frac{\tau}{r}
\end{equation}
we find
\begin{equation}
\begin{split}
   & \sum_{j = 0}^n a_j j! \oint_{\tau} d\tau \tau^{-j-1} (-1)^j (2\tau r + 2p\cdot n \tau^2 +2 r^2 p\cdot q)^{-2n+j-1} (r + 2p\cdot n \tau)^{2n -j} \\
    &= r^{-2n-2}  \sum_{j = 0}^n a_j j! \underbrace{\oint_{y} dy y^{-j-1} (-1)^j (2 y + 2p\cdot n y^2 +  2p\cdot q)^{-2n+j-1} (1 + 2p\cdot n y)^{2n -j}}_{C_j}.
    \end{split}
\end{equation}

For simplicity, let
\begin{equation}
L\equiv 2n-j+1 .
\end{equation}
Then
\begin{equation}
C_j=\oint dy y^{-j-1} (2 p\cdot q)^{-L}
\left(1+{y+ p \cdot n y^2\over p\cdot q}\right)^{-L}
(1+2 p\cdot ny)^{L-1}.
\end{equation}
The Taylor expansion of the first factor is
\begin{equation}
\left(1+{y+ p\cdot n y^2\over p\cdot q}\right)^{-L}
=
\sum_{\ell=0}^{j}
(-1)^\ell{(L)_\ell\over \ell!}
(p\cdot q)^{-\ell}y^\ell(1+p\cdot n y)^\ell,
\end{equation}
where the upper sum is truncated because we are only interested in the residue and we defined
\begin{equation}
    (L)_{\ell} = \frac{(L+\ell-1)!}{(L-1)!}.
\end{equation} 
Relabeling indices, we have for outgoing sources
\begin{equation}
C_j= {1\over 2^L (p\cdot q)^{2n+1}}
\sum_{k=0}^{j}
(-1)^{j-k}((p\cdot n)(p\cdot q))^k
{(L)_{j-k}\over (j-k)!}
\oint dx x^{-k-1} (1+x)^{j-k}(1+2x)^{L-1}.
\end{equation}
Then
\begin{equation}
\sum_{j = 0}^n a_j j!C_j =
{1\over (2 p\cdot q)^{2n+1}}
\sum_{k=0}^{n}
4^k\,{(2n-k)!\over (n-k)!}\,(p\cdot q p\cdot n)^k,
\end{equation}
where we have used the Vandermode identity
\begin{equation}
    \sum_{j = k}^n (-1)^{j - k} 2^j \frac{2n - j + 1}{(j+1)(n - j)!(j - k)!} \oint dx x^{-k-1} (1+x)^{j - k}(1 + 2x)^{2n - j} = \frac{4^k}{(n - k)!}.
\end{equation}
Setting
\begin{equation}
    z = -(2 p\cdot q) (p\cdot n)
\end{equation}
we obtain precisely the second equation in \eqref{eq:rdm2}.

\section{Massive sources in odd dimensions}
\label{app:massive-odd}

In this appendix we present additional details on the derivations and results presented in Section \ref{sec:odd-massive}. 

\subsection{Proof of eq. \eqref{eq:Jfs}}
\label{eq:idoddm}

We start with 
\begin{equation}
\label{eq:Jfs-app}
    p_{[\nu} \p_{\mu]} J(x,p) =  (x_{\mu}p_{\nu} - x_{\nu} p_{\mu})\frac{2}{d - 4}\left[\left(4 - d\right) \frac{(-x^2)^{1 - \frac{d}{2}}}{\Delta} H(z) + 4\frac{(-x^2)^{2 - \frac{d}{2}}}{\Delta^3} H(z)+ 4\frac{(x\cdot p)(-x^2)^{2 - \frac{d}{2}}}{\Delta^4} H'(z)\right],
\end{equation}
where $J$ and $H$ were defined in \eqref{eq:J} and \eqref{eq:H}. 

This expression can be readily simplified 
observing that
\begin{equation}
    z = \frac{x\cdot p + \frac{\Delta}{2}}{\Delta}, \quad -\frac{x^2}{\Delta^2} = z (z - 1), \quad \Delta = \frac{2 x\cdot p}{2z - 1}.
\end{equation}
and using the contiguous hypergeometric function identity
\begin{equation}
    \left[
q +2z(z-1)+(2z-1)z(z-1)\frac{d}{dz}
\right]
{}_2F_1(1,2q;q+1;z)
=
q\,{}_2F_1(1,2q-2;q;z), \quad q = (2 - \frac{d}{2})
\end{equation}
as follows
\begin{equation}
\begin{split}
    p_{[\nu} \p_{\mu]} J(x,p) &=  (x_{\mu}p_{\nu} - x_{\nu} p_{\mu})\frac{2}{d - 4} \frac{(-x^2)^{1-\frac{d}{2}}}{\Delta}\left[\left(4 - d\right)  H(z) + 4\frac{(-x^2)}{\Delta^2} H(z)+ 4\frac{(x\cdot p)}{\Delta}\frac{(-x^2)}{\Delta^2} H'(z)\right]\\
    &=(x_{\mu}p_{\nu} - x_{\nu} p_{\mu})\frac{4}{d - 4} \frac{(-x^2)^{1-\frac{d}{2}}}{\Delta}\left[\left(2 - \frac{d}{2}\right)  + 2z(z-1)+ (2z-1)z(z-1) \frac{d}{dz} \right]H(z)\\
    &=(x_{\mu}p_{\nu} - x_{\nu} p_{\mu})\frac{4}{d - 4} \frac{(-x^2)^{1-\frac{d}{2}}}{\Delta}(2-\frac{d}{2}){}_2 F_1\left[1, 2 - d; 2 - \frac{d}{2}; z \right]\\
    &=2(x_{\mu}p_{\nu} - x_{\nu} p_{\mu})\frac{(\tau_+\tau_-)^{1-\frac{d}{2}}}{\tau_--\tau_+}{}_2 F_1\left[1, 2 - d; 2 - \frac{d}{2}; \frac{\tau_-}{\tau_- - \tau_+} \right].
\end{split}
\end{equation}

\subsection{Solutions to Maxwell's equations}
\label{sec:sol-odd}

In this appendix we show that the field strength \eqref{eq:reg-mass-odd} obeys the free Maxwell equations away from the source trajectories and $x^2 = 0$. 

Start with the regularized field strength sourced by $M$ incoming and $N$ outgoing particles of constant momenta $p_i$ in odd $d$,
\begin{equation}
\label{eq:fsa}
    \begin{split}
    F_{\mu\nu} &= -\frac{(-1)^{\frac{d - 1}{2}}}{2 \pi^{\frac{d}{2}}}\Gamma\left( \frac{d - 2}{2} \right)\Big[ \sum_{i = 1}^{M+N} \Theta(-x^2) 2 Q_i \frac{(\tau_+ \tau_-)^{1 - \frac{d}{2}}}{\tau_- - \tau_+} {}_2 F_1\left[1, 2 - d; 2 - \frac{d}{2}; \frac{\tau_-}{\tau_- - \tau_+} \right](x_{\mu}p_{i,\nu} - x_{\nu} p_{i,\mu}) \\
    &+ \sum_{i = 1}^M 2(2 - d) Q_i  B(2 - \frac{d}{2}, d - 2) (\tau_- - \tau_+)^{1 - d}  (x_{\mu}p_{i,\nu} - x_{\nu} p_{i,\mu}) \Big].
    \end{split}
\end{equation}
The field strength depends on the following functions
\begin{equation}
    f_i(\tau_-,\Delta) = \frac{((\tau_- -\Delta) \tau_-)^{1 - \frac{d}{2}}}{\Delta} {}_2 F_1\left[1, 2 - d; 2 - \frac{d}{2}; \frac{\tau_-}{\Delta} \right],\quad f_o(\tau_-,\Delta) = \Delta^{1-d},\quad \Delta = \tau_- - \tau_+.
\end{equation}
In order to compute the divergence of \eqref{eq:fsa} for $\nu = u$, we will again use the identities \eqref{eq:identities-del-tpm} relevant for the even $d$ case, which we repeat here for convenience
\begin{equation}
\begin{split}
p \cdot \p \tau_{\pm} &= 1, \quad ~~ p \cdot \p \Delta = 0, \\
    x\cdot \p \tau_{\pm} &= \tau_{\pm}, \quad x \cdot \p \Delta = \Delta.
\end{split}
\end{equation}
These imply that all functions $f(\tau_-, \Delta)$ obey the identities 
\begin{equation}
    \partial^\mu \left[ f(\tau_-,\Delta) (x_\mu p_u-x_u p_\mu)\right] = \partial^\mu(x_\mu p_u-x_\nu p_\mu) f(\tau_-,\Delta) + p_u x\cdot \partial f - x_u p\cdot \partial f,
\end{equation}
where
\begin{equation}
    x\cdot\partial f= (x \cdot \partial \tau_-) \partial_{\tau_-} f + (x \cdot \partial \Delta)\partial_\Delta f= (\tau_- \partial_{\tau_-}+\Delta\partial_\Delta)f,\quad p\cdot \partial f = \partial_{\tau_-} f.
\end{equation}
Next, using the identities
\begin{equation}
    \begin{split}
       \partial^\mu(x_\mu p_u -x_u p_\mu) &= (d - 1)p_{u}, \\
      ( \tau_- \p_{\tau_-} + \Delta \p_{\Delta}) f_i &= -(d-1) f_i, \\
       ( \tau_- \p_{\tau_-} + \Delta \p_{\Delta}) f_o &= -(d-1) f_o,
    \end{split}
\end{equation}
we find
\begin{equation}
\label{eq:int-eq}
    \begin{split}
         \partial^\mu \left[ f_o(\tau_-,\Delta) (x_\mu p_u-x_u p_\mu)\right] &= 0, \\
          \partial^\mu \left[f_i(\tau_-,\Delta) (x_\mu p_u-x_u p_\mu)\right] &= -x_u \p_{\tau_-} f_i.
    \end{split}
\end{equation}
We note that these results and the conclusion below hold strictly away from the loci where  $f_i, f_o$ diverge, namely away from the source trajectories, as well as the massless limit $\Delta \rightarrow 0, \rho \rightarrow 0$ where both $f_i, f_o$ clearly diverge. By construction, the field strength \eqref{eq:fsa} obeys the Maxwell equations up to delta-function source terms which will be restored in Section \ref{sec:massive} and shown to be responsible for the non-vanishing memory effect.

Substituting \eqref{eq:int-eq} into \eqref{eq:fsa}, we obtain
\begin{equation*}
   \begin{split}
    \partial^\mu F_{\mu u}
    &= \frac{(-1)^{\frac{d + 1}{2}}}{2 \pi^{\frac{d}{2}}}\Gamma\left( \frac{d - 2}{2} \right)\Big[ \sum_{i = 1}^{M+N} \Theta(-x^2) 2 Q_i  \left[ \partial_{\tau_-}f_i(\tau_-,\Delta) \right] x_u \Big].
    \end{split}
\end{equation*}
We finally compute $\partial_{\tau_-}f_i(\tau_-,\Delta)$,
\begin{equation}
\begin{split}
    \partial_{\tau_-}f_i(\tau_-,\Delta) &= \frac{1}{\Delta^2}(\frac{d}{2} - 1)\left(\tau_- (\tau_- - \Delta) \right)^{-\frac{d}{2}}\\
    &\times \left( \Delta(\Delta - 2 \tau_-){}_2F_1\left[1, 2 - d; 2 - \frac{d}{2}; \frac{\tau_-}{\Delta}\right] + \frac{4\tau_-(\tau_- - \Delta)}{d - 4}{}_2F_1\left[2, 3 - d; 3 - \frac{d}{2};\frac{\tau_-}{\Delta} \right] \right),
    \end{split}
\end{equation}
whereas for odd positive integer $d$, 
\begin{equation}
\label{eq:last-id}
  \Delta(\Delta - 2 \tau_-){}_2F_1\left[1, 2 - d; 2 - \frac{d}{2}; \frac{\tau_-}{\Delta}\right] + \frac{4\tau_-(\tau_- - \Delta)}{d - 4}{}_2F_1\left[2, 3 - d; 3 - \frac{d}{2};\frac{\tau_-}{\Delta} \right] = \Delta^2. 
\end{equation}
The last identity \eqref{eq:last-id} follows from 
\begin{equation}
    (1 - 2z) F(z) + \frac{2z(z - 1)}{d - 2} F'(z) = 1,
\end{equation}
obeyed by 
\begin{equation}
    F(z) = {}_2 F_1\left[1, 2 - d; 2 - \frac{d}{2}; z \right],
\end{equation}
which is a consequence of the fact that $F(z)$ solves the hypergeometric equation
\begin{equation}
    z(1 - z) F'' + \left(2 - \frac{d}{2} - (4 - d)z \right) F' + (d - 2) F = 0.
\end{equation}

We conclude that
\begin{equation*}
\begin{split}
    \partial^\mu F_{\mu u} 
    &= \frac{(-1)^{\frac{d + 1}{2}}}{ \pi^{\frac{d}{2}}}\Gamma\left( \frac{d - 2}{2} \right)\Big[  \Theta(-x^2) \left[ \left( \frac{d}{2}-1 \right) \left( -x^2 \right)^{-\frac{d}{2}} \right] x_u  \sum_{i = 1}^{M+N}  Q_i  \Big] =0
\end{split}
\end{equation*}
due to charge conservation.

\bibliographystyle{utphys}
\bibliography{references}

\end{document}